\documentclass[10pt,conference]{IEEEtran}

\usepackage{amsmath,amsfonts}

\usepackage{hyperref}
\usepackage{subcaption}
\usepackage{adjustbox}
\newsavebox{\mybox}

\newcommand{\revision}[1]{\textcolor{black}{#1}}

\usepackage{amsthm}
\usepackage{mdframed}
\usepackage{centernot}
\usepackage{comment}

\usepackage{wrapfig,lipsum,booktabs}

\newcommand{\Aa}{\mathcal{A}}

\newcommand{\Dd}{\mathcal{D}}

\newcommand\vd[2]{d_{i, p}}
\newcommand{\set}[1]{\left\{ #1 \right\}}

\usepackage{tikz}

\newtheorem{definition}{Definition}[section]

\definecolor{gold}{rgb}{0.99,0.78,0.07}
\usetikzlibrary{arrows,shapes,snakes,automata,backgrounds,positioning,decorations.pathmorphing,
	decorations.markings,calc}

\tikzstyle{dtreenode}=[draw=blue!10!gray,rounded rectangle, minimum size=5mm,fill=blue!10!white]
\tikzstyle{dtreeleaf}=[draw=black!60,minimum width=1cm,minimum height=0.4cm,rectangle,fill=blue!50!white]
\tikzset{every loop/.style={looseness=7}}
\tikzset{
	gluon/.style={decorate,draw=black,
		decoration={coil,amplitude=1pt, segment length=5pt}}
}
\tikzset{
	gluon1/.style={decorate,draw=black,
		decoration={coil,amplitude=3pt, segment length=3pt}}
}
\tikzset{
	gluonew/.style={decorate,draw=black,
		decoration={coil,amplitude=1pt, segment length=2pt}}
}

\tikzset{bicolor/.style args={#1 and #2 and #3}{
		path picture={
			\tikzset{rounded corners=0}
			\fill [#1] (path picture bounding box.south west)
			rectangle
			($(path picture  bounding box.north west)!#3!(path picture bounding
			box.north east)$);
			\fill [#2]
			($(path picture bounding box.south west)!#3!(path picture bounding
			box.south east)$)
			rectangle (path picture bounding box.north east);
}}}

\tikzset{tricolor/.style args={#1 and #2 and #3 and #4 and #5}{
		path picture={
			\tikzset{rounded corners=0}
			\fill [#1] (path picture bounding box.south west)
			rectangle
			($(path picture  bounding box.north west)!#4!(path picture bounding
			box.north east)$);
			\fill [#2]
			($(path picture bounding box.south west)!#4!(path picture bounding
			box.south east)$)
			rectangle
			($(path picture  bounding box.north west)!#5!(path picture bounding
			box.north east)$);
			\fill [#3]
			($(path picture bounding box.south west)!#5!(path picture bounding
			box.south east)$)
			rectangle (path picture bounding box.north east);
}}}

\usepackage[many]{tcolorbox}
\tcbuselibrary{listings,skins}

\lstdefinestyle{mystyle}{
  xleftmargin=0pt,
   basicstyle={\footnotesize\ttfamily},
   aboveskip=3mm,
   belowskip=3mm,
   keywordstyle=\bfseries,
   showstringspaces=false,
  escapechar=?,
  language=Java
}
\definecolor{code_indent}{HTML}{CCCCCC}

\newtcblisting{mylisting}[2][]{
  arc=0pt, outer arc=0pt,
  listing only,
  listing style=mystyle,
  title={\large #2},
  #1
}

 \definecolor{dkgreen}{rgb}{0,0.6,0}
 \definecolor{gray}{rgb}{0.5,0.5,0.5}
 \definecolor{mauve}{rgb}{0.58,0,0.82}

\definecolor{cadmiumgreen}{rgb}{0.0, 0.42, 0.24}
\definecolor{verde}{rgb}{0.25,0.5,0.35}
\definecolor{jpurple}{rgb}{0.5,0,0.35}
\definecolor{darkgreen}{rgb}{0.0, 0.2, 0.13}

\usepackage[ruled,vlined,linesnumbered,lined]{algorithm2e}
\usepackage{algpseudocode}

\usepackage{mathtools,xparse}

\usepackage{tabu}
\usepackage{multirow}

\usepackage{pifont}
\usepackage{enumerate}
\usepackage[shortlabels]{enumitem}

\usepackage{pgfplotstable}
\usepackage{pgfplots}

\usepackage[noframe]{showframe}
\usepackage{framed}

 {\endMakeFramed}
 \definecolor{shadecolor}{gray}{0.85}

\definecolor{bgblue}{RGB}{245,243,253}
\definecolor{ttblue}{RGB}{91,194,224}

\mdfdefinestyle{mystyle}{%
  rightline=true,
  innerleftmargin=10,
  innerrightmargin=10,
  outerlinewidth=3pt,
  topline=false,
  rightline=false,
  bottomline=false,
  skipabove=\topsep,
  skipbelow=false%\topsep
}

\newtcolorbox{myboxi}[1][]{
  breakable,
  title=#1,
  colback=white,
  colbacktitle=white,
  coltitle=black,
  fonttitle=\bfseries,
  bottomrule=0pt,
  toprule=0pt,
  leftrule=3pt,
  rightrule=3pt,
  titlerule=0pt,
  arc=0pt,
  outer arc=0pt,
  colframe=black!50,
}

\newtcolorbox{myboxii}[1][style=mystyle]{
  breakable,
  freelance,
  colback=white,
  colbacktitle=white,
  coltitle=black,
  fonttitle=\bfseries,
  bottomrule=0pt,
  boxrule=0pt,
  colframe=white,
  after skip=0pt,
  overlay unbroken and first={
    \draw[white!75!black,line width=3pt]
    ([yshift=-9pt]frame.north west) --
    ([yshift=9pt]frame.south west);
  },
  }
\usepackage{colortbl}

\newcommand{\greencheck}{}%
\DeclareRobustCommand{\greencheck}{%
  \tikz\fill[scale=0.4, color=green]
  (0,.35) -- (.25,0) -- (1,.7) -- (.25,.15) -- cycle;%
}

\newcommand{\redx}{}%
\DeclareRobustCommand{\redx}{%
\tikz[scale=0.23, color=red] {
    \draw[line width=0.7,line cap=round] (0,0) to [bend left=6] (1,1);
    \draw[line width=0.7,line cap=round] (0.2,0.95) to [bend right=3] (0.8,0.05);
}}

\begin{document}

\title{Fairness Testing through Extreme Value Theory}

\author{
\IEEEauthorblockN{Verya Monjezi\footnote{}}
\IEEEauthorblockA{
vmonj@uic.edu \\
University of Illinois Chicago}
\\
\IEEEauthorblockN{Vladik Kreinovich}
\IEEEauthorblockA{
vladik@utep.edu \\
University of Texas at El Paso}
\and
\IEEEauthorblockN{Ashutosh Trivedi}
\IEEEauthorblockA{
ashutosh.trivedi@colorado.edu\\
University of Colorado Boulder}
\\
\IEEEauthorblockN{Saeid Tizpaz-Niari}
\IEEEauthorblockA{
saeid@uic.edu \\
University of Illinois Chicago}
}

\IEEEtitleabstractindextext{
\begin{abstract}
Data-driven software is increasingly being used as a critical component of automated decision-support systems.
Since this class of software learns its logic from historical data, it can encode or amplify discriminatory practices. Previous research on algorithmic fairness has focused on improving ``average-case'' fairness. 
On the other hand,  fairness at the extreme ends of the spectrum, which often signifies lasting and impactful shifts in societal attitudes, has received significantly less emphasis.

Leveraging the statistics of extreme value theory (EVT),
we propose a novel fairness criterion called \emph{extreme counterfactual discrimination} (ECD). This criterion estimates the worst-case amounts of disadvantage in outcomes for individuals solely based on their memberships in a protected group. Utilizing tools from search-based software engineering and generative AI, we present a randomized algorithm that samples a statistically significant set of points from the tail of ML outcome distributions even if the input dataset lacks a sufficient number of relevant samples. 

We conducted several experiments on four ML models (deep neural networks, logistic regression, and random forests) over 10 socially relevant tasks from the literature on algorithmic fairness. First, we evaluate
the generative AI methods and find that they generate sufficient samples to infer valid EVT distribution in 95\% of cases. Remarkably, we found that the prevalent bias mitigators reduce the average-case discrimination but increase the worst-case discrimination significantly in \revision{35\%} of cases. We also observed that even the tail-aware mitigation algorithm---MiniMax-Fairness---increased the worst-case discrimination in \revision{30\%} of cases. We propose a novel ECD-based mitigator that improves fairness in the tail in \revision{90\%} of cases with no degradation of the average-case discrimination. We hope that the EVT framework serves as a robust tool for evaluating fairness in both average-case and worst-case discrimination.
\end{abstract}
}

\maketitle
\IEEEdisplaynontitleabstractindextext

\section{Introduction}
\label{sec:intro}
Recent technological advancements in training large machine learning (ML) models, such as deep neural networks~\cite{goodfellow2016deep}, deep reinforcement learning~\cite{sutton2018reinforcement}, and large language models~\cite{devlin2018bert,chatgpt}, have led to a proliferation of data-driven software in almost every aspect of modern socioeconomic infrastructure. 
These data-driven systems, such as those that decide on recidivism~\cite{compas-article}, predict benefit eligibility~\cite{elyounes2020computer,Ranchords2021AutomatedGF}, or decide whether to audit a given taxpayer~\cite{dorothyIRS,tizpaz2023metamorphic}, learn their decision logic as ML models by mining simple patterns from historical data. 
However, these systems often codify and amplify the biases present in the historical data due to various systemic factors.
To address this challenge, the software engineering community has developed solutions to characterize, quantify, and mitigate bias in the ML models. We discuss their inadequacies and propose new tools and techniques for the tail of outcome distributions of data-driven software. 

\vspace{0.25em}
\noindent\textbf{Inadequacies of Average-Case Fairness.}
Although there is an increased participation of minorities (e.g., women) in the labor market (parity in average), they are considerably underrepresented in high-paying occupations and leadership positions~\cite{petrongolo2019gender} (disparity in the extreme). Additionally, the wage gap between privileged and unprivileged individuals continues to be more pronounced in high-paying jobs \cite{anderson1996racial,petrongolo2019gender}.
Considering these factors, it is indeed surprising that a notable gap exists in the literature regarding the evaluation of algorithmic fairness in the context of extreme outcomes. 

One broad class of fairness definitions is individual fairness~\cite{dwork2012fairness} which requires treating individuals similarly if they are deemed similar based on their non-protected attributes, regardless of their protected attributes. 
One popular individual fairness notion is \emph{counterfactual discrimination} which necessitates that algorithmic outcomes should be similar for an individual and any related counterfactual individual who differs only in protected attributes.
However, these fairness notions primarily focus on the average behavior (expected value or variance) of the model, which can create a false sense of fairness by ignoring the discrimination in socially influential edge cases. 
This paper presents a framework rooted in EVT to quantify AI fairness within the tail of ML outcomes.

\vspace{0.25em}
\noindent\textbf{Statistics of the Extreme: Extreme Value Theory.}
While statistics and machine learning typically focus on ``usual'' behavior, extreme value theory (EVT) \cite{coles2001introduction} is a branch of statistics that deals with unusual or extreme behaviors. 
EVT can be applied to model rare events such as the maximum temperature in the summer. Under appropriate assumptions, the statistics of extreme values follow the generalized extreme value (GEV) distribution, which is analogous to the central limit theorem for the statistics of averages or expected values. 

\vspace{0.25em}
\noindent\textbf{Fairness through Extreme Value Theory.} 
The primary focus of this paper centers on a narrow view of \emph{equality of opportunity}, which necessitates similar individuals to be treated similarly at the time of decision-making, as defined by Dwork et al.~\cite{dwork2012fairness}. 
We consider the distribution of ``counterfactual discrimination,'' which refers to the distribution of differences in the ML outcomes when a protected attribute, like race or gender, is altered from observed value A to a counterfactual B.
While previous studies have focused on the expected values from this distribution known as \revision{average causal discrimination} (ACD)~\cite{10.1145/3106237.3106277}, representing the ``average'' change in the outcome when a protected attribute is flipped,
this work quantifies the ``maximum'' change in the ML outcome when a protected attribute is flipped. We call this quantity \emph{extreme counterfactual discrimination} (ECD) and use GEV distributions to model and quantify it. 
By comparing the GEV distributions of different (sub-)groups, we quantify the fairness of ML models in the extreme tail of outcome distributions as well as the effectiveness of mitigation algorithms in reducing discrimination in the tail.

The  ECD metric, proposed in this paper, has a normative implication.
It tells us ``in the worst-case, how much (dis)advantages are experienced by individuals solely due to their memberships in (un)privileged groups at the time of ML decision-making.'' Our proposal complements \textsc{Themis}~\cite{10.1145/3106237.3106277} that shows the amounts of such discrimination on average. For example, an ACD of +0.05 vs. an ECD of +0.25
for an unprivileged group show that flipping their protected attributes to a privileged group increased their likelihood of receiving
favorable ML outcomes by 5\% on average, but up to 25\% in the worst-case. The statistics of EVT allow us to directly model GEV distributions that bring significant advantages. 
It directly models the tail distribution,
which allows us to investigate the validity of the tail or provide statistical guarantees on the returns/likelihood of extreme discrimination. Other metrics, like the conditional value at risk (CVaR)~\cite{williamson2019fairness}, model the tail of a usual distribution (e.g., normal distributions). Hence, they fail to reason about the validity of the tail and provide any statistical guarantees on extrapolations.

\noindent\textbf{Statistical ECD-Testing Framework.} 
In this paper, we have developed a randomized test-case generation algorithm that explores the tail of ML models and applies the exponentiality test~\cite{diebolt2007goodness,abella2017measurement} to convince statistical significance. The primary challenge stems from the statistical test's requirements for a certain size of tail samples and the scarcity of samples in the extreme tail (which is precisely why they are considered extreme). If only a subset of these tail samples is included in the analysis, it can result in low confidence in the model due to high variance. On the other hand, selecting a larger number of data points will lead to the erroneous inclusion of non-tail samples and the inference of mixture distributions that violate the asymptotic basis of extreme value theory. Rather than randomly generating the test-cases from the domain of variables~\cite{10.1145/3106237.3106277}, we leverage and evaluate various generative methods, such as GANs~\cite{CTGAN} and VAEs~\cite{8285168}, to synthesize samples with realistic combinations of features. 

\vspace{0.25em}
\noindent\textbf{Experiments.}
We conducted experiments on nine fairness-sensitive datasets with four popular classifiers (an overall 40 training scenarios). Our findings indicate that EVT fits well to the tail of counterfactual bias distributions in 95\% of cases that enable us to derive worst-case guarantees. 
In 25\% of scenarios, the worst-case and average-case CD differ significantly across different groups. We also evaluated the characteristics of fairness in the tail over \revision{four} mitigation algorithms: exponentiated gradient (EG)~\cite{agarwal2018reductions}, Fair-SMOTE~\cite{10.1145/3468264.3468537}, \revision{MAAT~\cite{10.1145/3540250.3549093}, and STEALTH~\cite{10109333}}.
Our results over the mitigated model show that the worst-case and average-case CD differ significantly across different groups in 52\% of cases.
In addition, the average-based mitigated models significantly increase worst-case discrimination in \revision{35\% of the cases, while preserving or improving average fairness in 63\% of cases. With tail-based methods, we implement an in-process mitigation strategy that outperforms MiniMax-Fairness~\cite{10.1145/3461702.3462523} and reduces the discrimination in the tail for 90\% of cases while improving average fairness in 45\% of cases.}

% 26\% of cases and only decrease it in 30\% of cases, while in tail-based, MiniMax-Fairness performs slightly better results with 22\% and 30\%. To improve fairness in the tail, we implement an in-process mitigation strategy with success in 82\% of cases.
\vspace{0.25em}
\noindent\textbf{Contributions.} In this paper, we
\begin{itemize}
    \item introduce a metric to measure unfairness in the tail of ML outcome distributions,
    \item present a fairness testing method that generates realistic test-cases and provides statistical guarantees in the tail,
    \item evaluate the worst-case discrimination for a large set of well-established algorithms and bias mitigators, and
    \item propose and evaluate a novel tail-aware mitigator. 
\end{itemize}

\section{Overview}
\label{sec:overview}
We first give a background overview for extreme value theory. We then go through our approach step by step using an example of adult census income dataset, trained using a DNN algorithm.

\vspace{0.25em}
\noindent \textbf{Extreme Value Theory.} Given a set of independent and identically distributed random variables $\set{z_1,\ldots,z_n}$, the extreme
value theory is concerned with the max statistics of a random process, i.e., $M_n = \max(\set{z_1,\ldots,z_n})$. Under some mild assumptions, it has been proved (e.g., see Leadbetter et al.~\cite{leadbetter2012extremes}) that 
$M_n$ belongs to a family of distributions called the \emph{generalized extreme value (GEV)}. There are two basic approaches to infer the parameters of GEV distributions: block maximum and threshold approach~\cite{coles2001introduction}. In this paper, we use the threshold approach where extreme events that exceed some high threshold $u$, i.e., $\{z_i: z_i > u\}$, are extreme values. The GEV distribution has three parameters: a location parameter, a scale parameter, and a shape parameter. When the shape is close to zero or negative, the statistical guarantees on the worst-case discrimination may be feasible. 

\noindent \textit{Threshold Selection.}
A proper choice of threshold value $u$ is critical to analyze the behavior of GEV.
Low values of threshold $u$ might include non-tail samples and lead to mixture distributions that violate the asymptotic basis of the model. 
On the other hand, high values of threshold $u$ might include only a few tail samples and lead to low confidence in the model due to high variance.
In this work, we use coefficient of variation (CV) and provide statistical guarantees
in picking thresholds. 

\noindent \textit{Return Level.}
A \textit{return level} describes by the set of points
{$(m,\delta_m)$} where $m$ is the time period (e.g., the number of queries to the ML software) and the level $\delta_m$ is expected to observed during the $m$ period (e.g., maximum discrimination after $m$ interactions).

\begin{figure*}
    \centering
    \includegraphics[width = 0.24\textwidth]{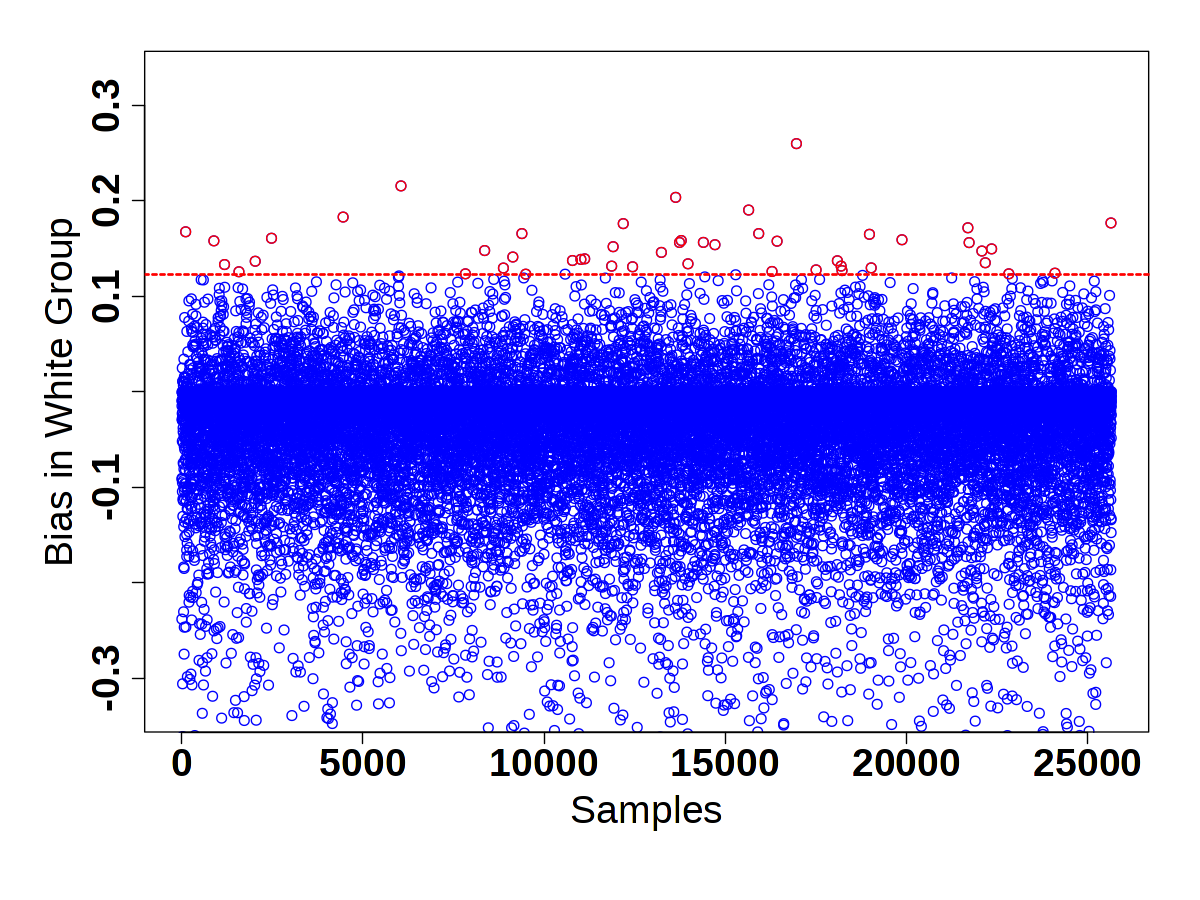}
    \includegraphics[width = 0.24\textwidth]{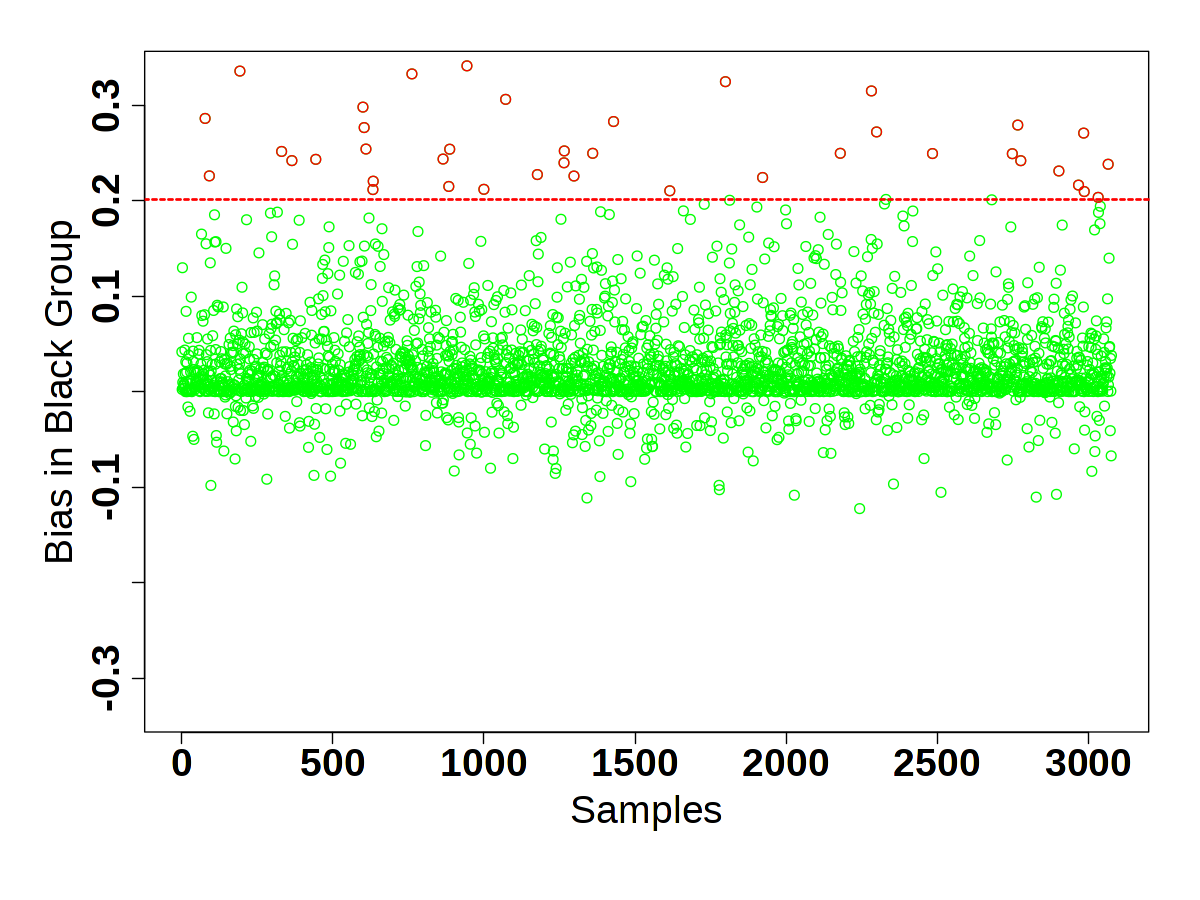}
    \includegraphics[width = 0.24\textwidth]{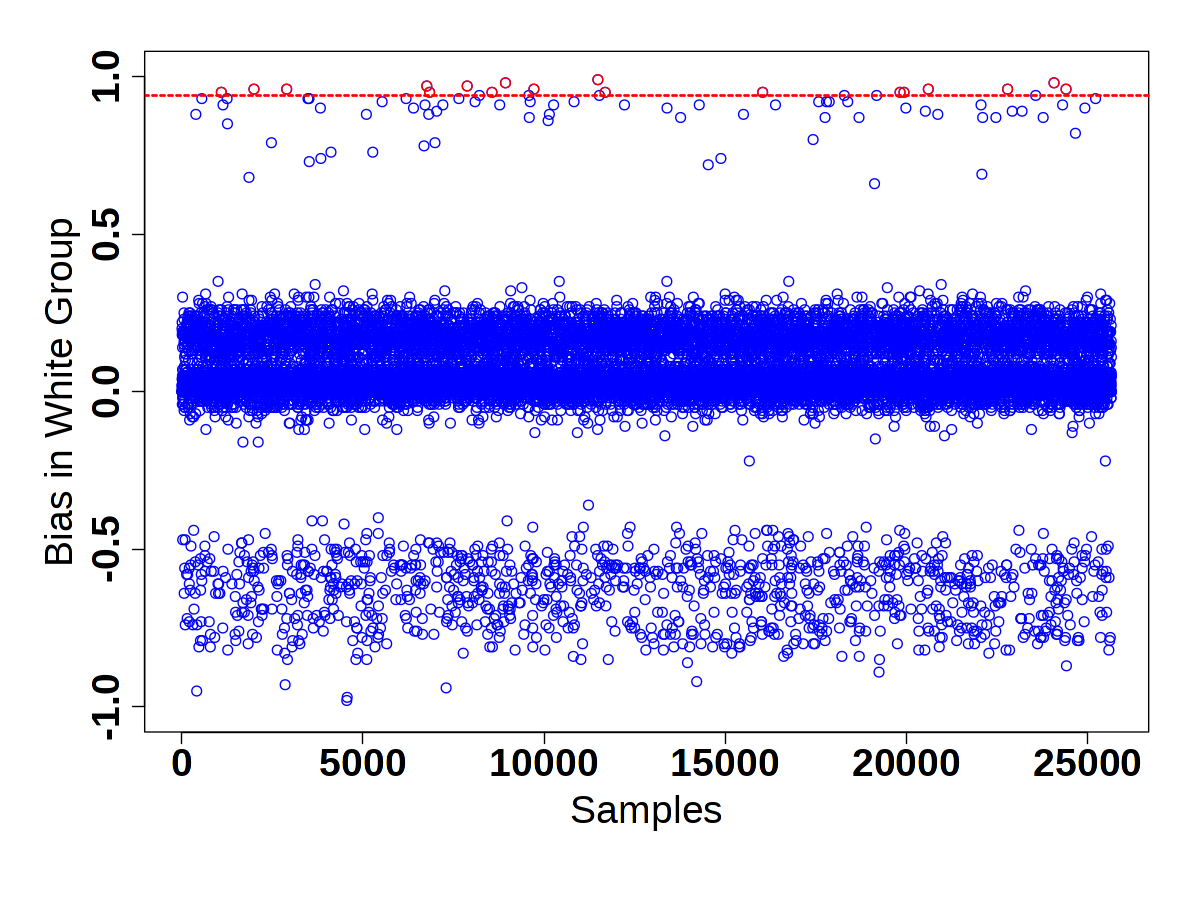}
    \includegraphics[width = 0.24\textwidth]{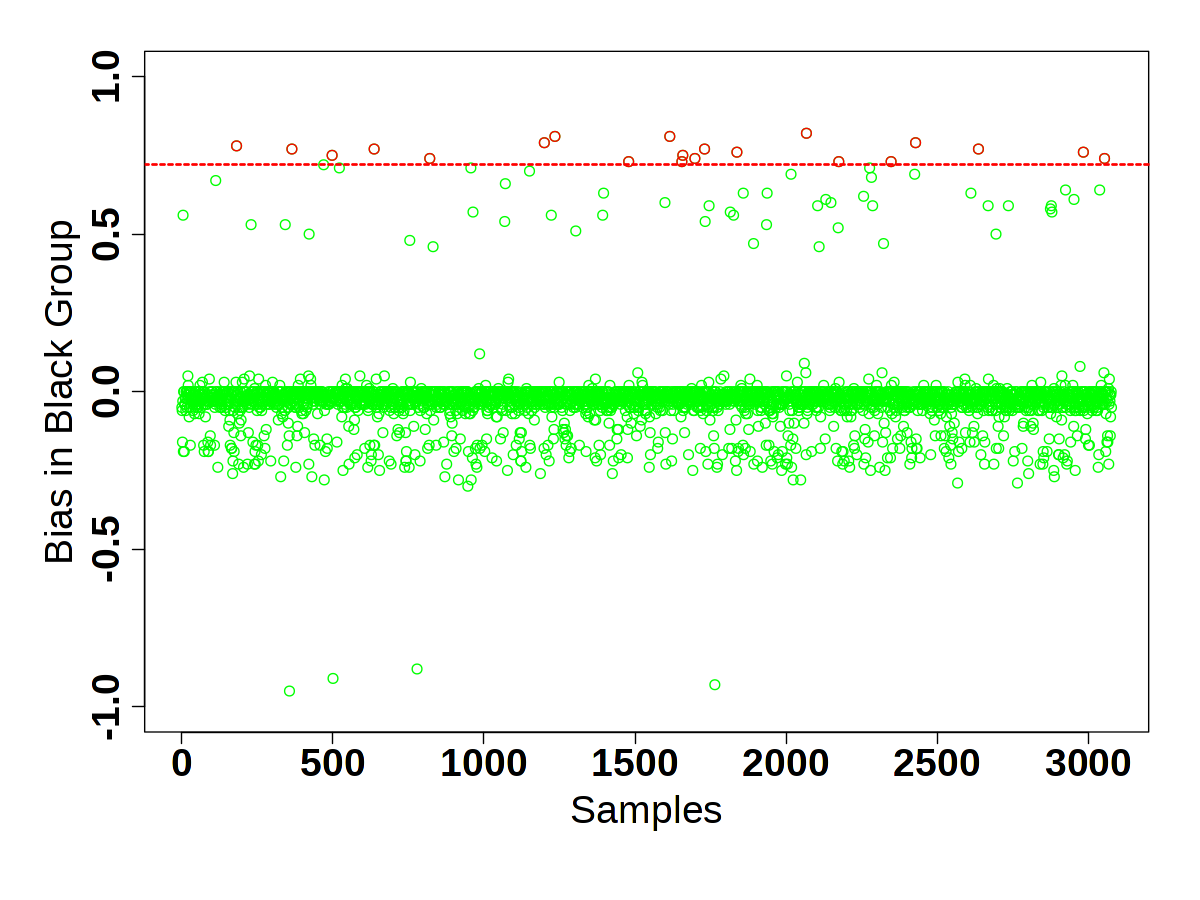}
    \vspace{-1.0em}
    \caption{\textit{Counterfactual discrimination of DNN}: (Left) The observed $CD$ for white with the threshold red line at 0.12; (Mid-Left) The observed $CD$ for black with the threshold line at 0.20. \textit{CD of Mitigated DNN}: (Mid-Right) The observed $CD$ for white with the threshold line sets at 0.81; and (Right) The observed $CD$ for black with the threshold line sets at 0.72.}
    \label{fig:dnn_adult_threshold}
    \label{fig:mitigated_dnn_adult_qq}
    \vspace{-1.0em}
\end{figure*}

\vspace{0.25em}
\noindent \textbf{Dataset.}
The Adult Census Income dataset~\cite{Dua:2019-census} is a binary classification dataset used to predict whether an individual has an annual income over $50K$. It consists of $48,842$ instances and $14$ attributes. In our study, we consider \textit{race} as the protected attribute and compare the outcomes between white and black individuals.

\vspace{0.25em}
\noindent \textbf{ML Model and Typical Fairness.} We used the same neural network architecture as previous literature on fairness testing~\cite{zhang2020white,9793943,10.1145/3460319.3464820}, which is a six-layered fully-connected neural network with 128 neurons that produces probabilities from the raw logit scores. The model was trained on the Adult Census Income dataset using the \texttt{Adam} optimizer with a learning rate of 0.001. The accuracy of the model on the test data is 84\%. The true positive rates for white and black individuals are 0.75 and 0.65, respectively. This yields an average odd difference (AOD) of 0.05.

\vspace{0.25em}
\noindent\textbf{Test-case Generations.} 
% First, we run our search algorithm for 20 mins. 
% The algorithm uses generative adversarial network (GAN) to
% generate more test cases when the current set of data does not satisfy the required statistical test.
% For this scenario, 
The search algorithm samples 25,658 and 3,076 test-cases for white and black groups, respectively. 
For each sub-group, we compute the likelihood of a favorable outcome for the original sample
and its counterfactual, i.e., \emph{counterfactual discrimination} (CD).
% The difference between these two probabilities represents the amount of , where positive (negative) values indicate that the model disadvantages (advantages) the original sample due to its protected group membership. 
Figure~\ref{fig:dnn_adult_threshold} (left part) shows the CD of these samples for the white and black sub-groups. The mean and standard deviation of CD are -0.04 (+/- 0.05) and 0.03 (+/- 0.04) for the white and black groups.

\begin{figure*}
    \centering
    \includegraphics[width = 0.24\textwidth]{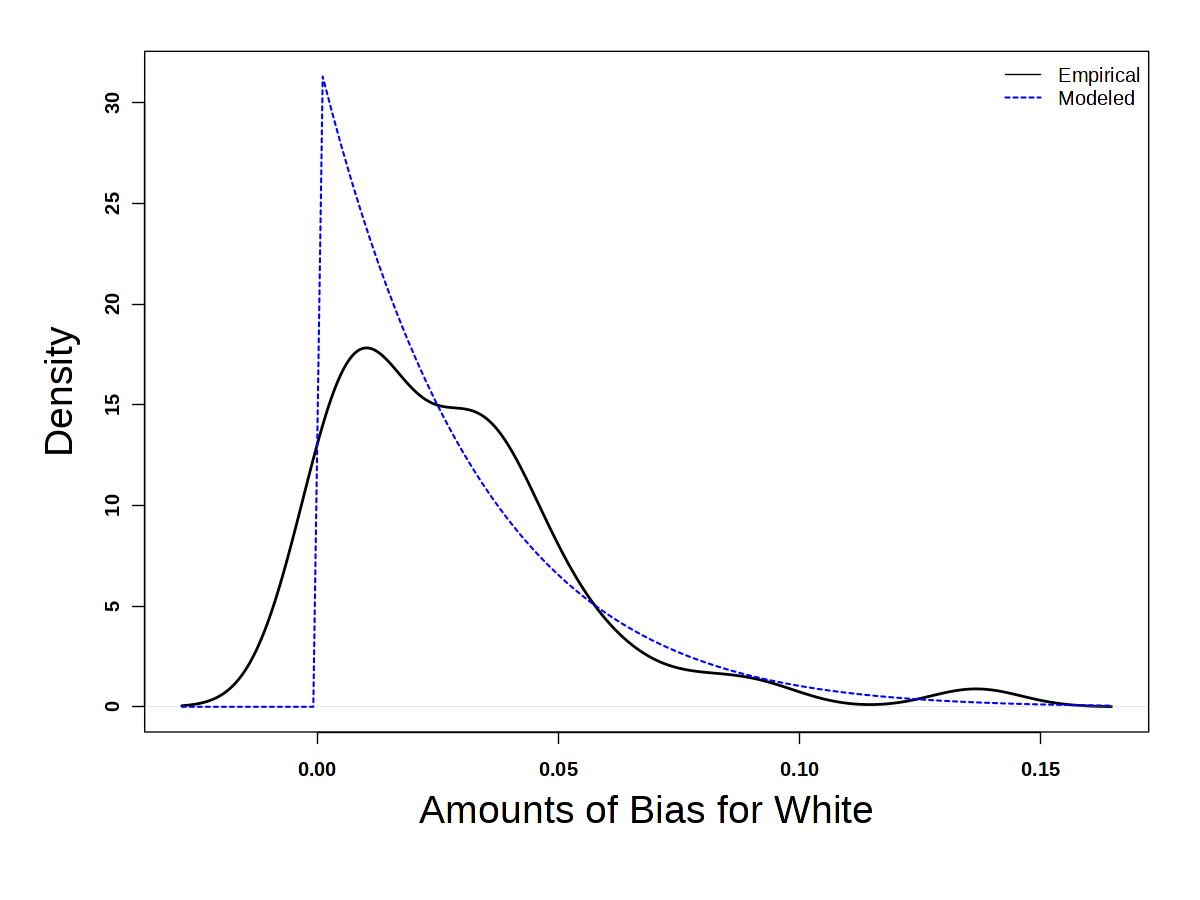}
    \includegraphics[width = 0.24\textwidth]{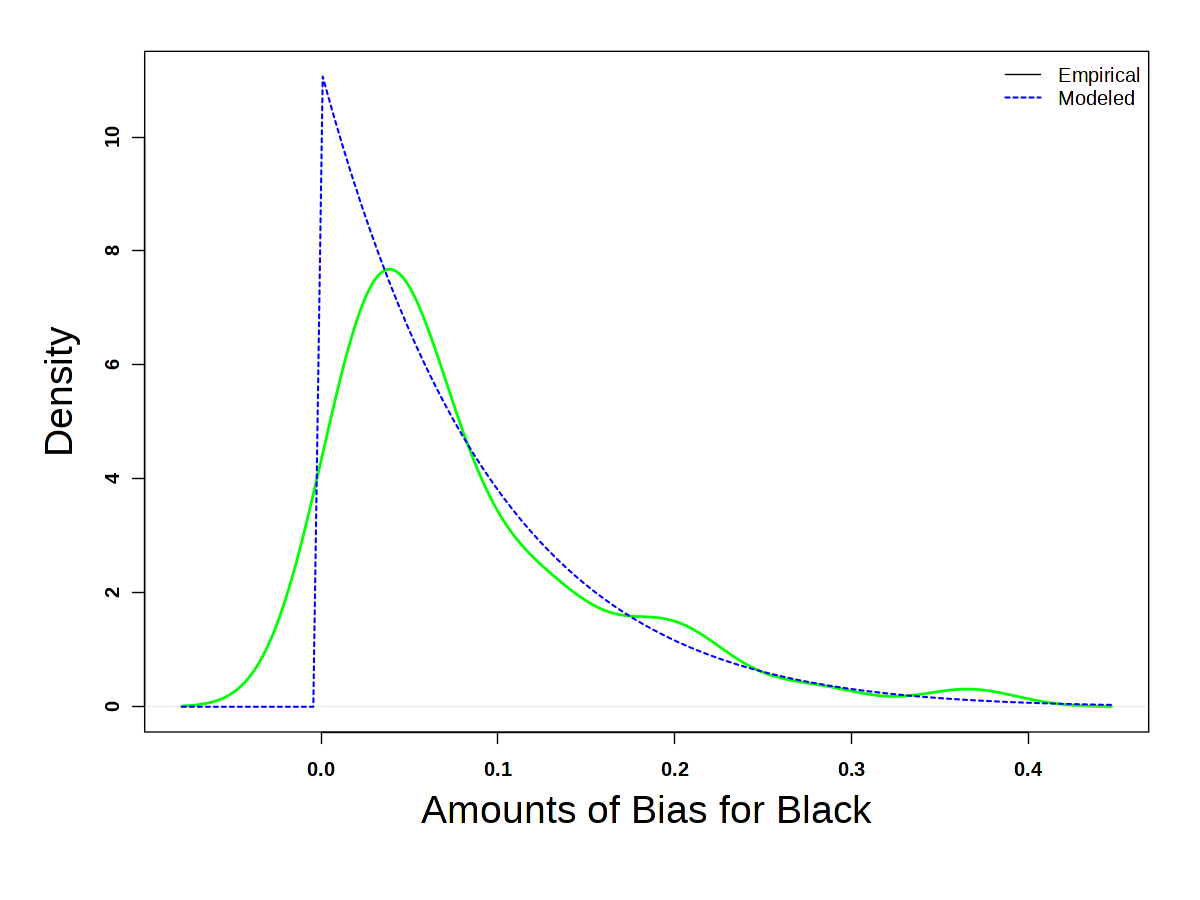} 
    \includegraphics[width = 0.24\textwidth]{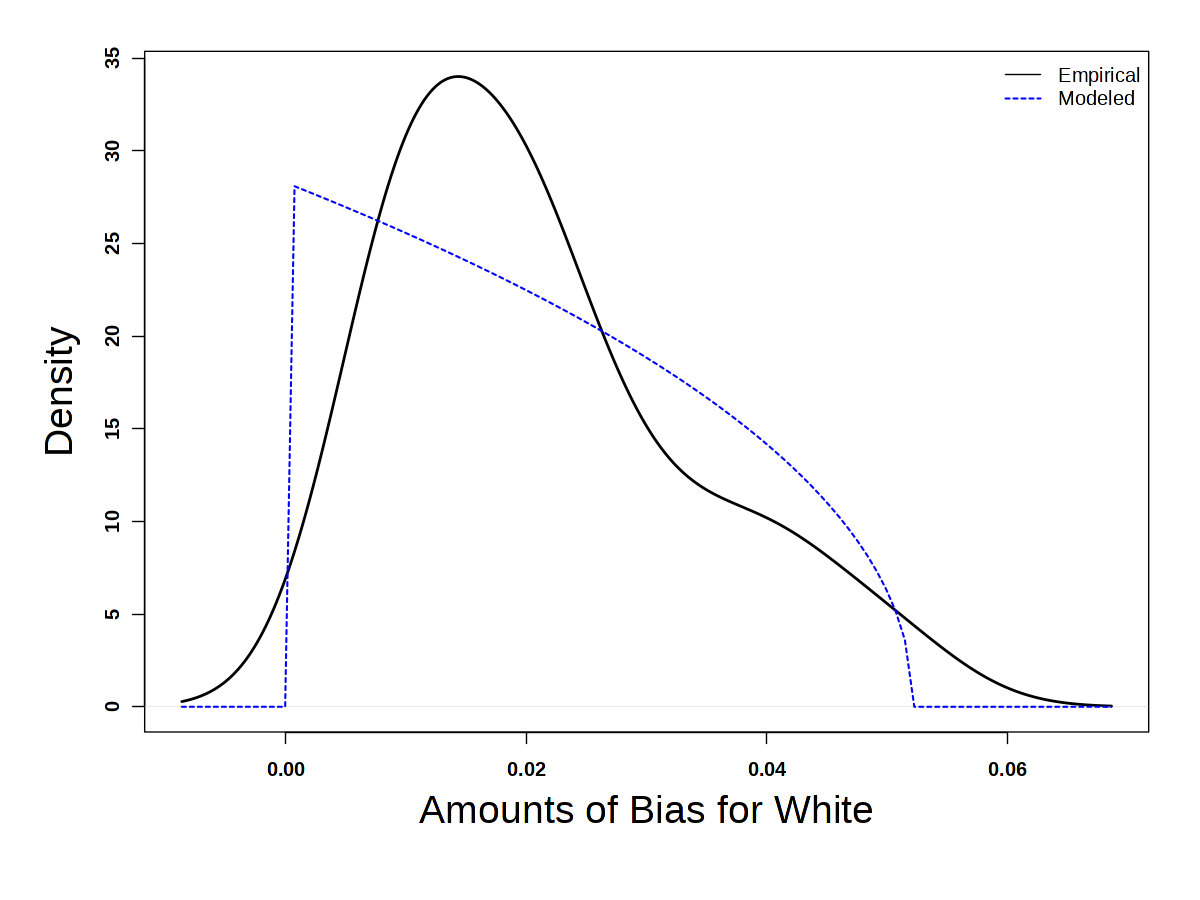}
    \includegraphics[width = 0.24\textwidth]{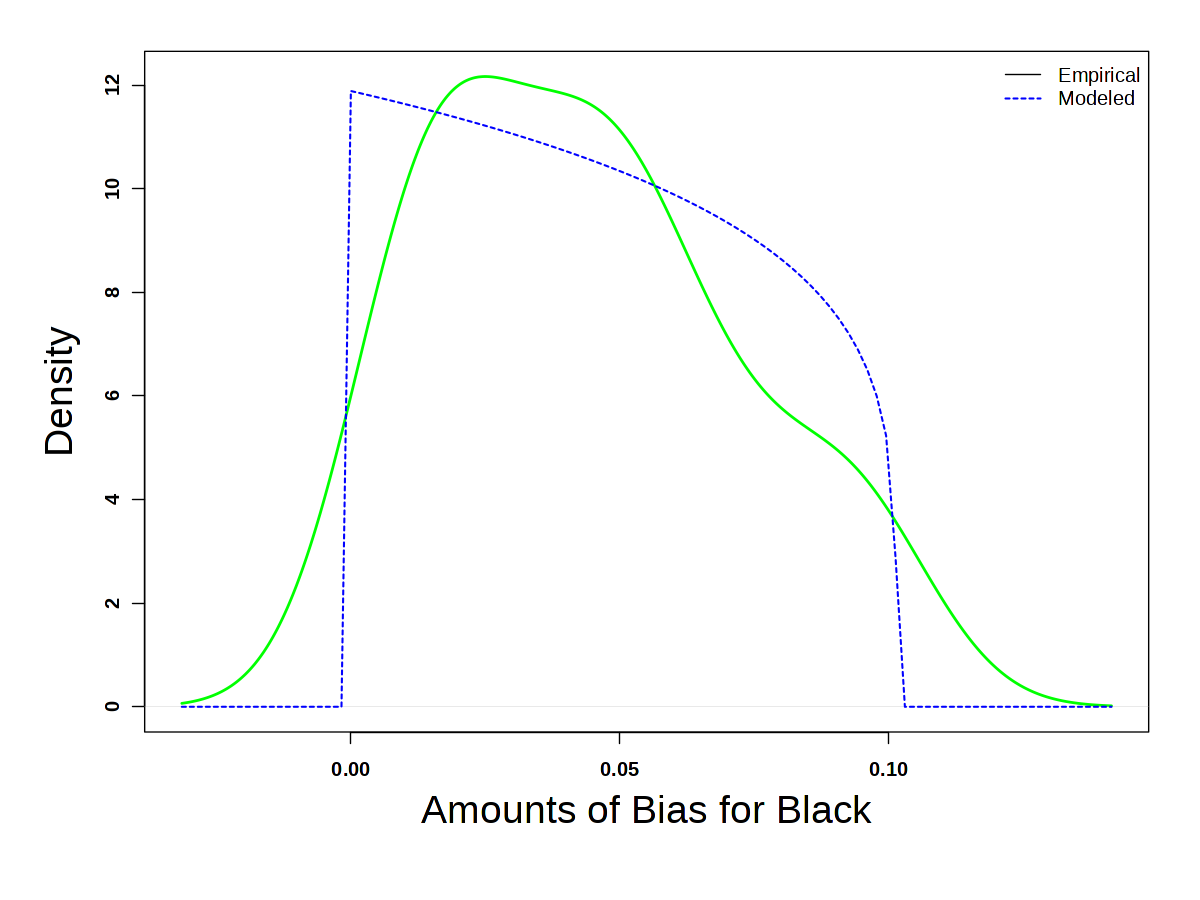} 
    \caption{\textit{The density of GEV for DNN}: (Left) The density for white with the location of 0.15 and scale of 0.03; (Mid-Left) The density for black with the location of 0.28 and the scale of 0.08. \textit{The density of GEV for Mitigated DNN}: (Mid-Right) The density of GEV distributions for white with the location of 0.84 and scale of 0.02; (Right) The density of GEV distributions for black with the location of 0.77 and the scale of 0.04.}
    \vspace{-1.0em}
    \label{fig:adult_density}
    \label{fig:adult_density_2}
    \vspace{-0.5em}
\end{figure*}

\vspace{0.25em}
\noindent\textbf{Inferring Extreme Value Distributions.} 
Figure~\ref{fig:dnn_adult_threshold} (left part) shows CD of samples with the threshold (red) lines for white and black groups where red points are extreme values. 
To infer the parameters of GEV distribution,
we set the threshold to 0.12 and 0.2 for white and black groups allowing only $50$ samples to exceed the threshold~\cite{abella2017measurement}. 
Given this criterion, the estimated location, scale, and shape are 0.15 (+/- 0.01), 0.03 (+/- 0.01), and -0.08 (+/- 0.01) for white, and 0.28 (+/- 0.01), 0.08 (+/- 0.02), and -0.08 (+/- 0.02) for black, respectively. 
% Each point in the plot represents a particular quantile (eg., $5$-{th} percentile) from the data vs. that predicted by the model. An ideal fit is denoted by a $45^{\circ}$ line in the plot. 

\vspace{0.25em}
\noindent\textbf{Fairness Measures through Extreme Value Distributions.} 
We use the characteristics of GEV distributions to measure the amounts of discrimination between two groups in the tail
of the DNN outcomes. Figure~\ref{fig:adult_density} shows the GEV density plot for white (left) and black (mid-left) sub-groups. While the average of CDs for these two groups differ by 0.07 (favoring white); the expected extreme CDs differ by 0.13.
Crucially, GEV distributions allow us to compute the expected return levels (RL) for a given number of interactions. Table~\ref{tab-RL-DNN-census-race} (original DNN) shows the RLs. For instance, the table indicates that the worst-case CDs of 0.14 and 0.42 are expected in the next 1,000 interactions for white/black sub-groups, respectively.

\vspace{0.25em}
\noindent\textbf{Validating Prevalent Fairness Mitigators.}
We validate the behaviors of popular in-process and \revision{pre-process} bias mitigators---exponentiated gradient (EG)~\cite{agarwal2018reductions} and Fair-SMOTE~\cite{10.1145/3468264.3468537}---in the tail. When using EG, the average odd difference is 0.02 which shows improvements in the average-case fairness with only 2\% accuracy loss. 
Figure~\ref{fig:mitigated_dnn_adult_qq} (right parts) shows CD values for the mitigated DNN with a threshold (red) line set at 0.81 and 0.72 for white and black, respectively.
The estimated location, scale, and shape of GEV are 0.84 (+/- 0.02), 0.02 (+/- 0.01), and -0.10 (+/- 0.09) for white; and 0.77 (+/- 0.02), 0.04 (+/- 0.02), and -0.05 (+/- 0.05) for Black, respectively, a significant increase in the tail discrimination, see Figure~\ref{fig:adult_density_2} (right parts).
The RLs of the mitigated models have also significantly increased, as shown in Table~\ref{tab-RL-DNN-Mitigated-census-race}. Within the white sub-group, in the next 2,000 interactions, we expect an extreme bias of 0.16 in the DNN, whereas an RL of 0.95 is expected for the mitigated DNN.  For the black sub-group, the expected RL has increased from 0.47 to 0.81.

\begin{table}
\centering
 \caption{Return Levels of ECD for original vs. mitigated.}
\vspace{-1.0em}
\resizebox{0.5\textwidth}{!}{
 \begin{tabular}{ | c | c  c | c  c  | } 
  \hline
  \textbf{Num.} & \multicolumn{2}{c|}{\textbf{Original DNN}}  & \multicolumn{2}{c|}{\textbf{Mitigated DNN}}
  \\
 \textbf{Interactions} & \textbf{RL} (white) & \textbf{RL} (black) &  \textbf{RL} (white) & \textbf{RL} (black) \\ 
 \hline 
 500 & 0.12 (+/- 0.01) & 0.37 (+/- 0.01) & 0.90 (+/- 0.02) & 0.78 (+/- 0.1)  \\
 \hline
 1,000 & 0.14 (+/- 0.03) & 0.42 (+/- 0.03) & 0.93 (+/- 0.03) & 0.80 (+/- 0.18) \\
 \hline
 2,000 & 0.16 (+/- 0.04) & 0.47 (+/- 0.05) & 0.95 (+/- 0.04) & 0.81 (+/- 0.19)  \\
 \hline
 % 5,000 & 0.25 (+/- 0.06) & 0.30 (+/- 0.07)  \\
 % \hline
 \end{tabular}
 }
 \label{tab-RL-DNN-census-race}
 \label{tab-RL-DNN-Mitigated-census-race}
\vspace{-1.0em} 
\end{table}

\vspace{0.25em}
\noindent\textbf{Tail-Aware Bias Reduction Algorithms.} We first validate the MiniMax-Fainess~\cite{10.1145/3461702.3462523}, a tail-aware bias reduction algorithm. Compared to the EG, the MiniMax-Fairness reduces the ECD to -0.27 and degrades the average-based fairness by 0.04. We guide an in-process bias mitigator that reduces the ECD to 0.02 while degrading average-based fairness only by 0.01.

\section{Extreme Counterfactual Discrimination}
\label{sec:problem}
We consider machine learning classifiers with a set of input variables $\Aa$, which are divided into a protected set of variables $Z$ (e.g., race, sex, and age) and non-protected variables $X$ (e.g., profession, income, and education). A learning problem can be defined as identifying a mapping from the inputs to a probabilistic score of the favorable outcome, inferred from a fixed training dataset $\Dd_T = \set{({\bf (x_i,z_i)}, {\bf y_i})}_{i=1}^N$, such that the ML model generalizes well to previously unseen situations based on a test dataset $\Dd_* = \set{({\bf (x^*_i,z^*_i)}, {\bf y^*_i})}_{i=1}^M$.

We abstractly express a machine learning classifier as a function $ML: X {\times} Z \to [0,1]$. The accuracy of model is measured for the fraction of points in $\Dd_*$ that satisfy the predicate $ML(x^*_i,z^*_i)\ge 0.5 == y^*_i$.
As a convention, we let $Z_i=1$ indicate membership in a privileged group, and $Z_i=0$ in the unprivileged group.

\begin{definition}[CD]
\label{def:counterfactual-bias}
Given an individual with non-protected value $X=x$ and protected attribute $Z=z$, the amount of discrimination over the protected attribute $i$ based on the \revision{causal fairness} notion~\cite{10.1145/3510003.3510137,10.1145/3106237.3106277,zhang2020white,9793943} defines as the difference between ML outcomes over the individual and its counterfactual, i.e., $CD(x,z) = ML(x,z')-ML(x,z)$ where  $z'_i=1-z_i$ with $z'_j=z_j$ for all protected attributes $1 \leq j\ne i \leq r$ and $-1 \leq CD \leq +1$.
The positive values indicate that the ML model disadvantages the individual $x$ in the group $z$ whereas the negative values show unfair advantages for the membership in $Z=z$. 
\end{definition}

Considering individuals with $z_i{=}\{0,1\}$, the average \revision{causal} discrimination for sub-groups ({\bf p}rivileged and {\bf u}nprivileged) is:
$ACD_p=\mathop{\mathbb{E}_{z_i=1}~CD(x,z)}$ and $ACD_u=\mathop{\mathbb{E}_{z_i=0}~CD(x,z)}$.

Previously, \textsc{Themis}~\cite{10.1145/3106237.3106277} used z-score testing with an assumption about the normal distribution of counterfactual outcomes to deem discrimination between two groups. From a practical standpoint, it is crucial to ensure fairness on average outcomes (e.g., \textsc{Themis}~\cite{10.1145/3106237.3106277}) as well as in the tail.

\begin{definition}[ECD]
    \label{def-problem}
    % \begin{tcolorbox}[boxrule=1pt,left=1pt,right=1pt,top=1pt,bottom=1pt]
     Given an $ML$ model and a protected attribute $Z_i$, our goal is to (1) model the statistics of extreme counterfactual discrimination for each group, i.e., $M_p = \max_{z_i=1} CD(x,z)$ and $M_u = \max_{z_i=0}CD(x,z)$; (2) compute whether the difference between two groups ($M_p$ and $M_u$) is statistically significant to detect a discrimination in the tail; (3) provide worst-case guarantees on the amounts of discrimination; and (4) mitigate biases in the tail.
    % \end{tcolorbox}
\end{definition}

\section{Approach}
\label{sec:approach}
We are interested in determining the maximum values of counterfactual discrimination, denoted as $M_p$ for privileged groups and $M_u$ for unprivileged groups. Since these values for different individuals are independent of each other, we can consider the estimation over a large number of independent and identically distributed random variables.

Extreme value theory is the field of study that examines the limit distributions of such extreme values and the convergence towards these distributions. Our objective, therefore, is to estimate the worst-case counterfactual discrimination by comparing the statistical characteristics of GEV distributions between privileged and unprivileged sub-groups.

However, analyzing extreme values necessitates having an adequate number of samples from the tail behavior of ML models for any given group to have confidence in the results. Our approach comprises three key steps:
1) \emph{Learning the underlying distributions} of the target population to generate valid samples for any sub-group;
2) \emph{Collecting tail samples with statistical guarantees} through a randomized test-case generation algorithm; and
3) \emph{Inferring the tail distributions of counterfactual discrimination} by fitting GEV distributions to each group and comparing the results to determine statistically significant discrimination in worst-case scenarios.

\vspace{0.25 em}
\noindent\textbf{Learning the underlying distributions.} 
The scarcity of samples for some protected groups in datasets can result in statistical uncertainties in extreme value distributions.
For instance, in the heart dataset~\cite{Heart-disease}, the number of samples for male individuals is notably limited. The conventional approach of sampling data points uniformly at random from the domain of each variable without considering the relationships between variables has the risk of producing samples that do not represent the target group~\cite{10.1145/3106237.3106277, 10.1145/3238147.3238165, 10.1145/3510003.3510123, 10.1145/3460319.3464820}. For example, random generation could result in an income level that is out of line with the general age distribution. 

Generative Adversarial Networks (GANs) and Variational Autoencoders (VAE) have been shown to effectively learn and reproduce actual data distributions, making them suitable for generating synthetic data that closely resembles the real-world distributions of sensitive groups~\cite{pmlr-v157-zhao21a,math11040977,10.1007/978-3-031-35891-3_26,8285168,Zhang2015LearningCF,ISLAM2021105950}.
In the GAN paradigm, during the training phase, the generator's primary function is to produce synthetic data samples, while the discriminator is tasked with distinguishing between real and synthetic samples. 
After multiple rounds of training, the generator learns to generate data so indistinguishable from the original samples. VAEs on the other hand, are trained by encoding input data into a latent representation and recovering it afterward. The decoder then reconstructs the input using the sampled latent points. Training involves optimizing two essential components: the reconstruction loss and the Kullback-Leibler (KL) divergence.

However, in addition to making sure to learn the target distribution of each demographic to alleviate
the risk of unrealistic data generation, we also need to ensure that they generate samples proportional
to the representation of groups in the original dataset. In doing so, we explore Conditional Tabular GAN (CTGAN)~\cite{CTGAN} and Triplet-based Variational Autoencoder (TVAE)~\cite{CTGAN}. CTGAN and TVAE are specialized for tabular data, capable of handling mixed variable types and complex relationships, unlike traditional GANs and VAEs which focus more on image data. \revision{Previosly, Xiao et al.~\cite{xiao2023latent} used CTGAN~\cite{CTGAN} to generate natural test cases in fairness testing as well. CTGAN allowed them to improve the naturalness of test cases by 20\% on average, compared to the baseline~\cite{10.1145/3510003.3510123,10.1145/3510003.3510137}.} In our application, the generator learns to produce samples for a given protected group as a target
that closely reflects the underlying distribution of the group.

\begin{algorithm}[t!]
\DontPrintSemicolon
    \KwIn{Decision-Support ML Software $ML$, Generative Adversarial Network $GAN$, Training Dataset $\mathcal{D}$, Test Samples $\mathcal{D}_{*}$, Target Group $\mathcal{G}$, Counterfactual Group $\mathcal{G}'$, Low-Bounds on Exp. Test
    $k_{min}$, Upper-Bounds on Exp. Test $k_{max}$, Num. of GAN Samples $m$, and Timeout $\mathcal{T}$.
    }
    
    $Done$ $\gets$ \texttt{False}
    
    \While{$t \leq \mathcal{T}$ OR $Passed$}{
        $Y$ $\gets$ $ML$($\mathcal{D}_{*}$, $\mathcal{G}$)

        $Y'$ $\gets$ $ML$($\mathcal{D}_{*}$, $\mathcal{G'}$)

        $\Delta$ $\gets$ $Y'$ - $Y$

        $HQ\_Samples$ $\gets$ \texttt{True}

        \For{$k$ $\gets$ $k_{min}$ to $k_{max}$}
        {
            \If{\texttt{size}($\Delta$) $<$ $k_{max}$}{
                $HQ\_Samples$ $\gets$ \texttt{False}

                \texttt{Break}
            }
            $\Theta$ $\gets$ \texttt{Select\_Top\_k}($\Delta$, $k$)

            $\overline{\Theta}$, $\sigma(\Theta)$  $\gets$ \texttt{average}($\Theta$), \texttt{std}($\Theta$)

             $CV_k$ $\gets$ $\frac{\sigma(\Theta)}{\overline{\Theta}}$
             
             \If{$CV_k$ $\geq$ $1.0 + (\frac{1}{4*k})$}{
                $HQ\_Samples$ $\gets$ \texttt{False}                
                
                \texttt{Break}
             }
             \Else{
                $Passed$, $HQ\_Samples$ $\gets$ \texttt{True}, \texttt{True}
             }
        }
        \If{$Not$ $HQ\_Samples$}{
            % \stsays{We need to say how many samples: should depends on the num. test samples?}
            $\mathcal{D}_{*}$ $\gets$ $\mathcal{D}_{*}$ ++ \texttt{Data\_Generator}($\mathcal{D}$, $\mathcal{G}$, m)
        }
    }
    \Return  $\mathcal{D}_{*}$, $\Delta$

\caption{Tail Sample Generations.}
\label{alg:algorithm1}
\end{algorithm}

\vspace{0.25 em}
\noindent\textbf{Collecting tail samples with statistical guarantees.}
Algorithm~\ref{alg:algorithm1} shows our approach to assess the fairness of ML models in the tail. 
Given a dataset, its protected attribute, the target group, and a ML model; we first initialize the test-case
samples $\mathcal{D}_{*}$ to all samples from the target group (e.g all samples with race white) and compute the likelihood
of favorable outcomes of the ML model for this target group (line 3). We do the same for the counterfactual group by flipping the value of the protected attribute (e.g., white to black), (line 4). We set the counterfactual discrimination ($\Delta$) for the target group as the differences in the ML outcomes between the counterfactual and original group (line 5). 
Then we start our search algorithm to collect enough samples to fit the EVT distributions on the tail.
In doing so, we perform the exponential test, adopted from~\cite{diebolt2007goodness,abella2017measurement} on the current
samples $\mathcal{D}_{*}$ (line 7-18). This test utilized the Coefficient of Variation (CV) to determine the type of extreme value distribution. Specifically, the test goes over $k$ highest values of the counterfactual discrimination and calculates the CV value
where $k$ ranges from $k\_{min}$ to $k\_{max}$ (line 13). If for all values of $k \in [k\_{min},k\_{max}]$, the CV is less than
($1.0 + \frac{1}{4*k}$), then we are statistically confident that we have enough samples from the tail to infer valid extreme value
distribution with exponential or light tails~\cite{castillo2014methods}, noting that the extra term ($\frac{1}{4*k}$) is
to correct the bias in the estimation of CV due to small sample size in the tail~\cite{sokal1995biometry} (line 17-18). 
Otherwise, if any values of CV are greater than ($1.0 + \frac{1}{4*k}$), we may not be able to fit an EVT distribution in the tail under the current samples $\mathcal{D}_{*}$ (line 14-16).
Only in this case, we use synthetic data generation methods (CTGAN and TVAE)
to generate $m$ data samples, similar to the
training data samples of the target group (lines 19-20). We repeat the search until we pass the CV or a timeout occurs. 

\vspace{0.25 em}
\noindent\textbf{Inferring the tail distributions of counterfactual discrimination.}
Given the generated tail samples $D_{*}$ and the counterfactual discrimination measurements $\Delta$;
our final goal is to infer the parameters of GEV distributions to estimate counterfactual
discrimination on the tail for each group. Following~\cite{abella2017measurement},
we initially set the threshold of extreme values to $M_{k\_{max}}$, i.e., only $k\_{max}$ measurements
exceed the threshold. Then, we fit the GEV distribution and analyze the shape of the distributions
to decide the validity. If we are statistically confident that
the shape is zero or negative ($\xi <= 0$), then the GEV belongs to the type I (exponential)
or type III (light), and we compare
the expected worst-case values ($\mu$) and the scale ($\sigma$). 

\textit{Given valid GEVs for privileged and unprivileged groups, we measure the
amounts of discrimination between them with $\mu_u - \mu_p$, that is the expected worst-case
discrimination for a unprivileged group $\mu_u$ minus the privileged group $\mu_p$.}

\vspace{0.25 em}
\noindent\textbf{Tail-Aware Bias Mitigation.}
Our approach to mitigating bias in the tail employs in-process bias reduction algorithm via hyperparameter optimization. Our approach extends \textsc{Parfait-ML}~\cite{tizpaz2022fairness} where we change the objective of optimization from
the AOD to ECD while keeping the accuracy constraints the same.

\section{Experiments}
\label{sec:experiment}
In this section, we first formulate the research questions (RQs). Then, we overview datasets, ML models, bias mitigation algorithms, and our implementations. Finally, we carefully analyze and answer the research questions. 

\begin{enumerate}[start=1,label={\bfseries RQ\arabic*},leftmargin=3em]
\item (\textbf{Generating realistic test cases}) Can the previously proposed algorithm generate realistic data from the underlying distribution of the real dataset? 

\item (\textbf{Feasibility + Usefulness + Guarantee}) Can extreme value theory (EVT) model and quantify the counterfactual discrimination in the tail of ML outcome distributions with statistical guarantees? 

\item (\textbf{Average-based Bias Mitigators}) Can we validate the efficacy of the prevalent bias mitigation algorithm~\cite{bird2020fairlearn,10.1145/3468264.3468537} via EVT?

\item (\textbf{Tail-based Bias Mitigators}) What are the performance of existing tail-aware bias reductions? How does an EVT-based mitigator compare to them?
\end{enumerate}

{\footnotesize
\begin{table*}[!t]
\caption{Datasets used in our experiments.}
\centering
\resizebox{0.75\textwidth}{!}{
\begin{tabu}{|l|l|l|ll|ll|}
  \hline
  \multirow{2}{*}{\textbf{Dataset}} & \multirow{2}{*}{\textbf{$|$Instances$|$}} & \multirow{2}{*}{\textbf{$|$Features$|$}} & \multicolumn{2}{c|}{\textbf{Protected Groups}} & \multicolumn{2}{c|}{\textbf{Outcome Label}} \\
   &  & & \textit{Group~1} & \textit{Group~2} & \textit{Label 1} & \textit{Label 0}  \\
  \hline
  Adult \textit{Census}~\cite{Dua:2019-census} & \multirow{2}{*}{$32,561$} & \multirow{2}{*}{$14$} & Sex-Male & Sex-Female & \multirow{2}{*}{High Income} & \multirow{2}{*}{Low Income} \\ \cline{4-5}
  Income &  &    &  Race-White & Race-Black  &   &    \\
  \hline
  German \textit{Credit}~\cite{Dua:2019-credit}  & $1,000$ & $20$ & Sex-Male & Sex-Female & Good Credit & Bad Credit \\
  \hline
  \textit{Bank} Marketing~\cite{Dua:2019-bank}  & $45,211$ & $17$ & Age-Young & Age-Old & Subscriber & Non-subscriber \\
  \hline
  
  \multirow{1}{*}{\textit{Compas}~\cite{compas-dataset}}  & \multirow{1}{*}{$7,214$} & \multirow{1}{*}{$28$} & Race-Caucasian & Race-Non Caucasian  & \multirow{1}{*}{Did not Reoffend} & \multirow{1}{*}{Reoffend} \\  \cline{4-5}

  \hline
  \textit{Default}~\cite{Default-Credit}  & $30,000$ & $23$ & Sex-Male & Sex-Female & Default & Not Default \\
  \hline
  \textit{Heart}~\cite{Heart-disease}  & $297$ & $12$ & Sex-Male & Sex-Female & Disease & Not Disease \\
  \hline
  \textit{Meps15}~\cite{MEP} & $15,830$ & $137$ & Sex-Male & Sex-Female & Utilized Benefits & Not Utilized Benefits \\
  \hline
  \textit{Meps16}~\cite{MEP} & $15,675$ & $137$ & Sex-Male & Sex-Female & Utilized Benefits & Not Utilized Benefits \\
  \hline
  \textit{Student Performance}~\cite{Student-performance}  & $1,044$ & $32$ & Sex-Male & Sex-Female & Pass & Not Pass\\
  \hline

\end{tabu}
}
\label{table:dataset}
\vspace{-1.0em}
\end{table*}
}

\vspace{0.25 em}
\noindent \textbf{Dataset.} We consider $9$ socially critical datasets from the literature of algorithmic fairness.
These datasets and their properties are described in Table~\ref{table:dataset}.
We assume that group 1 is privileged and group 2 is unprivileged.

\vspace{0.25 em}
\noindent \textbf{Training Algorithms and ML Models.}
We consider $4$ popular ML models from the literature. We use a six-layer DNN, following~\cite{10.1145/3460319.3464820,9793943,zhang2020white}. We trained DNN in TensorFlow~\cite{tensorflow2015-whitepaper} and used the same hyperparameters for all tasks with num\_epochs, batch\_size, and learning\_rate are set to $25$, $32$, and $0.001$, respectively. We use the LR, SVM, and Random Forest algorithms from scikit-learn library~\cite{scikit-learn} with the default hyperparameter configuration, similar to~\cite{10.1145/3106237.3106277,udeshi2018automated,chakraborty2020fairway}. 

\vspace{0.25 em}
\noindent \textbf{Average-based Bias Mitigation Algorithm.} We consider
\revision{four} commonly used (average-based) bias mitigation algorithms, exponentiated gradient (EG)~\cite{agarwal2018reductions} (implemented in both AI Fairness 360~\cite{bellamy2019ai} and Fairlearn~\cite{bird2020fairlearn}), Fair-SMOTE~\cite{10.1145/3468264.3468537}, \revision{ MAAT~\cite{10.1145/3540250.3549093}, and STEALTH~\cite{10109333}.}
\revision{EG~\cite{agarwal2018reductions} algorithm adapts Lagrange methods to find the multipliers
that balance accuracy and fairness. Fair-SMOTE looks for bias in the training data and aims to balance the statistics of sensitive features by generating synthetic samples. MAAT employs a fairness model alongside a performance model to infer the final decision. STEALTH employs a surrogate
model to use in predictions and explanations. }
\revision{For evaluating fairness in average-based scenarios, in addition to AOD and EOD metrics, we also included Statistical Parity Difference (SPD), and Disparate Impact (DI) which compare the probabilities of favorable outcomes among protected groups~\cite{bellamy2019ai}.}

\vspace{0.25 em}
\noindent \textbf{Tail-aware Bias Reduction.} We utilize Minimax-Fairness~\cite{10.1145/3461702.3462523} that takes an iterative game-theoretical approach to reduce the maximum error for protected groups. To investigate the usefulness of the ECD-based mitigator, we adopt a hyperparameter optimization technique, \textsc{Parfait-ML}~\cite{tizpaz2022fairness} that finds the configurations of ML algorithms to minimize the bias of resultant ML models in the tail. We set ECD as the objective search criteria and run the tool for $1$ hour on each benchmark.    

\vspace{0.5 em}
\noindent \textbf{Implementation and Technical Details.}
We run all the experiments on an Ubuntu 20.04.4 LTS server equipped with an AMD Ryzen Threadripper PRO 3955WX CPU and two NVIDIA GeForce RTX 3090 GPUs.
We split the dataset into \revision{training ($60\%$), validating ($20\%$), and test ($20\%$) data }where accuracy, F1, and \revision{fairness measures} are reported over the test data.
We use Fairlearn~\cite{bird2020fairlearn} to quantify the fairness. To measure counterfactual bias, we sample data instances independently and at random for each sub-group. 
We use the implementation of EG in Fairlearn~\cite{bird2020fairlearn} to study the common bias mitigation algorithm. 
We repeated each query $100$ times and took the average to control the stochastic behavior of the EG with high precision. We obtained the implementation of Minimax-Fairness~\cite{10.1145/3461702.3462523} from their GitHub repository. We also modify the implementation to support training on GPU. We set the $error\_type$, $numsteps$, and $epochs$ to $0/1$ $loss$, $2000$, and $50$, respectively.
We implemented the EVT algorithms in \texttt{R} using \texttt{evd} and \texttt{extRemes} libraries~\cite{gilleland2013software}. 
In Algorithm~\ref{alg:algorithm1}, we set $k_{min}$, $k_{max}$, $m$, $\mathcal{T}$
to $10$, $50$, $1$, and $1200$(s), respectively. This choice of $k_{min}$ and $k_{max}$
provides 95\% confidence on the feasibility of worst-case guarantees via EVT~\cite{abella2017measurement}. \revision{We obtained the implementation of Fair-SMOTE~\cite{10.1145/3468264.3468537}, MAAT~\cite{10.1145/3540250.3549093}, and STEALTH~\cite{10109333} from their GitHub repository and used the recommended configuration to achieve their best results. We repeated each experiment 20 times and conducted 4,400 runs in total. For the statistical tests, we follow prior work~\cite{10109333,10.1145/3540250.3549093,Hess_cliff,6235961} and perform a nonparametric test using the Scott-Knott procedure. This involved applying Cliff's Delta and a bootstrap test to assess the results.
In our Scott-Knott ranking, we classify results as wins, ties, or losses based on statistically significant improvements, indistinguishable performance, or significant degradations, respectively, compared to the original baseline (vanilla) model. We compare  different methods to each other based on number of wins, ties, and losses.}
\textbf{The replication package is available at} \url{https://figshare.com/s/5b4fe7b676e1f7f7b107}.

\subsection{Evaluating Synthetic Data Generation (RQ1)}
We assess the performance of Conditional Tabular GAN (CTGAN)~\cite{CTGAN} and Triplet-based Variational
Autoencoder (TVAE)~\cite{CTGAN} by comparing their synthetic data against the original dataset, focusing on statistical similarities and distribution characteristics. We aim to determine which model better generates representative test cases for target demographic groups. We also included datasets generated independently at randomly from the domain of variables. For quality assessment, we considered two criteria: similarity to the dataset in several statistical properties and the performance of a downstream ML model trained on generated data versus the actual dataset~\cite{8983215, Theis2015ANO, chundawat2022tabsyndex}.

\setlength{\tabcolsep}{15pt} % Default value: 6pt
\begin{table}[t!]
    \centering
    \caption{Data generation techniques.
      Legend:
      \textbf{Algorithm}: Generating method,
      \textbf{FID}: Fréchet inception distance,
      \textbf{KL-D}: Kullback–Leibler divergence,
      \textbf{LG-D}: logistic regression detection,
      \textbf{Acc}: accuracy difference,
      \textbf{F1} : Downstream F1 loss.     
    }
    \label{tab:RQ0}
    \resizebox{0.48\textwidth}{!}{
      \begin{tabular}{|c|c|c|c|c|c|} 
      \hline
          
    & &\multicolumn{3}{c|}{\textbf{Similarity}} & \textbf{ML Perf} \\ \cline{3-6} 
     \textbf{Dataset} & \textbf{Algorithm} & FID & KL-D & LG-D   & F1 loss \\
    \hline
    \multirow{3}{*}{Adult}&CTGAN&.02 & \textbf{.93}&.74  & .03 \\
    &TVAE&\textbf{.01} & \textbf{.93}&\textbf{.78}  & \textbf{.01} \\
    &RND&.11 & .19&.01 & .25 \\
    \hline
    \multirow{3}{*}{Compas}&CTGAN&\textbf{.08} & .97&.61  & \textbf{.0}\\
    &TVAE&\textbf{.08} & \textbf{.98}&\textbf{.63 } & \textbf{.0} \\
    &RND&.30 & .76&.02 & .51 \\
    \hline
    \multirow{3}{*}{Credit}&CTGAN&\textbf{.05} & .13&.39  & .2 \\
    &TVAE&.06 & \textbf{.15}&\textbf{.54}  & \textbf{.0} \\
    &RND&.07 &\textbf{ .15}&.05  & .17 \\
    \hline
    \multirow{3}{*}{Bank}&CTGAN&\textbf{.02} & \textbf{.88}&\textbf{.73}  & \textbf{.0} \\
    &TVAE&.03 & .81&.67  & \textbf{.0} \\
    &RND&.20 & .12&.01  & .26 \\
    \hline
    \multirow{3}{*}{Default}&CTGAN&\textbf{.03 }& \textbf{.82}&\textbf{.65}  & \textbf{.01} \\
    &TVAE&.04 & .72&.47  & .03 \\
    &RND&.34 & .33&.00  & .13 \\
    \hline
    \multirow{3}{*}{Heart}&CTGAN&\textbf{.09 }& \textbf{.93}&\textbf{.54 } & .16 \\
    &TVAE&.13 & .77&.35  & \textbf{.09} \\
    &RND&.13 & .31&.15  & .39 \\
    \hline
     \multirow{3}{*}{MEPS15}&CTGAN&.26 & \textbf{.91}&.05 & .07 \\
    &TVAE&\textbf{.06} & .88&\textbf{.42 } & \textbf{.01} \\
    &RND&.78 & .89&.00 &  .45 \\
    \hline
     \multirow{3}{*}{MEPS16}&CTGAN&.26 & .9&.06  & .10 \\
    &TVAE&\textbf{.06} & .89&\textbf{.39}  & \textbf{.01} \\
    &RND&.77 & \textbf{.91}&.00  & .34 \\
    \hline
     \multirow{3}{*}{Students}&CTGAN&\textbf{.09} & .90&\textbf{.22 } & .18 \\
    &TVAE&\textbf{.09} & \textbf{.97}&.18  & \textbf{.02} \\
    &RND&.10 & .87&.02  & .76 \\
    \hline    
    \end{tabular}
}
\end{table}

Table~\ref{tab:RQ0} shows the results of  experiments and the evaluation of metrics for each benchmark.
Column~\texttt{FID} reports the Fréchet Inception Distance~\cite{FID} (FID) is an inception score-based metric to measure the resemblance between generated and actual datasets. The normalized KL-Divergence, shown in the \texttt{KL-D} column, measures the disparity in the informational content between two distributions, with a value of 1.0 indicating minimal divergence. Column~\texttt{LG-D} indicates the logistic regression detection score~\cite{SDV} that calculates how difficult it is to distinguish real from synthetic data based on the average ROC AUC scores across cross-validation splits.
Column \texttt{F1 loss} highlights the performance disparities between models trained on actual and synthetic datasets, with values near zero indicating comparable ML performance across both datasets. In this downstream evaluation, we trained two logistic regression models on the actual dataset and the generated data, and then assess their F1 score against the identical test set from the dataset. 

\begin{wrapfigure}{t}{0.25\textwidth}
  \begin{center}
  \vspace{-2.0 em}
    \includegraphics[width = 0.25\textwidth]{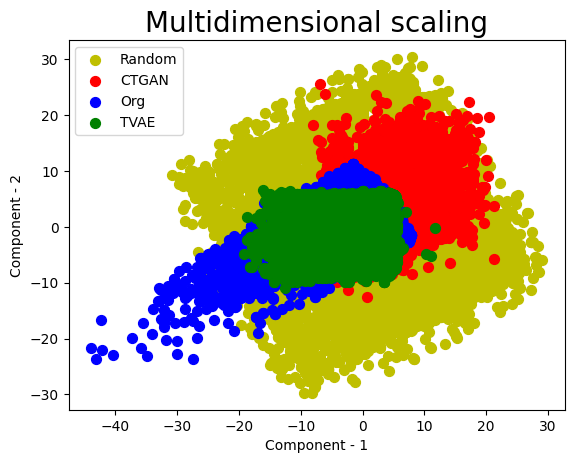}
  \vspace{-2.0 em}
  \end{center}
  \caption{MDS plot}
  \label{fig:RQ0_plot}
\end{wrapfigure}
Our results indicate that both CTGAN and TVAE are effective in learning and replicating the actual data distribution. However, their ability to capture complex feature relationships varies across datasets. For instance, with the Compas dataset, CTGAN's performance stands out: it achieves a KL-divergence of 0.98. Conversely, TVAE shows its strength with the Adult dataset, as supported by all four evaluation metrics. As Table~\ref{tab:RQ0} reveals, data generated randomly tend to deviate significantly from the actual data distribution.

We also employ Multidimensional Scaling (MDS)~\cite{borg2005modern} to visualize these methods. By reducing data to two principal dimensions, MDS provides a visual and analytical means to assess the accuracy with which different generation techniques replicate the characteristics of real dataset. Figure~\ref{fig:RQ0_plot} displays this comparison for the Compas dataset, particularly highlighting the alignment of TVAE-generated data with the actual dataset's distribution.

\begin{tcolorbox}[boxrule=1pt,left=1pt,right=1pt,top=1pt,bottom=1pt]
\textbf{Answer RQ1:}
CTGAN and TAVE demonstrate their ability to accurately replicate the distribution of actual datasets. In our experiments, they generated data with a KL-Divergence as high as 0.98 and an inception distance as low as 0.008. But, we found that their effectiveness is dataset-dependent. 
\end{tcolorbox}

% \footnotesize{
\setlength{\tabcolsep}{28.0pt} % Default value: 6pt

\begin{table*}[tbp!]
    \Huge
    \centering
    \caption{Characteristics of Extreme Value Distributions. 
      Legend: (\textit{Dataset})
      \textbf{P}: Protected Attribute,
      \textbf{\#N}: Number of test cases,
      \textbf{ACD} = ACD$_u$ - ACD$_p$: Average \revision{Causal} Discrimination Difference,
      \textbf{ECD} = $\mu_u-\mu_p$: The Amounts of ECD Tail Discrimination,
      (\textit{EVT Characteristics})
      ($\mu, \sigma, \xi$, type) for parameters of distributions,
      \textbf{$\tau$}: threshold.
      (\textit{Feasibility}) EVT-based extrapolation based on the type of EVT, Q-Q Plot, and its horizon for extrapolations (B). ($\epsilon<0.01$).
    }
    
    \label{tab:EVT-validity-guarantee}
    
    \resizebox{1.0\textwidth}{!}{ 
    % \renewcommand{\arraystretch}{+.2}
      % \begin{tabular}{|p{+.1em}|m{+.1em}m{+.1em}m{1.2em}|m{+.1em}m{+.1em}m{+.3em}|m{5em}m{5em}m{6em}m{+.1em}m{4em}|m{5em}m{+.5em}|}
      \begin{tabular}{|l|ccc|ccc|ccccc|cc|}
      \hline
      \multicolumn{4}{|c|}{Name}& \multicolumn{3}{c|}{Fairness} & \multicolumn{5}{c|}{EVT Characteristics} & \multicolumn{2}{c|}{Feasibility}
      \\
    \multicolumn{1}{|l}{Alg}& Dataset  & P & \#N(k) & ACD~\cite{10.1145/3106237.3106277} & CVaR~\cite{williamson2019fairness} & ECD & $\mu$ & $\sigma$ & $\xi$ & $\tau$ & Type & Q-Q Plot & B  \\ \hline
    \multirow{14}{*}{DNN} & \multirow{2}{*}{Census} &  White &  2.56/0&\cellcolor{gray!20} &\multirow{2}{*}{-.23}&\cellcolor{gray!40}& \cellcolor{gray!60}.15 (+/- $.01$) & .03 (+/- $.01$)&-.08 (+/- $.01$)&.12 &  III-Finite & Linear (\greencheck) & $\infty$ \\
    &  &  Black & 3.1/0&\multirow{-2}{*}{\cellcolor{gray!20}.07}&&\multirow{-2}{*}{.13}\cellcolor{gray!40}& \cellcolor{gray!100}.28 (+/- $.02$)&.08 (+/- $.02$)&-.08 (+/- $.02$)&.2 & III-Finite &  Linear (\greencheck) & $\infty$ \\ \cline{2-14}
     & \multirow{2}{*}{Census} &  Male  & 20.1/0&\cellcolor{gray!5}&\cellcolor{gray!20}& \multirow{2}{*}{.21}\cellcolor{gray!80}& .03 (+/- $\epsilon$)&.01 (+/- $\epsilon$)&.04 (+/- $\epsilon$)&.02 & I-Log &  Skewed-Right & 500 \\ 
    &  &  Female &  10.2/0&\multirow{-2}{*}{.05}\cellcolor{gray!5}&\multirow{-2}{*}{\cellcolor{gray!20}.08}& \multirow{-2}{*}{\cellcolor{gray!80}.21}&\cellcolor{gray!80}.24 (+/- $.02$)&.08 (+/- $.02$)&-.1 (+/- $.02$)&.16 & III-Finite &  Linear (\greencheck) & $\infty$ \\\cline{2-14}
     & \multirow{2}{*}{Credit} &  Male & .3/1.0&\cellcolor{gray!5}&\multirow{2}{*}{-.00}&\cellcolor{gray!20}& .0 (+/- $\epsilon$)&.00 (+/- $\epsilon$)&-141.17 (+/- $\epsilon$)&.00&  III-Finite & Linear (\greencheck) & $\infty$ \\
    &  &  Female & .7/0&\multirow{-2}{*}{.05}\cellcolor{gray!5}&& \multirow{-2}{*}{.09}\cellcolor{gray!20}&\cellcolor{gray!20}.09 (+/- $\epsilon$)&.01 (+/- $\epsilon$)&-1.05 (+/- $\epsilon$)&.07 & III-Finite & Linear (\greencheck) & $\infty$ \\ \cline{2-14}
     & \multirow{2}{*}{Bank} &  Young & 9.4/0&\multirow{2}{*}{.00}&\multirow{2}{*}{-.05}&\multirow{2}{*}{-.01}&\cellcolor{gray!20}.08 (+/- $\epsilon$)&.03 (+/- $\epsilon$)&.17 (+/- $\epsilon$)&.05 &  II-Infinite & Heavy-Tail (\redx) & 0 \\
    &  &  Old & 1.5/0& &&&\cellcolor{gray!00} .07 (+/- $.02$) &.03 (+/- $.02$)&-.03 (+/- $.02$)&.04 & I-Log & Skewed-Left & 5,000\\ \cline{2-14}
     % & \multirow{2}{*}{Compas} &  Male &  4.0/1.0&\multirow{2}{*}{.00}&\multirow{2}{*}{.00}&\multirow{2}{*}{.04}&.03 (+/- $.01$)&.02 (+/- $.01$)&.00 (+/- $.01$)&.01 &  I-Log &  Skewed-Left & 2,000  \\
    % &  &  Female &  1.4/.5 & & & &\cellcolor{gray!5} .07 (+/- $.01$)&.04 (+/- $.01$)&.03 (+/- $.01$)&.03 & I-Log & Skewed-Left & 1,000 \\ \cline{2-14}    
     & \multirow{2}{*}{Compas} &  Caucasian   & 1.4/0&\multirow{2}{*}{.02}&\multirow{2}{*}{-.01}&\cellcolor{gray!40}& .02 (+/- $\epsilon$) & .01 (+/- $\epsilon$)&.04 (+/- $\epsilon$)&.02 &  I-Log & Skewed-Left & 500  \\  
      &  &   Other & 3.0/0&&&\multirow{-2}{*}{.11}\cellcolor{gray!40}& \cellcolor{gray!40} .13 (+/- $.01$)&.02 (+/- $.01$)&.03 (+/- $.01$)&.1 & I-Log & Skewed-Left & 1,000\\  \cline{2-14}        
     % & \multirow{2}{*}{Compas} &  Young & 3.2/0& \multirow{2}{*}{.08}&\multirow{2}{*}{-.02}&\multirow{2}{*}{.11}& .0 (+/- $\epsilon$)&.01 (+/- $\epsilon$)&.92 (+/- $\epsilon$)&.00&   II-Infinite & Heavy-Tail (\redx) & 0\\
    % &  &  Old & .8/3.0&& & & \cellcolor{gray!25}.11 (+/- $\epsilon$) & .02 (+/- $\epsilon$)&-.02 (+/- $\epsilon$)&.09 & I-Log &  Skewed-Right & 1,000 \\  \cline{2-14}
    & \multirow{2}{*}{Default} &  Male & 11.4/0&\multirow{2}{*}{.03}&\multirow{2}{*}{-.00}&\cellcolor{gray!40}&.01 (+/- $\epsilon$)&.01 (+/- $\epsilon$)&.29 (+/- $\epsilon$)&.01 &   II-Infinite& Heavy-Tail (\redx) & 0 \\     
      &  &   Female & 17.2/0&&&\multirow{-2}{*}{.10}\cellcolor{gray!40} &\cellcolor{gray!40} .11 (+/- $\epsilon$) & .02 (+/- $\epsilon$)&-.07 (+/- $\epsilon$)&.1 & III-Finite & Linear (\greencheck) & $\infty$ \\ \cline{2-14}
    & \multirow{2}{*}{Heart} &  Male & .1/1.0&\multirow{2}{*}{-.02}&\multirow{2}{*}{-.00}&\multirow{2}{*}{-.01}&.01 (+/- $\epsilon$)&.00 (+/- $\epsilon$)&-.14 (+/- $\epsilon$)&.01 & III-Finite & Linear (\greencheck)  & $\infty$ \\     
      &  &   Female & .2/1.0&&&&.00 (+/- $\epsilon$)&.00 (+/- $\epsilon$)&-2.72 (+/- $\epsilon$)&.00&  III-Finite & Linear  (\greencheck)  & $\infty$ \\ \cline{2-14}
      & \multirow{2}{*}{Meps15} &  Male & 8.2/0&\cellcolor{gray!5}&\multirow{2}{*}{-.00}&\cellcolor{gray!60}& \cellcolor{gray!60} .16 (+/- $.01$)&.04 (+/- $.01$)&-.07 (+/- $.01$)&.11 & III-Finite & Linear (\greencheck)  & $\infty$ \\     
      &  &   Female & 7.6/0&\multirow{-2}{*}{.05}\cellcolor{gray!5}& &\multirow{-2}{*}{.18}\cellcolor{gray!60}&\cellcolor{gray!110} .34 (+/- $.01$) & .05 (+/- $.01$)&-.09 (+/- $.01$)&.29 &  III-Finite & Linear (\greencheck)  & $\infty$ \\ \cline{2-14}
      & \multirow{2}{*}{Meps16} &  Male & 8.3/0&\cellcolor{gray!20}&\multirow{2}{*}{.00}&\cellcolor{gray!80}&\cellcolor{gray!60}.18 (+/- $.01$) &.06 (+/- $.01$)&.02 (+/- $.01$)&.12 &I-Log &  Skewed-Left & 5,000 \\     
      &  &   Female & 7.4/0&\multirow{-2}{*}{.07}\cellcolor{gray!20}&&\multirow{-2}{*}{.20}\cellcolor{gray!80} &\cellcolor{gray!110}.38 (+/- $.02$)&.06 (+/- $.02$)&-.24 (+/- $.02$)&.31 &  III-Finite & Linear  (\greencheck)  & $\infty$ \\ \cline{2-14}
      & \multirow{2}{*}{Students} &  Male & .6/0&\multirow{2}{*}{.02}&\multirow{2}{*}{-.00}&\cellcolor{gray!20}&.00 (+/- $\epsilon$)&.00 (+/- $\epsilon$)&-347.33 (+/- $\epsilon$)&.00& III-Finite & Linear  (\greencheck) & $\infty$ \\     
      &  &   Female &  .5/0&&& \multirow{-2}{*}{.09}\cellcolor{gray!20}&\cellcolor{gray!20}.09 (+/- $.01$)&.04 (+/- $.01$)&-.27 (+/- $.01$)&.04 &  III-Finite & Linear  (\greencheck)  & $\infty$ \\ \cline{2-14}\hline

    \multirow{14}{*}{LR} & \multirow{2}{*}{Census} &  White &  25.7/0&\cellcolor{gray!20}&\multirow{2}{*}{-.19}&\cellcolor{gray!40}&.0 (+/- $\epsilon$)&.00 (+/- $\epsilon$)&.00 (+/- $\epsilon$)&.00&  I-Log & Skewed-Left & 500 \\
    &  &  Black & 3.1/0&\multirow{-2}{*}{.09}\cellcolor{gray!20}&&\multirow{-2}{*}{.11}\cellcolor{gray!40}& \cellcolor{gray!40}.11 (+/- $\epsilon$)&.00 (+/- $\epsilon$)&-1.24 (+/- $\epsilon$)&.11 & III-Finite & Linear (\greencheck)  & $\infty$ \\ \cline{2-14}
     & \multirow{2}{*}{Census} &  Male  &20.1/0&\cellcolor{gray!20}&\cellcolor{gray!80}&\cellcolor{gray!20}& .0 (+/- $\epsilon$) & .0 (+/- $\epsilon$)&-1.0 (+/- $\epsilon$)&.00& III-Finite & Linear (\greencheck)  & $\infty$\\ 
    &  &  Female & 10.2/0&\multirow{-2}{*}{.06}\cellcolor{gray!20}&\multirow{-2}{*}{.22}\cellcolor{gray!80}&\multirow{-2}{*}{.07}\cellcolor{gray!20} &\cellcolor{gray!20} .07 (+/- $\epsilon$)&.00 (+/- $\epsilon$)&-1.62 (+/- $\epsilon$)&.07 & III-Finite & Linear (\greencheck)  & $\infty$ \\\cline{2-14}
     & \multirow{2}{*}{Credit} &  Male & .3/.5& \cellcolor{gray!20}&\multirow{2}{*}{-.01}&\cellcolor{gray!40}& -.02 (+/- $\epsilon$)&.01 (+/- $\epsilon$)&-.74 (+/- $\epsilon$)&-.03 &  III-Finite & Linear (\greencheck)  & $\infty$ \\
    &  &  Female & .7/0&\multirow{-2}{*}{.09}\cellcolor{gray!20}&&\multirow{-2}{*}{.1}\cellcolor{gray!40}& \cellcolor{gray!20}.08 (+/- $\epsilon$)&.00 (+/- $\epsilon$)&-1.0 (+/- $\epsilon$)&.08 & III-Finite & Linear (\greencheck)  & $\infty$ \\ \cline{2-14}
     & \multirow{2}{*}{Bank} &  Young & 9.4/0&\multirow{2}{*}{.0}&\multirow{2}{*}{-.09}&\multirow{2}{*}{.01}&.00 (+/- $\epsilon$)&.00 (+/- $\epsilon$)&-1.0 (+/- $\epsilon$)&.00&  III-Finite & Linear (\greencheck)  & $\infty$ \\
    &  &  Old & 1.5/0&&&&.01 (+/- $\epsilon$)&.00 (+/- $\epsilon$)&-1.33 (+/- $\epsilon$)&.01 & III-Finite & Linear (\greencheck)  & $\infty$ \\ \cline{2-14}
     % & \multirow{2}{*}{Compas} &  Male & 3,527&\multirow{2}{*}{-.01}&\multirow{2}{*}{.0}&\multirow{2}{*}{-.02}&.02 (+/- $\epsilon$)&.00 (+/- $\epsilon$)&-.55 (+/- $\epsilon$)&.01 &  III-Finite & Linear (\greencheck)  & $\infty$ \\
    % &  &  Female & 906&&&&.00 (+/- $\epsilon$)&.00 (+/- $\epsilon$)&-.49 (+/- $\epsilon$)&.00& III-Finite & Linear (\greencheck)  & $\infty$ \\ \cline{2-14}    
     & \multirow{2}{*}{Compas} &  Caucasian & 1.5/0&\multirow{2}{*}{-.04}&\multirow{2}{*}{.0}&\multirow{2}{*}{-.06}&.00 (+/- $\epsilon$)&.00 (+/- $\epsilon$)&-.73 (+/- $\epsilon$)&.00&  III-Finite & Linear (\greencheck)  & $\infty$ \\  
      &  &   Other & 3.0/0&&&&\cellcolor{gray!20}.06 (+/- $\epsilon$)&.01 (+/- $\epsilon$)&-.54 (+/- $\epsilon$)&.06 & III-Finite & Linear (\greencheck)  & $\infty$ \\  \cline{2-14}        
     % & \multirow{2}{*}{Compas} &  Young &2,559&\multirow{2}{*}{.04}&\multirow{2}{*}{.0}&\multirow{2}{*}{.06}&.00 (+/- $\epsilon$)&.00 (+/- $\epsilon$)&-.42 (+/- $\epsilon$)&.00&   III-Finite & Linear (\greencheck)  & $\infty$ \\
    % &  &  Old & 839&&&&.06 (+/- $\epsilon$)&.01 (+/- $\epsilon$)&-.6 (+/- $\epsilon$)&.06 & III-Finite & Linear (\greencheck)  & $\infty$ \\  \cline{2-14}
    & \multirow{2}{*}{Default} &  Male & 11.4/0&\multirow{2}{*}{.04}&\multirow{2}{*}{-.07}&\multirow{2}{*}{.03}&.00 (+/- $\epsilon$)&.00 (+/- $\epsilon$)&-.66 (+/- $\epsilon$)&.00& III-Finite & Linear (\greencheck)  & $\infty$ \\     
      &  &   Female & 17.2/0&&&&.03 (+/- $\epsilon$)&.00 (+/- $\epsilon$)&-1.07 (+/- $\epsilon$)&.03 &  III-Finite & Linear (\greencheck)  & $\infty$ \\ \cline{2-14}
    & \multirow{2}{*}{Heart} &  Male & .1/1.0& \multirow{2}{*}{-.02}&\multirow{2}{*}{.02}&\multirow{2}{*}{-.01}&.01 (+/- $\epsilon$)&.01 (+/- $\epsilon$)&-1.0 (+/- $\epsilon$)&.00& III-Finite & Linear (\greencheck)  & $\infty$ \\     
      &  &   Female & .2/.5&&&&.00 (+/- $\epsilon$)&.00 (+/- $\epsilon$)&-1.0 (+/- $\epsilon$)&.00&  III-Finite & Linear (\greencheck)  & $\infty$ \\ \cline{2-14}
      & \multirow{2}{*}{Meps15} &  Male &  8.2/0&\cellcolor{gray!20}&\multirow{2}{*}{-.1}&\cellcolor{gray!20}&.00 (+/- $\epsilon$)&.00 (+/- $\epsilon$)&-.44 (+/- $\epsilon$)&.00& III-Finite & Linear (\greencheck)  & $\infty$ \\     
      &  &   Female & 7.6/0&\multirow{-2}{*}{.06}\cellcolor{gray!20}&&\multirow{-2}{*}{.07}\cellcolor{gray!20}& \cellcolor{gray!20} .07 (+/- $\epsilon$)&.00 (+/- $\epsilon$)&-1.65 (+/- $\epsilon$)&.07 &  III-Finite & Linear (\greencheck)  & $\infty$\\ \cline{2-14}
      & \multirow{2}{*}{Meps16} &  Male & 8.3/0&\cellcolor{gray!40}&\multirow{2}{*}{-.13}&\cellcolor{gray!40}& .0 (+/- $\epsilon$)&.00 (+/- $\epsilon$)&-.14 (+/- $\epsilon$)&.00& III-Finite & Linear (\greencheck)  & $\infty$ \\     
      &  &   Female & 7.4/0&\multirow{-2}{*}{.1}\cellcolor{gray!40}&&\multirow{-2}{*}{.12}\cellcolor{gray!40}& \cellcolor{gray!25} .12 (+/- $\epsilon$)&.00 (+/- $\epsilon$)&-.99 (+/- $\epsilon$)&.12 &  III-Finite & Linear (\greencheck)  & $\infty$\\ \cline{2-14}
      & \multirow{2}{*}{Students} &  Male & .6/0&\multirow{2}{*}{.04}&\multirow{2}{*}{.0}&\cellcolor{gray!40}&.00 (+/- $\epsilon$)&.00 (+/- $\epsilon$)&-1.06 (+/- $\epsilon$)&.00& III-Finite & Linear (\greencheck)  & $\infty$ \\     
      &  &   Female &  .5/0&&&\multirow{-2}{*}{.1}\cellcolor{gray!40}&\cellcolor{gray!40}.1 (+/- $\epsilon$)&.01 (+/- $\epsilon$)&-2.92 (+/- $\epsilon$)&.08 & III-Finite & Linear (\greencheck)  & $\infty$\\ \cline{2-14}\hline

    \multirow{14}{*}{SVM} & \multirow{2}{*}{Census} &  White &  25.7/0&\multirow{2}{*}{.04}&\multirow{2}{*}{-.11}&\multirow{2}{*}{.03}& \cellcolor{gray!100} .28 (+/- $.02$)&.04 (+/- $.02$)&-.52 (+/- $.02$)&.23 &  III-Finite & Linear (\greencheck) & $\infty$ \\
    &  &  Black & 3.1/0&&&& \cellcolor{gray!110} .31 (+/- $.02$)&.06 (+/- $.02$)&-.33 (+/- $.02$)&.24 & III-Finite & Linear (\greencheck) & $\infty$\\ \cline{2-14}
     & \multirow{2}{*}{Census} &  Male  & 20.1/0&\multirow{2}{*}{.04}&\cellcolor{gray!20}&\cellcolor{gray!60}&.01 (+/- $\epsilon$)&.00 (+/- $\epsilon$)&-.75 (+/- $\epsilon$)&.01 & III-Finite & Linear (\greencheck) & $\infty$ \\
    &  &  Female & 10.2/0&&\multirow{-2}{*}{.08}\cellcolor{gray!20}&\multirow{-2}{*}{.15}\cellcolor{gray!60}& \cellcolor{gray!60} .16 (+/- $.01$)&.02 (+/- $.01$)&-.06 (+/- $.01$)&.14 & III-Finite & Linear (\greencheck) & $\infty$\\ \cline{2-14}
     & \multirow{2}{*}{Credit} &  Female &.3/.5&\multirow{2}{*}{.01}&\multirow{2}{*}{-.03}&\multirow{2}{*}{.02}&.01 (+/- $\epsilon$)&.01 (+/- $\epsilon$)&-.07 (+/- $\epsilon$)&.01&  III-Finite & Linear (\greencheck) & $\infty$ \\
    &  &  Male & .7/0&&&&.03 (+/- $\epsilon$)&.00 (+/- $\epsilon$)&-.41 (+/- $\epsilon$)&.02 & III-Finite & Linear (\greencheck) & $\infty$ \\ \cline{2-14}
     & \multirow{2}{*}{Bank} &  Young &  9.4/0&\multirow{2}{*}{.01}&\multirow{2}{*}{-.09}&\cellcolor{gray!20}&\cellcolor{gray!40} .1 (+/- $.02$)&.05 (+/- $.02$)&.01 (+/- $.02$)&.05 & I-Log & Skewed-Left & 1,000\\
    &  &  Old & 1.4/0&&&\multirow{-2}{*}{.07}\cellcolor{gray!20}&\cellcolor{gray!60} .17 (+/- $.01$)&.05 (+/- $.01$)&.03 (+/- $.01$)&.12 &  I-Log & Skewed-Left & 1,000 \\ \cline{2-14}
    %  & \multirow{2}{*}{Compas} &  Male & 3,527&\multirow{2}{*}{.0}&\multirow{2}{*}{.0}&\multirow{2}{*}{.0}&.01 (+/- $\epsilon$)&.01 (+/- $\epsilon$)&.81 (+/- $\epsilon$)&.00&  II-Infinite& Heavy-Tail (\redx) & 0 \\
    % &  &  Female & 906&&&&.01 (+/- $\epsilon$)&.01 (+/- $\epsilon$)&.04 (+/- $\epsilon$)&.00&  I-Log & Skewed-Left & 2,000  \\ \cline{2-14}    
     & \multirow{2}{*}{Compas} &  Caucasian  & 1.4/0&\multirow{2}{*}{.0}&\multirow{2}{*}{.0}&\multirow{2}{*}{.01}&.00 (+/- $\epsilon$)&.00 (+/- $\epsilon$)&.12 (+/- $\epsilon$)&.00&  II-Infinite& Heavy-Tail (\redx) & 0 \\ 
      &  &   Other & 3.0/0&&&&.01 (+/- $\epsilon$)&.01 (+/- $\epsilon$)&.4 (+/- $\epsilon$)&.01 &  I-Log & Skewed-Left & 2,000  \\  \cline{2-14}
    %  & \multirow{2}{*}{Compas} &  Young & 2,559&\multirow{2}{*}{.0}&\multirow{2}{*}{.0}&\multirow{2}{*}{.0}&.01 (+/- $\epsilon$)&.01 (+/- $\epsilon$)&.55 (+/- $\epsilon$)&.01&   II-Infinite& Heavy-Tail (\redx) & 0 \\
    % &  &  Old & 839&&&&.01 (+/- $\epsilon$)&.00 (+/- $\epsilon$)&-.09 (+/- $\epsilon$)&.00& III-Finite & Linear  (\greencheck) & $\infty$ \\  \cline{2-14}
    & \multirow{2}{*}{Default} &  Male & 11.4/0&\multirow{2}{*}{.0}&\multirow{2}{*}{-.01}&\multirow{2}{*}{.01}&.03 (+/- $\epsilon$)&.00 (+/- $\epsilon$)&.00 (+/- $\epsilon$)&.02 & I-Log &  Skewed-Left & 2,000  \\     
      &  &   Female & 17.2&&&&.04 (+/- $\epsilon$)&.01 (+/- $\epsilon$)&-.33 (+/- $\epsilon$)&.03 &  III-Finite & Linear  (\greencheck) & $\infty$ \\ \cline{2-14}
    & \multirow{2}{*}{Heart} &  Male & .1/1.0&\multirow{2}{*}{.0}&\multirow{2}{*}{.04}&\multirow{2}{*}{-.01}&.01 (+/- $\epsilon$)&.00 (+/- $\epsilon$)&-.18 (+/- $\epsilon$)&.00& III-Finite & Linear  (\greencheck) & $\infty$ \\     
      &  &   Female & .2/.5&&&&.00 (+/- $\epsilon$)&.00 (+/- $\epsilon$)&.00 (+/- $\epsilon$)&.00&  I-Log &  Skewed-Left & 5,000  \\ \cline{2-14}
      & \multirow{2}{*}{Meps15} &  Male &  8.2/0&\multirow{2}{*}{.02}&\multirow{2}{*}{-.07}&\cellcolor{gray!20}&.01 (+/- $\epsilon$)&.00 (+/- $\epsilon$)&-.07 (+/- $\epsilon$)&.00& III-Finite & Linear  (\greencheck) & $\infty$ \\     
      &  &   Female & 7.6/0&&&\multirow{-2}{*}{.07}\cellcolor{gray!20}& \cellcolor{gray!20} .08 (+/- $.01$)&.02 (+/- $.01$)&-.35 (+/- $.01$)&.06 &  III-Finite & Linear  (\greencheck) & $\infty$\\ \cline{2-14}
      & \multirow{2}{*}{Meps16} &  Male & 8.3/0&\multirow{2}{*}{.03}&\multirow{2}{*}{-.13}&\cellcolor{gray!40}&.00 (+/- $\epsilon$)&.00 (+/- $\epsilon$)&-.26 (+/- $\epsilon$)&.00& III-Finite & Linear  (\greencheck) & $\infty$ \\     
      &  &   Female & 7.4/0&&& \multirow{-2}{*}{.12}\cellcolor{gray!40}&\cellcolor{gray!40} .12 (+/- $\epsilon$)&.01 (+/- $\epsilon$)&-.5 (+/- $\epsilon$)&.11 &  III-Finite & Linear  (\greencheck) & $\infty$ \\ \cline{2-14}
      & \multirow{2}{*}{Students} &  Male &  .6/0&\multirow{2}{*}{.0}&\multirow{2}{*}{.0}&\multirow{2}{*}{.0}&.01 (+/- $\epsilon$)&.00 (+/- $\epsilon$)&.36 (+/- $\epsilon$)&.00& II-Infinite& Heavy-Tail (\redx) & 0\\     
      &  &   Female &  .5/0&&&&.01 (+/- $\epsilon$)&.01 (+/- $\epsilon$)&.03 (+/- $\epsilon$)&.00&  I-Log & Skewed-Left & 10,000  \\ \cline{2-14}\hline

      \multirow{14}{*}{RF} & \multirow{2}{*}{Census} &  White &  25.7/0&\multirow{2}{*}{.04}&\multirow{2}{*}{-.11}&\multirow{2}{*}{.01}& \cellcolor{gray!120} .54 (+/- $.02$) & .07 (+/- $.02$)&-.16 (+/- $.02$)&.47 &   III-Finite & Linear  (\greencheck)  & $\infty$ \\
    &  &  Black & 3.1/0&&& &\cellcolor{gray!120} .55 (+/- $.03$)&.1 (+/- $.03$)&-.41 (+/- $.03$)&.42 & III-Finite & Linear (\greencheck)  & $\infty$ \\ \cline{2-14}
     & \multirow{2}{*}{Census} &  Male  & 20.1/0&\multirow{2}{*}{-.04}&\cellcolor{gray!20} &\multirow{2}{*}{-.18} &\cellcolor{gray!110} .36 (+/- $.01$)&.07 (+/- $.01$)&.00 (+/- $.01$)&.29 & I-Log &  Skewed-Left & 1,000 \\ 
    &  &  Female &  10.3/0&&\multirow{-2}{*}{.08}\cellcolor{gray!20} && \cellcolor{gray!120} .54 (+/- $.02$)&.08 (+/- $.02$)&-.27 (+/- $.02$)&.44 & III-Finite & Linear (\greencheck)  & $\infty$ \\\cline{2-14}
     & \multirow{2}{*}{Credit} &  Male & .3/1.0&\multirow{2}{*}{.01}&\multirow{2}{*}{-.03}&\multirow{2}{*}{.03}& \cellcolor{gray!20} .07 (+/- $.01$)&.02 (+/- $.01$)&-.34 (+/- $.01$)&.04&  III-Finite & Linear (\greencheck)  & $\infty$ \\
    &  &  Female & .7/0&& &&\cellcolor{gray!40} .1 (+/- $.01$)&.03 (+/- $.01$)&.01 (+/- $.01$)&.07 & I-Log & Skewed-Right& 500\\ \cline{2-14}
     & \multirow{2}{*}{Bank} &  Young &  9.5/0&\multirow{2}{*}{-.01}&\multirow{2}{*}{-.09}&\multirow{2}{*}{-.02}&\cellcolor{gray!60} .15 (+/- $.01$)&.03 (+/- $.01$)&-.49 (+/- $.01$)&.11 &  III-Finite & Linear (\greencheck)  & $\infty$ \\
    &  &  Old & 1.5/0&&&&\cellcolor{gray!40} .13 (+/- $.01$)&.04 (+/- $.01$)&-.23 (+/- $.01$)&.08 & III-Finite & Linear  (\greencheck)  & $\infty$ \\ \cline{2-14}
    %  & \multirow{2}{*}{Compas} &  Male & 3,527&\multirow{2}{*}{-.02}&\multirow{2}{*}{.0}&\multirow{2}{*}{-.18}&\cellcolor{gray!55} .48 (+/- $.02$)&.11 (+/- $.02$)&.00 (+/- $.02$)&.37 &  I-Log &  Skewed-Left & 5000\\
    % &  &  Female & 906&&&& \cellcolor{gray!45} .30 (+/- $.07$) & .2 (+/- $.07$)&-.56 (+/- $.07$)&.06 & III-Finite & Linear  (\greencheck) & $\infty$ \\ \cline{2-14}    
     & \multirow{2}{*}{Compas} &  Caucasian    & 1.5/0&\multirow{2}{*}{.02}&\multirow{2}{*}{.0}&\cellcolor{gray!60}& \cellcolor{gray!110} .36 (+/- $.02$) & .08 (+/- $.02$)&-.74 (+/- $.02$)&.23 &  III-Finite & Linear (\greencheck)  & $\infty$ \\
      &  &   Other & 3.0/0&&& \multirow{-2}{*}{.18}\cellcolor{gray!60}&\cellcolor{gray!120} .54 (+/- $.03$)&.06 (+/- $.03$)&-.69 (+/- $.03$)&.45 & III-Finite & Linear  (\greencheck) & $\infty$ \\  \cline{2-14}        
    %  & \multirow{2}{*}{Compas} &  Young & 2,559&\multirow{2}{*}{.01}&\multirow{2}{*}{.01}&\multirow{2}{*}{-.06}&\cellcolor{gray!45} .38 (+/- $.03$)&.1 (+/- $.03$)&-.23 (+/- $.03$)&.27&   III-Finite & Linear (\greencheck)  & $\infty$ \\
    % &  &  Old & 839&& &&\cellcolor{gray!45} .32 (+/- $.07$)&.17 (+/- $.07$)&-.69 (+/- $.07$)&.09 & III-Finite & Linear (\greencheck)  & $\infty$ \\  \cline{2-14}
    & \multirow{2}{*}{Default} &  Male &  11.4/0&\multirow{2}{*}{-.01}&\multirow{2}{*}{-.01}&\multirow{2}{*}{-.02}&\cellcolor{gray!120} .65 (+/- $.03$)&.09 (+/- $.03$)&-.21 (+/- $.03$)&.54 & III-Finite & Linear (\greencheck)  & $\infty$ \\     
      &  &   Female & 17.2/0&&&&  \cellcolor{gray!120} .63 (+/- $.02$) & .06 (+/- $.02$)&-.43 (+/- $.02$)&.56 &  III-Finite & Linear (\greencheck) & $\infty$  \\ \cline{2-14}
    & \multirow{2}{*}{Heart} &  Male & .1/2.0&\multirow{2}{*}{.0}&\multirow{2}{*}{.04}&\multirow{2}{*}{.0}&\cellcolor{gray!5}.05 (+/- $.01$)&.03 (+/- $.01$)&-.21 (+/- $.01$)&.01 & III-Finite & Linear (\greencheck)  & $\infty$ \\     
      &  &   Female & .2/1.5&&& &\cellcolor{gray!5}  .05 (+/- $.01$)&.04 (+/- $.01$)&-.12 (+/- $.01$)&.01 &III-Finite & Linear  (\greencheck) & $\infty$ \\ \cline{2-14}
      & \multirow{2}{*}{Meps15} &  Male &  8.2/0&\multirow{2}{*}{-.02}&\multirow{2}{*}{-.07}&\multirow{2}{*}{-.15}& \cellcolor{gray!120} .51 (+/- $\epsilon$)&.13 (+/- $\epsilon$)&-1.04 (+/- $\epsilon$)&.34 & III-Finite & Linear (\greencheck)  & $\infty$ \\     
      &  &   Female & 7.6/0&&& & \cellcolor{gray!110} .36 (+/- $.04$)&.13 (+/- $.04$)&-.21 (+/- $.04$)&.22 &  III-Finite & Linear (\greencheck)  & $\infty$ \\ \cline{2-14}
      & \multirow{2}{*}{Meps16} &  Male & 8.3/0&\multirow{2}{*}{-.01}&\multirow{2}{*}{-.13}&\multirow{2}{*}{-.03}& \cellcolor{gray!110} .38 (+/- $.05$)&.14 (+/- $.05$)&-.4 (+/- $.05$)&.22 & III-Finite & Linear  (\greencheck) & $\infty$ \\     
      &  &   Female & 7.4/0&& && \cellcolor{gray!110} .35 (+/- $.03$) &.08 (+/- $.03$)&-.38 (+/- $.03$)&.24 &  III-Finite & Linear (\greencheck)  & $\infty$ \\ \cline{2-14}
      & \multirow{2}{*}{Students} &  Male & .6/0&\multirow{2}{*}{.0}&\multirow{2}{*}{.0}&\multirow{2}{*}{.01}&.03 (+/- $.01$)&.02 (+/- $.01$)&-.38 (+/- $.01$)&.01 & III-Finite & Linear (\greencheck)  & $\infty$ \\     
      &  &   Female &  .5/0&&&&.04 (+/- $.01$)&.02 (+/- $.01$)&-.29 (+/- $.01$)&.02 & III-Finite & Linear (\greencheck)  & $\infty$ \\ \cline{2-14}\hline

    \end{tabular}
 }
 % \vspace{-.5em}
\end{table*}
% }

\vspace{0.25em}
\subsection{Feasibility, Usefulness, and Guarantee of EVT (RQ2)}
\label{sec:rq1}
One important investigation of this paper is to find out whether Extreme Value Theory (EVT) can effectively model the tail of ML outcome distributions. In Table~\ref{tab:EVT-validity-guarantee}, we present $80$ experiments with their corresponding EVT characteristics and the feasibility of EVT to provide fairness guarantees. The number of test cases generated for each group is shown in column $\#N$, determined by the exponential testing  in Algorithm~\ref{alg:algorithm1}. 
The numbers reported in this column include both the original sample size from the dataset and the additional synthetic samples required to pass the test. For instance, a value of 0.1/1.0 indicates that there are 100 original samples with 1000 additional synthetic samples. 
Columns ACD~\cite{10.1145/3106237.3106277}, CVaR~\cite{williamson2019fairness}, and ECD show average, conditional value at risk, and extreme counterfactual discrimination. 
In the columns ($\mu, \sigma, \xi, \tau$, type) of Table~\ref{tab:EVT-validity-guarantee}, we detail the characteristics of the GEV distribution for each benchmark that informs ECD. Here, $\mu$ represents the mean of the extreme value distribution at a specific threshold $\tau$ for each combination of algorithm, dataset, and subgroup. For instance, in the DNN application to the Census dataset with sex as protected attribute, we observe an ACD of 0.05, CVaR of 0.08, and ECD of 0.21 where $\mu_{M}$ and $\mu_{F}$ is 0.03 and 0.24, respectively,
implying a significant counterfactual discrimination toward female in the tail of DNN's outcome.

The shape  $\xi$ indicates the tail behavior of the GEV. A shape $\xi$ around zero or negative suggests that GEV can extrapolate for a long finite (based on Q-Q Plot) or infinite interactions with statistical guarantees, shown with $B$.
In 62 out of 80 scenarios (78\%), EVT results in a type III distribution with a negative shape, indicating a finite tail and enabling extrapolation for an unlimited number of queries. For 14 cases (18\%), EVT produces a type I distribution with a near-zero shape, implying an infinite but exponentially decaying tail, suitable for extrapolation within bounded queries $B$. 
Overall, the worst-case guarantees are achievable in 76 cases (95\%). 

We examine the relevance of extreme counterfactual discrimination in ML model fairness by employing EVT to measure tail biases, comparing them to established fairness metrics like ACD and CVaR. For instance, in the DNN model trained on the Compas dataset, an ACD of 0.02 and a CVaR of -0.01 indicate fairness in both average and tail cases, yet an ECD of 0.11 suggests a tail-bias toward Caucasians. 
We classify any ECD difference exceeding 0.05 as discrimination, with its significance indicated by the grayscale in the ECD column.
Out of 40 cases, ECD-based discrimination occurs in 19 (48\%). In contrast, average-case discrimination (ACD) is observed in 10 out of 40 cases (25\%). Notably, in 13 cases (33\%), ECD is significantly greater than ACD.
In 18 out of 40 experiments, ECD found significant discrimination against the unprivileged group in the tail that missed by the CVaR metric. 

\begin{tcolorbox}[boxrule=1pt,left=1pt,right=1pt,top=1pt,bottom=1pt]
\textbf{Answer RQ2:}
EVT effectively models extreme counterfactual discrimination (ECD), in 95\% of cases, allowing for valid extrapolation of worst-case discrimination. In 33\% of cases, ECD shows significantly higher discrimination than the average-case one (ACD~\cite{10.1145/3106237.3106277}). In 18 out of 40 experiments, ECD found significant discrimination against the unprivileged group in the tail that missed by prevalent tail-based metric (CVaR~\cite{williamson2019fairness}).

\end{tcolorbox}

\subsection{Validation of Prevalent Bias Mitigation Algorithms (RQ3)}
\label{sec:rq3}
\setlength{\tabcolsep}{14.0pt} % Default value: 6pt
\begin{table*}[tbp!]
    \caption{Average based bias mitigation.
    Legend:
      \textbf{P}: Protected Attribute,
      \textbf{Acc loss}:Accuracy loss in mitigation ,
      \textbf{AOD, EOD, SPD, DI}: Average-based Fairness Measures,
      \textbf{ECD} = $\mu_u-\mu_p$: The Amounts of ECD Tail Discrimination,
      \textbf{NV}: Not Valid.
    %   Legend:
    % \textbf{P}: Protected Attribute,
    % \textbf{Acc$'$}: Accuracy of mitigated model,
    % \textbf{EOD$'$}: EOD of mitigated model,
    % \textbf{AOD$'$}: AOD of mitigated model,
    % \textbf{ACD$'$}: ACD$^2$ of mitigated model,
    % \textbf{ECD$'$}: ECD$^2$ of mitigated model, 
    % \textbf{RLD}$'_{n}$: RLD for the next $n$ queries for mitigated model.           
    }
    \label{tab:EVT-Bias-Mitigation}
    \centering
    \resizebox{0.92\textwidth}{!}{
    {\color{black}\begin{tabular}{ | c | l  l |  c c c c c  c  | c  c c c c  c  | }
      \hline
      \multicolumn{3}{|c|}{\color{black}{\textbf{Name}}}  &  \multicolumn{6}{c|}{\color{black}{\textbf{Exponentiated Gradient (EG)~\cite{agarwal2018reductions}}}} & \multicolumn{6}{c|}{\color{black}{\textbf{Fair-SMOTE~\cite{10.1145/3468264.3468537}}}}
      \\ 
    \multicolumn{1}{|c}{\color{black}{Algorithm}}  & \color{black}{Dataset}  & \color{black}{P} & \color{black}{$Acc$} \color{black}{$loss$}  & \color{black}{$AOD$} & $EOD$ & $SPD$  & $DI$ & \color{black}{$ECD$} & \color{black}{Acc loss}  & \color{black}{$AOD$} & $EOD$ & $SPD$  & $DI$  & \color{black}{$ECD$} 
    \\ \hline
    \multirow{9}{*}{\color{black}{Avg-based}} & \multirow{2}{*}{\color{black}{census}} & \color{black}{race} & $-0.01$  &\cellcolor{gray!25}$0.04$ & \cellcolor{gray!25}$0.08$ & \cellcolor{gray!25}$0.06$ & $0.72$ & \cellcolor{gray!50}$0.08$   &$-0.04$ &\cellcolor{gray!50}$0.08$ & \cellcolor{gray!25}$0.11$ & \cellcolor{gray!75}$0.13$ & \cellcolor{gray!50}$1.49$ & \cellcolor{gray!75}$0.1$ \\
     & & \color{black}{sex} & $-0.03$&$0.01$ & $0.02$ & \cellcolor{gray!75}$0.09$ & \cellcolor{gray!25}$0.49$ & \cellcolor{gray!75}$0.11$    & $-0.07$ & \cellcolor{gray!50}$0.03$ & \cellcolor{gray!50}$0.05$ & \cellcolor{gray!100}$0.19$ & \cellcolor{gray!75}$0.68$ & \cellcolor{gray!100}$0.2$  \\ \cline{2-3}
   &  \color{black}{credit} & \color{black}{sex} & $-0.09$ & $0.02$ & $0.05$ & $0.02$ & $0.55$ & \cellcolor{gray!100}$0.65$   & $-0.01$ &\cellcolor{gray!50}$0.09$ & \cellcolor{gray!50}$0.15$ & \cellcolor{gray!50}$0.09$ & $0.4$ & \cellcolor{gray!75}$0.11$    \\ \cline{2-3}
    \color{black}{DNN} & \color{black}{bank} & \color{black}{age} & $-0.01$&\cellcolor{gray!25}$0.03$ & \cellcolor{gray!25}$0.05$ & \cellcolor{gray!25}$0.01$ & \cellcolor{gray!25}$0.4$ & \cellcolor{gray!50}$0.09$    &$0.0$ & \cellcolor{gray!25}$0.03$ & \cellcolor{gray!25}$0.06$ & \cellcolor{gray!50}$0.02$ & \cellcolor{gray!25}$0.26$ & \cellcolor{gray!25}$0.06$    \\ \cline{2-3}
    
     & \multirow{1}{*}{\color{black}{compas}} & \color{black}{race}&$-0.01$ &$0.03$ & $0.0$ & \cellcolor{gray!25}$0.08$ & \cellcolor{gray!25}$0.14$ & \cellcolor{gray!25}$0.09$    &$-0.02$ &  $0.03$ & \cellcolor{gray!50}$0.01$ & \cellcolor{gray!25}$0.07$ & \cellcolor{gray!25}$0.13$ & \cellcolor{gray!50}$0.15$ \\ \cline{2-3}

    % & \multirow{3}{*}{Compas} & sex & .0&.01&-.02& -.07&.0  & .02 & -.01 & \cellcolor{gray!45}.12 \\
    % &  &  race &  .0&.01&-.03& -.40&-.01 & .02 & .01 & \cellcolor{gray!45}.12 \\
    % & &  age & .0&.01&.04& \cellcolor{gray!45}.45&.0  & .02 & .0 & -.08 \\ \cline{2-3}
     & \color{black}{default} & \color{black}{sex} & $0.0$&$0.01$ & $0.01$ & $0.02$ & $0.18$ & \cellcolor{gray!100}$0.3$ &$-0.02$ & \cellcolor{gray!25}$0.03$ & \cellcolor{gray!75}$0.04$ & \cellcolor{gray!25}$0.04$ & \cellcolor{gray!25}$0.25$ & \cellcolor{gray!25}$0.06$      \\ \cline{2-3}
     & \color{black}{heart} & \color{black}{sex} & $-0.06$&$0.02$ & $0.03$ & $0.06$ & $0.23$ & \cellcolor{gray!75}$0.17$     &$-0.08$ & \cellcolor{gray!25}$0.15$ & \cellcolor{gray!25}$0.26$ & \cellcolor{gray!50}$0.18$ & \cellcolor{gray!25}$0.45$ & \cellcolor{gray!75}$0.12$   \\ \cline{2-3}
     & \color{black}{meps15} & \color{black}{sex} & $-0.04$&\cellcolor{gray!25}$0.02$ & $0.02$ & \cellcolor{gray!25}$0.06$ & \cellcolor{gray!25}$0.61$ & \cellcolor{gray!50}$0.05$     & $-0.03$ &  \cellcolor{gray!25}$0.04$ & \cellcolor{gray!50}$0.05$ & \cellcolor{gray!25}$0.06$ & \cellcolor{gray!25}$0.65$ & \cellcolor{gray!75}$0.09$  \\ \cline{2-3}
     & \color{black}{meps16} & \color{black}{sex} &$0.01$ &$0.01$ & $0.01$ & \cellcolor{gray!25}$0.05$ & $0.57$ & \cellcolor{gray!100}$0.42$  & $-0.03$ & \cellcolor{gray!75}$0.05$ & \cellcolor{gray!75}$0.08$ & \cellcolor{gray!50}$0.07$ & \cellcolor{gray!25}$0.87$ & \cellcolor{gray!50}$0.04$  \\ \cline{2-3}
     & \color{black}{students} & \color{black}{sex} &$-0.07$ & $0.02$ & $0.02$ & $0.03$ & $0.03$ & \cellcolor{gray!50}$0.09$ &$0.02$ & \cellcolor{gray!50}$0.1$ & \cellcolor{gray!25}$0.05$ & \cellcolor{gray!25}$0.06$ & \cellcolor{gray!25}$0.07$ & \cellcolor{gray!25}$0.05$ \\ \cline{2-3}
    \hline
    \multirow{9}{*}{\color{black}{Avg-based}} & \multirow{2}{*}{\color{black}{census}} & \color{black}{race} &$-0.03$ & \cellcolor{gray!25}$0.04$ & \cellcolor{gray!25}$0.07$ & \cellcolor{gray!25}$0.05$ & $0.53$ & \cellcolor{gray!75}$0.47$   &$0.03$ & \cellcolor{gray!100}$0.3$ & \cellcolor{gray!100}$0.47$ & \cellcolor{gray!125}$0.24$ & \cellcolor{gray!75}$8.14$ & \cellcolor{gray!50}$0.4$     \\
       & & \color{black}{sex} & $-0.01$&$0.02$ & $0.03$ & \cellcolor{gray!25}$0.07$ & $0.47$ & \cellcolor{gray!50}$0.19$   &$0.02$ & \cellcolor{gray!75}$0.12$ & \cellcolor{gray!50}$0.13$ & \cellcolor{gray!125}$0.21$ & \cellcolor{gray!50}$0.71$ & \cellcolor{gray!75}$0.29$  \\ \cline{2-3}
    & \color{black}{credit} & \color{black}{sex} & $0.0$ &\cellcolor{gray!25}$0.07$ & $0.1$ & \cellcolor{gray!25}$0.07$ & $0.5$ & \cellcolor{gray!25}$0.02$ &$0.0$  & \cellcolor{gray!25}$0.11$ & \cellcolor{gray!25}$0.17$ & \cellcolor{gray!25}$0.08$ & $0.41$ & \cellcolor{gray!75}$0.09$    \\ \cline{2-3}
    \color{black}{LR} & \color{black}{bank} & \color{black}{age} & $-0.11$&\cellcolor{gray!25}$0.05$ & \cellcolor{gray!25}$0.07$ & \cellcolor{gray!75}$0.03$ & \cellcolor{gray!25}$0.44$ & \cellcolor{gray!50}$0.13$ &$0.0$ & \cellcolor{gray!75}$0.14$ & \cellcolor{gray!75}$0.25$ & \cellcolor{gray!100}$0.06$ & \cellcolor{gray!75}$2.28$ & \cellcolor{gray!75}$0.22$   \\ \cline{2-3}
    
     & \multirow{1}{*}{\color{black}{compas}} & \color{black}{race} & $-0.05$&$0.01$ & \cellcolor{gray!25}$0.01$ & $0.05$ & $0.09$ & \cellcolor{gray!75}$0.22$   &$-0.01$ & \cellcolor{gray!25}$0.03$ & \cellcolor{gray!25}$0.0$ & \cellcolor{gray!25}$0.07$ & $0.13$ & \cellcolor{gray!25}$0.07$     \\ \cline{2-3}

    % & \multirow{3}{*}{Compas} & sex & .0&.01&-.02& -.07&.0  & .02 & -.01 & \cellcolor{gray!45}.12 \\
    % &  &  race &  .0&.01&-.03& -.40&-.01 & .02 & .01 & \cellcolor{gray!45}.12 \\
    % & &  age & .0&.01&.04& \cellcolor{gray!45}.45&.0  & .02 & .0 & -.08 \\ \cline{2-3}
     & \color{black}{default} & \color{black}{sex} & $-0.05$&$0.02$ & $0.02$ & \cellcolor{gray!25}$0.02$ & \cellcolor{gray!25}$0.28$ & \cellcolor{gray!50}$0.1$   &$0.02$ & \cellcolor{gray!75}$0.08$ & \cellcolor{gray!50}$0.12$ & \cellcolor{gray!75}$0.05$ & \cellcolor{gray!25}$0.29$ & \cellcolor{gray!75}$0.19$  \\ \cline{2-3}
     & \color{black}{heart} & \color{black}{sex} & $-0.06$&$0.12$ & $0.24$ & $0.09$ & $0.45$ & \cellcolor{gray!100}$0.37$ & $0.19$ &\cellcolor{gray!25}$0.19$ & \cellcolor{gray!25}$0.33$ & \cellcolor{gray!25}$0.17$ & $0.48$ & \cellcolor{gray!50}$0.06$  \\ \cline{2-3}
     & \color{black}{meps15} & \color{black}{sex} & $-0.03$& $0.02$ & $0.03$ & \cellcolor{gray!50}$0.05$ & $0.58$ & \cellcolor{gray!25}$0.09$   &$-0.03$ & \cellcolor{gray!25}$0.03$ & \cellcolor{gray!25}$0.05$ & \cellcolor{gray!75}$0.05$ & \cellcolor{gray!25}$0.75$ & $0.03$   \\ \cline{2-3}
     & \color{black}{meps16} & \color{black}{sex} &$0.01$ &  $0.01$ & $0.02$ & $0.04$ & $0.63$ & \cellcolor{gray!100}$0.79$ &$-0.03$ &\cellcolor{gray!75}$0.08$ & \cellcolor{gray!50}$0.13$ & \cellcolor{gray!50}$0.08$ & \cellcolor{gray!75}$1.38$ & \cellcolor{gray!75}$0.26$    \\ \cline{2-3}
     & \color{black}{students} & \color{black}{sex} &$-0.03$ &$0.04$ & \cellcolor{gray!25}$0.03$ & \cellcolor{gray!25}$0.06$ & \cellcolor{gray!25}$0.07$ & \cellcolor{gray!50}$0.05$   &$-0.01$ & \cellcolor{gray!25}$0.08$ & \cellcolor{gray!50}$0.05$ & \cellcolor{gray!25}$0.05$ & \cellcolor{gray!25}$0.07$ & NV     \\ \cline{2-3}
     \hline

    \multirow{9}{*}{\color{black}{Avg-based}} & \multirow{2}{*}{\color{black}{census}} & \color{black}{race} & $-0.05$&$0.02$ & $0.04$ & $0.04$ & $0.54$ & \cellcolor{gray!25}$0.31$ &$-0.01$ & \cellcolor{gray!50}$0.18$ & \cellcolor{gray!50}$0.29$ & \cellcolor{gray!50}$0.16$ & \cellcolor{gray!50}$3.47$ & $0.08$         \\
    & & \color{black}{sex} &$-0.04$ &$0.01$ & $0.03$ & $0.07$ & $0.53$ & \cellcolor{gray!75}$0.72$  &$0.07$ & \cellcolor{gray!50}$0.11$ & \cellcolor{gray!50}$0.13$ & \cellcolor{gray!75}$0.19$ & \cellcolor{gray!50}$0.72$ & \cellcolor{gray!50}$0.43$ \\\cline{2-3}
    & \color{black}{credit} & \color{black}{sex} &$-0.08$ &$0.04$ & $0.06$ & $0.05$ & $0.23$ & \cellcolor{gray!50}$0.23$  &$-0.02$  & \cellcolor{gray!25}$0.09$ & \cellcolor{gray!25}$0.13$ & $0.07$ & \cellcolor{gray!25}$0.65$ & \cellcolor{gray!25}$0.04$    \\\cline{2-3}
    \color{black}{SVM} & \color{black}{Bank} & \color{black}{age} & $-0.06$&$0.01$ & $0.02$ & $-0.0$ & $0.21$ & \cellcolor{gray!25}$0.03$  &$0.01$ & \cellcolor{gray!25}$0.05$ & \cellcolor{gray!25}$0.1$ & \cellcolor{gray!25}$0.02$ & \cellcolor{gray!50}$1.07$ & \cellcolor{gray!75}$0.24$   \\\cline{2-3}

    & \multirow{1}{*}{\color{black}{compas}} & \color{black}{race} & $-0.03$&$0.03$ & $0.0$ & $0.05$ & $0.08$ & $0.02$  & $0.01$ & $0.03$ & $0.0$ & $0.07$ & $0.13$ & \cellcolor{gray!25}$0.06$ \\  \cline{2-3}

    % & \multirow{3}{*}{Compas} & sex &.0&.02&-.04 &-.08&.01  & .02 & .0 & .0 \\  
    % &  &  race &  .01&.01&-.02 &-.06&.01  & .01 & .0 & -.04 \\
    % & &  age & .0&.0&-.01 &-.03&.01  & .02 & .0 & -.01 \\\cline{2-3} 
    
     & \color{black}{default} & \color{black}{sex} & $-0.02$&$0.02$ & $0.03$ & \cellcolor{gray!25}$0.02$ & \cellcolor{gray!50}$0.29$ & \cellcolor{gray!25}$0.02$   &$-0.02$ & $0.02$ & $0.04$ & \cellcolor{gray!25}$0.02$ & $0.15$ & \cellcolor{gray!50}$0.1$  \\\cline{2-3}
     & \color{black}{heart} & \color{black}{sex} & $0.03$& $0.06$ & $0.19$ & $0.13$ & $0.32$ & \cellcolor{gray!50}$0.1$   &$0.08$ & \cellcolor{gray!25}$0.2$ & \cellcolor{gray!25}$0.33$ & $0.19$ & \cellcolor{gray!25}$0.55$ & \cellcolor{gray!25}$0.06$   \\\cline{2-3}
     & \color{black}{meps15} & \color{black}{sex} & $0.0$&$0.02$ & $0.02$ & \cellcolor{gray!25}$0.04$ & $0.53$ & \cellcolor{gray!50}$0.3$    &-$0.02$ &$0.02$ & $0.04$ & \cellcolor{gray!25}$0.04$ & $0.54$ & \cellcolor{gray!25}$0.1$      \\\cline{2-3}
     & \color{black}{meps16} & \color{black}{sex} & $-0.04$&$0.02$ & $0.03$ & $0.03$ & \cellcolor{gray!25}$0.56$ & \cellcolor{gray!75}$0.52$     &$-0.03$ &$0.02$ & $0.04$ & \cellcolor{gray!25}$0.04$ & \cellcolor{gray!50}$0.7$ & \cellcolor{gray!50}$0.13$     \\\cline{2-3}
     & \color{black}{students} & \color{black}{sex} & $-0.09$& $0.02$ & \cellcolor{gray!25}$0.04$ & $0.02$ & $0.03$ & \cellcolor{gray!75}$0.72$ &$0.03$ & \cellcolor{gray!25}$0.08$ & $0.02$ & \cellcolor{gray!25}$0.06$ & \cellcolor{gray!25}$0.06$ & \cellcolor{gray!50}$0.03$  \\\cline{2-3}\hline

     \multirow{9}{*}{\color{black}{Avg-based}} & \multirow{2}{*}{\color{black}{census}} & \color{black}{race} &$-0.03$ &$0.09$ & $0.11$ & \cellcolor{gray!25}$0.13$ & $1.58$ & $0.07$  &$-0.01$ & $0.1$ & $0.13$ & \cellcolor{gray!25}$0.14$ & $1.85$ & \cellcolor{gray!25}$0.17$      \\
    & & \color{black}{sex} & $-0.04$&$0.08$ & $0.07$ & \cellcolor{gray!50}$0.18$ & \cellcolor{gray!25}$0.69$ & $0.04$ &$0.08$ & \cellcolor{gray!25}$0.1$ & \cellcolor{gray!25}$0.11$ & \cellcolor{gray!50}$0.18$ & \cellcolor{gray!50}$0.71$ & \cellcolor{gray!25}$0.07$     \\\cline{2-3}
    & \color{black}{credit} & \color{black}{sex} &$-0.05$ &$0.06$ & $0.17$ & $0.08$ & $0.31$ & NV   & $0.01$ & $0.08$ & $0.12$ & $0.08$ & $0.53$ & \cellcolor{gray!50}$0.15$   \\\cline{2-3}
    \color{black}{RF} & \color{black}{bank} & \color{black}{age} &$0.01$ &\cellcolor{gray!25}$0.01$ & \cellcolor{gray!25}$0.03$ & \cellcolor{gray!25}$0.01$ & $0.18$ & \cellcolor{gray!75}$0.17$   &$0.01$ & \cellcolor{gray!50}$0.02$ & \cellcolor{gray!25}$0.04$ & \cellcolor{gray!25}$0.01$ & $0.27$ & \cellcolor{gray!25}$0.05$     \\\cline{2-3}

    & \multirow{1}{*}{\color{black}{compas}} &  \color{black}{race} & $-0.02$&$0.02$ & \cellcolor{gray!50}$0.01$ & $0.07$ & $0.11$ & $0.03$ &$0.0$ & $0.03$ & \cellcolor{gray!25}$0.01$ & $0.07$ & $0.13$ & \cellcolor{gray!25}$0.05$   \\  \cline{2-3}

    % & \multirow{3}{*}{Compas} & sex & -.01&.02&-.03 &-.03&.0  & .0 & -.02 & .03 \\  
    % &  &  race &  -.02&.0&-.03 &.0&.02  & .0 & -.04 & -.28 \\
    % & &  age & -.02&.01&-.02 &.0&.0  & .01 & .05 & -.03\\ \cline{2-3} 
     & \color{black}{default} & \color{black}{sex} & $-0.07$&$0.01$ & $0.02$ & \cellcolor{gray!25}$0.03$ & $0.18$ & $0.02$     &$0.08$ & \cellcolor{gray!25}$0.02$ & \cellcolor{gray!25}$0.03$ & \cellcolor{gray!25}$0.03$ & $0.23$ & \cellcolor{gray!25}$0.06$   \\\cline{2-3}
     & \color{black}{heart} & \color{black}{sex} & $-0.09$&$0.07$ & $0.08$ & \cellcolor{gray!25}$0.18$ & $0.43$ & \cellcolor{gray!25}$0.06$     &$-0.08$ & \cellcolor{gray!25}$0.23$ & \cellcolor{gray!25}$0.36$ & \cellcolor{gray!25}$0.18$ & \cellcolor{gray!25}$0.64$ & \cellcolor{gray!50}$0.12$       \\\cline{2-3}
     & \color{black}{meps15} & \color{black}{sex} & $-0.05$&\cellcolor{gray!25}$0.05$ & $0.07$ & \cellcolor{gray!25}$0.07$ & \cellcolor{gray!50}$0.86$ & \cellcolor{gray!25}$0.04$   &$-0.02$ & $0.03$ & $0.05$ & $0.04$ & $0.44$ & \cellcolor{gray!50}$0.16$  \\\cline{2-3}
     & \color{black}{meps16} & \color{black}{sex} & $-0.02$&$0.02$ & $0.03$ & $0.04$ & $0.57$ & \cellcolor{gray!25}$0.01$   &$-0.03$ & $0.02$ & $0.04$ & $0.04$ & $0.52$ & \cellcolor{gray!25}$0.01$   \\\cline{2-3}
     & \color{black}{students} & \color{black}{sex} &$-0.01$ &$0.07$ & $0.01$ & $0.02$ & $0.03$ & NV &$0.0$ & $0.05$ & \cellcolor{gray!25}$0.04$ & \cellcolor{gray!25}$0.05$ & \cellcolor{gray!25}$0.06$ & $0.01$   \\\cline{2-3}\hline

    \end{tabular}}
 }
 % \vspace{-1.0 em}
\end{table*}

\setlength{\tabcolsep}{14.0pt} % Default value: 6pt
\begin{table*}[tbp!]
    \caption{\revision{Average-based bias mitigation (STEALTH~\cite{10109333} and MAAT~\cite{10.1145/3540250.3549093}).
    Legend is similar to Table~\ref{tab:EVT-Bias-Mitigation}
    %   Legend:
    % \textbf{P}: Protected Attribute,
    % \textbf{Acc$'$}: Accuracy of mitigated model,
    % \textbf{EOD$'$}: EOD of mitigated model,
    % \textbf{AOD$'$}: AOD of mitigated model,
    % \textbf{ACD$'$}: ACD$^2$ of mitigated model,
    % \textbf{ECD$'$}: ECD$^2$ of mitigated model, 
    % \textbf{RLD}$'_{n}$: RLD for the next $n$ queries for mitigated model.           
    }}
    \centering
    \label{tab:stealth_maat}
    \resizebox{0.92\textwidth}{!}{
     {\color{black}\begin{tabular}{ | c | l  l |  c c c c c  c  | c  c c c c  c  | }
      \hline
      \multicolumn{3}{|c|}{\textbf{Name}}  &  \multicolumn{6}{c|}{\textbf{STEALTH~\cite{10109333}}} & \multicolumn{6}{c|}{\textbf{MAAT~\cite{10.1145/3540250.3549093}}}
      \\
    \multicolumn{1}{|c}{Algorithm}  & Dataset  & P & $Acc$ $loss$ & $AOD$ & $EOD$ & $SPD$  & $DI$& $ECD$ & Acc loss  & $AOD  $& $EOD$ & $SPD$  &$DI$  & $ECD$ 
    \\ \hline
    \multirow{9}{*}{Avg-based} & \multirow{2}{*}{census} & race & $-0.05$ & \cellcolor{gray!50}$0.08$ & \cellcolor{gray!50}$0.12$ & \cellcolor{gray!50}$0.09$ & \cellcolor{gray!75}$2.74$ & \cellcolor{gray!25}$0.03$ & $0.02$ & $0.02$ & $0.04$ & \cellcolor{gray!75}$0.14$ & \cellcolor{gray!25}$1.17$ & \cellcolor{gray!75}$0.13$    \\
     & & sex & $0.05$ & \cellcolor{gray!100}$0.18$ & \cellcolor{gray!100}$0.27$ & \cellcolor{gray!100}$0.17$ & \cellcolor{gray!100}$0.87$ & \cellcolor{gray!50}$0.07$ & $0.11$ & \cellcolor{gray!75}$0.1$ & \cellcolor{gray!75}$0.2$ & \cellcolor{gray!50}$0.08$ & $0.43$ & \cellcolor{gray!50}$0.08$ \\ \cline{2-3}
    & credit & sex& $0.01$ & \cellcolor{gray!25}$0.06$ & \cellcolor{gray!25}$0.1$ & \cellcolor{gray!25}$0.06$ & $0.43$ & \cellcolor{gray!75}$0.1$ & $0.19$ & \cellcolor{gray!50}$0.1$ & \cellcolor{gray!75}$0.2$ & \cellcolor{gray!25}$0.07$ & $0.47$ & \cellcolor{gray!25}$0.02$   \\ \cline{2-3}
    DNN & bank & age& $0.0$ & $0.02$ & $0.03$ & $0.0$ & \cellcolor{gray!25}$0.34$ & $0.02$ & $0.06$ & \cellcolor{gray!25}$0.03$ & \cellcolor{gray!25}$0.06$ & \cellcolor{gray!50}$0.02$ & \cellcolor{gray!25}$0.32$ & \cellcolor{gray!25}$0.05$  \\ \cline{2-3}
    
     & \multirow{1}{*}{compas} & race& $-0.01$ & $0.03$ & $0.0$ & \cellcolor{gray!25}$0.07$ & \cellcolor{gray!25}$0.13$ & $0.03$ & $0.01$ & $0.02$ & \cellcolor{gray!25}$0.0$ & \cellcolor{gray!25}$0.07$ & \cellcolor{gray!25}$0.13$ & \cellcolor{gray!50}$0.11$  \\ \cline{2-3}

    % & \multirow{3}{*}{Compas} & sex & .0&.01&-.02& -.07&.0  & .02 & -.01 & \cellcolor{gray!45}.12 \\
    % &  &  race &  .0&.01&-.03& -.40&-.01 & .02 & .01 & \cellcolor{gray!45}.12 \\
    % & &  age & .0&.01&.04& \cellcolor{gray!45}.45&.0  & .02 & .0 & -.08 \\ \cline{2-3}
     & default & sex &  $-0.03$ & $0.02$ & \cellcolor{gray!50}$0.03$ & $0.02$ & \cellcolor{gray!25}$0.23$ & $0.01$ & $0.09$ &\cellcolor{gray!50}$0.04$ & \cellcolor{gray!100}$0.07$ & \cellcolor{gray!75}$0.06$ & \cellcolor{gray!75}$0.41$ & \cellcolor{gray!75}$0.11$  \\ \cline{2-3}
     & heart & sex & $0.0$ & \cellcolor{gray!25}$0.14$ & \cellcolor{gray!25}$0.24$ & \cellcolor{gray!25}$0.12$ & \cellcolor{gray!50}$0.61$ & \cellcolor{gray!50}$0.08$ & $0.06$ &\cellcolor{gray!50}$0.19$ & \cellcolor{gray!50}$0.38$ & \cellcolor{gray!25}$0.1$ & \cellcolor{gray!25}$0.47$ & \cellcolor{gray!25}$0.02$   \\ \cline{2-3}
     & meps15 & sex& $-0.03$ &  \cellcolor{gray!25}$0.03$ & \cellcolor{gray!50}$0.05$ & $0.04$ & \cellcolor{gray!25}$0.66$ & $0.01$& $0.08$ & \cellcolor{gray!25}$0.03$ & \cellcolor{gray!50}$0.06$ & \cellcolor{gray!50}$0.08$ & \cellcolor{gray!25}$0.73$ & \cellcolor{gray!100}$0.12$   \\ \cline{2-3}
     & meps16 & sex & $-0.04$ & \cellcolor{gray!25}$0.02$ & \cellcolor{gray!25}$0.04$ & $0.04$ & $0.65$ & $0.01$ & $0.08$ & \cellcolor{gray!25}$0.03$ & \cellcolor{gray!50}$0.05$ & \cellcolor{gray!75}$0.08$ & \cellcolor{gray!25}$0.83$ & \cellcolor{gray!75}$0.08$  \\ \cline{2-3}
     & students & sex & $0.05$ & $0.03$ & $0.03$ & $0.02$ & $0.03$ & $0.02$ & $0.1$ & \cellcolor{gray!25}$0.08$ & $0.03$ & \cellcolor{gray!25}$0.06$ & \cellcolor{gray!25}$0.08$ & $0.02$\\ \cline{2-3}
    \hline
    \multirow{9}{*}{Avg-based} & \multirow{2}{*}{census} & race & $-0.06$ & \cellcolor{gray!75}$0.17$ & \cellcolor{gray!75}$0.25$ & \cellcolor{gray!75}$0.17$ & \cellcolor{gray!50}$4.59$ & \cellcolor{gray!25}$0.12$ & $0.06$ &\cellcolor{gray!100}$0.29$ & \cellcolor{gray!100}$0.52$ & \cellcolor{gray!100}$0.22$ & \cellcolor{gray!50}$5.0$ & \cellcolor{gray!25}$0.16$  \\
    & & sex &$0.04$ & \cellcolor{gray!100}$0.19$ & \cellcolor{gray!75}$0.29$ & \cellcolor{gray!100}$0.18$ & \cellcolor{gray!100}$0.87$ & $0.06$ & $0.09$ & \cellcolor{gray!25}$0.05$ & \cellcolor{gray!50}$0.11$ & \cellcolor{gray!50}$0.1$ & \cellcolor{gray!25}$0.52$ & \cellcolor{gray!50}$0.18$   \\ \cline{2-3}
    & credit & sex & $0.01$ & \cellcolor{gray!25}$0.09$ & \cellcolor{gray!25}$0.17$ & \cellcolor{gray!25}$0.06$ & $0.47$ & \cellcolor{gray!25}$0.03$ & $0.14$ & \cellcolor{gray!25}$0.1$ & \cellcolor{gray!25}$0.19$ & \cellcolor{gray!25}$0.07$ & $0.68$ & \cellcolor{gray!50}$0.04$   \\ \cline{2-3}
    LR & bank & age&  $-0.01$ & $0.0$ & $0.01$ & $0.0$ & \cellcolor{gray!25}$0.6$ & $0.02$ & $0.05$ & \cellcolor{gray!50}$0.12$ & \cellcolor{gray!75}$0.22$ & \cellcolor{gray!75}$0.04$ & \cellcolor{gray!50}$0.93$ & \cellcolor{gray!50}$0.13$  \\ \cline{2-3}
    
     & \multirow{1}{*}{compas} &  race &$-0.01$ & \cellcolor{gray!25}$0.03$ & \cellcolor{gray!25}$0.01$ & \cellcolor{gray!25}$0.07$ & $0.13$ & $0.04$ & $-0.01$ & \cellcolor{gray!25}$0.02$ & \cellcolor{gray!25}$0.01$ & \cellcolor{gray!25}$0.07$ & $0.12$ & \cellcolor{gray!50}$0.1$  \\  \cline{2-3} 
    
    % & \multirow{3}{*}{Compas} & sex & .0&.02&-.03&-.08&.01 & .02 & -.03 & -.04 \\  
    % &  &  race &  -.02&.02&-.08& -.44&.01  & .01 & -.03 & -.05 \\
    % & &  age & .0&.01&-.02 &-.05&.0 & .01 & .02 & .01 \\ \cline{2-3} 
     & default & sex  &$-0.05$ & $0.02$ & $0.03$ & $0.01$ & \cellcolor{gray!25}$0.22$ & $0.02$ & $0.06$ & \cellcolor{gray!50}$0.05$ & \cellcolor{gray!50}$0.09$ & $0.01$ & $0.08$ & \cellcolor{gray!50}$0.12$ \\ \cline{2-3}
     & heart & sex & $0.09$ & \cellcolor{gray!25}$0.21$ & \cellcolor{gray!25}$0.37$ & \cellcolor{gray!25}$0.16$ & \cellcolor{gray!25}$0.68$ & \cellcolor{gray!25}$0.04$ & $0.13$ & \cellcolor{gray!25}$0.18$ & \cellcolor{gray!25}$0.36$ & \cellcolor{gray!25}$0.13$ & $0.48$ & $0.02$    \\ \cline{2-3}
     & meps15 & sex& $-0.03$ & $0.02$ & $0.04$ & $0.04$ & $0.56$ & $0.02$ & $0.05$ & $0.03$ & \cellcolor{gray!25}$0.06$ & \cellcolor{gray!100}$0.06$ & \cellcolor{gray!25}$0.66$ & \cellcolor{gray!25}$0.08$ \\ \cline{2-3}
     & meps16 & sex &$-0.03$ & \cellcolor{gray!25}$0.03$ & \cellcolor{gray!25}$0.05$ & $0.04$ & \cellcolor{gray!25}$0.74$ & $0.03$ & $0.06$ & $0.02$ & $0.03$ & \cellcolor{gray!25}$0.07$ & \cellcolor{gray!25}$0.77$ & \cellcolor{gray!50}$0.1$  \\ \cline{2-3}
     & students & sex & $0.0$ & \cellcolor{gray!25}$0.08$ & \cellcolor{gray!25}$0.04$ & \cellcolor{gray!25}$0.05$ & \cellcolor{gray!25}$0.06$ & \cellcolor{gray!25}$0.02$ & $0.08$ & \cellcolor{gray!25}$0.07$ & \cellcolor{gray!25}$0.03$ & \cellcolor{gray!25}$0.05$ & \cellcolor{gray!25}$0.07$ & \cellcolor{gray!25}$0.02$    \\ \cline{2-3}
     \hline

    \multirow{9}{*}{Avg-based} & \multirow{2}{*}{census} & race & $-0.06$ & \cellcolor{gray!25}$0.11$ & \cellcolor{gray!25}$0.18$ & \cellcolor{gray!25}$0.09$ & \cellcolor{gray!50}$4.98$ & $0.06$& $-0.06$ &  \cellcolor{gray!25}$0.1$ & \cellcolor{gray!25}$0.16$ & \cellcolor{gray!50}$0.16$ & \cellcolor{gray!25}$1.84$ & $0.08$     \\
    & & sex & $0.04$ & \cellcolor{gray!75}$0.16$ & \cellcolor{gray!75}$0.27$ & \cellcolor{gray!50}$0.14$ & \cellcolor{gray!75}$0.92$ & $0.07$ & $0.09$ & \cellcolor{gray!25}$0.03$ & \cellcolor{gray!25}$0.07$ & \cellcolor{gray!25}$0.12$ & \cellcolor{gray!25}$0.56$ & \cellcolor{gray!25}$0.13$  \\\cline{2-3}
    & credit & sex & $0.02$ & $0.05$ & $0.09$ & $0.04$ & \cellcolor{gray!25}$0.64$ & $0.01$ & $0.06$&\cellcolor{gray!25}$0.1$ & \cellcolor{gray!25}$0.17$ & $0.05$ & $0.28$ & \cellcolor{gray!25}$0.03$  \\\cline{2-3}
    SVM & bank & age & $-0.0$ & $0.01$ & $0.02$ & $0.0$ & \cellcolor{gray!25}$0.75$ & $0.0$ & $0.03$ & \cellcolor{gray!25}$0.05$ & \cellcolor{gray!25}$0.1$ & \cellcolor{gray!25}$0.02$ & \cellcolor{gray!25}$0.56$ & \cellcolor{gray!50}$0.15$    \\\cline{2-3}

    & \multirow{1}{*}{compas} & race &  $-0.01$ & $0.03$ & $0.0$ & $0.07$ & $0.13$ & $0.01$ & $-0.0$ & $0.02$ & $0.0$ & $0.07$ & $0.13$ & \cellcolor{gray!50}$0.11$    \\  \cline{2-3}

    % & \multirow{3}{*}{Compas} & sex &.0&.02&-.04 &-.08&.01  & .02 & .0 & .0 \\  
    % &  &  race &  .01&.01&-.02 &-.06&.01  & .01 & .0 & -.04 \\
    % & &  age & .0&.0&-.01 &-.03&.01  & .02 & .0 & -.01 \\\cline{2-3} 
    
     & default & sex &$-0.05$ & $0.01$ & $0.02$ & $0.01$ & \cellcolor{gray!25}$0.21$ & $0.01$ & $0.04$ & \cellcolor{gray!25}$0.02$ & \cellcolor{gray!25}$0.05$ & \cellcolor{gray!25}$0.02$ & $0.12$ & \cellcolor{gray!50}$0.12$   \\\cline{2-3}
     & heart & sex & $0.09$ & \cellcolor{gray!25}$0.18$ & \cellcolor{gray!25}$0.35$ & $0.15$ & \cellcolor{gray!25}$0.62$ & $0.02$ & $0.14$ & \cellcolor{gray!25}$0.19$ & \cellcolor{gray!25}$0.38$ & $0.12$ & $0.4$ & $0.02$  \\\cline{2-3}
     & meps15 & sex & $-0.03$ &  $0.02$ & $0.04$ & $0.03$ & $0.64$ & $0.0$ & $0.03$ & \cellcolor{gray!25}$0.03$ & \cellcolor{gray!25}$0.08$ & \cellcolor{gray!50}$0.05$ & $0.47$ & \cellcolor{gray!25}$0.12$       \\\cline{2-3}
     & meps16 & sex & $0.04$ &$0.02$ & $0.03$ & $0.03$ & \cellcolor{gray!25}$0.57$ & $0.0$ & $0.02$ & \cellcolor{gray!25}$0.05$ & \cellcolor{gray!25}$0.12$ & \cellcolor{gray!25}$0.04$ & $0.43$ & \cellcolor{gray!25}$0.03$   \\\cline{2-3}
     & students & sex & $-0.01$ &\cellcolor{gray!25}$0.08$ & $0.02$ & \cellcolor{gray!25}$0.05$ & \cellcolor{gray!25}$0.06$ & $0.0$ & $0.1$ & \cellcolor{gray!25}$0.07$ & $0.02$ & \cellcolor{gray!25}$0.05$ & \cellcolor{gray!25}$0.07$ & \cellcolor{gray!25}$0.01$   \\\cline{2-3}\hline

     \multirow{9}{*}{Avg-based} & \multirow{2}{*}{census} & race & $-0.05$ & $0.08$ & $0.15$ & $0.08$ & \cellcolor{gray!25}$3.14$ & $0.06$ & $0.08$ &\cellcolor{gray!25}$0.14$ & \cellcolor{gray!25}$0.26$ & \cellcolor{gray!50}$0.16$ & $1.7$ & $0.06$      \\
    & & sex& $0.05$ & \cellcolor{gray!50}$0.15$ & \cellcolor{gray!75}$0.24$ & \cellcolor{gray!25}$0.13$ & \cellcolor{gray!75}$0.87$ & $0.05$ & $0.11$ & $0.09$ & \cellcolor{gray!50}$0.17$ & $0.09$ & $0.46$ & \cellcolor{gray!25}$0.06$  \\\cline{2-3}
    & credit & sex & $0.01$ &$0.07$ & $0.12$ & $0.05$ & \cellcolor{gray!25}$0.87$ & $0.03$ & $0.21$ & \cellcolor{gray!25}$0.13$ & \cellcolor{gray!25}$0.23$ & $0.07$ & $0.39$ & $0.02$     \\\cline{2-3}
    RF & bank & age & $-0.01$ & $0.0$ & $0.01$ & $0.0$ & \cellcolor{gray!25}$0.88$ & $0.01$ & $0.04$ & \cellcolor{gray!50}$0.03$ & \cellcolor{gray!50}$0.05$ & \cellcolor{gray!25}$0.01$ & $0.24$ & \cellcolor{gray!50}$0.07$   \\\cline{2-3}

    & \multirow{1}{*}{compas} &  race & $0.0$ & $0.02$ & $0.0$ & $0.07$ & $0.13$ & $0.03$ & $0.0$ & $0.02$ & $0.0$ & $0.07$ & $0.13$ & \cellcolor{gray!50}$0.09$      \\  \cline{2-3}

    % & \multirow{3}{*}{Compas} & sex & -.01&.02&-.03 &-.03&.0  & .0 & -.02 & .03 \\  
    % &  &  race &  -.02&.0&-.03 &.0&.02  & .0 & -.04 & -.28 \\
    % & &  age & -.02&.01&-.02 &.0&.0  & .01 & .05 & -.03\\ \cline{2-3} 
     & default & sex  &$-0.03$ & $0.01$ & $0.02$ & $0.01$ & $0.23$ & $0.02$ & $0.07$ & \cellcolor{gray!25}$0.02$ & \cellcolor{gray!25}$0.03$ & \cellcolor{gray!50}$0.04$ & \cellcolor{gray!25}$0.32$ & \cellcolor{gray!50}$0.12$  \\\cline{2-3}
     & heart & sex & $0.09$ & \cellcolor{gray!25}$0.17$ & \cellcolor{gray!25}$0.3$ & \cellcolor{gray!25}$0.14$ & \cellcolor{gray!50}$0.86$ & $0.03$ & $0.16$ & \cellcolor{gray!25}$0.21$ & \cellcolor{gray!25}$0.43$ & $0.11$ & $0.44$ & $0.02$     \\\cline{2-3}
     & meps15 & sex & $-0.03$ &$0.03$ & $0.05$ & $0.04$ & \cellcolor{gray!25}$0.65$ & $0.0$ & $0.06$ & $0.03$ & $0.05$ & \cellcolor{gray!25}$0.07$ & \cellcolor{gray!25}$0.67$ & \cellcolor{gray!50}$0.11$      \\\cline{2-3}
     & meps16 & sex &$-0.03$ & $0.02$ & $0.04$ & $0.04$ & $0.52$ & $0.01$ & $0.06$ &$0.02$ & \cellcolor{gray!25}$0.06$ & \cellcolor{gray!25}$0.06$ & \cellcolor{gray!25}$0.68$ & \cellcolor{gray!50}$0.05$ \\\cline{2-3}
     & students & sex & $-0.01$ &$0.06$ & \cellcolor{gray!25}$0.04$ & \cellcolor{gray!25}$0.05$ & \cellcolor{gray!25}$0.06$ & $0.01$ & $0.06$ &  $0.07$ & $0.01$ & \cellcolor{gray!50}$0.08$ & \cellcolor{gray!50}$0.1$ & $0.01$ \\\cline{2-3}\hline
    \end{tabular}}
 }
 % \vspace{-1.0 em}
\end{table*}

In this analysis, we leverage EVT to assess the effectiveness of prevalent mitigation algorithms like exponentiated gradient (EG)~\cite{agarwal2018reductions} and Fair-SMOTE~\cite{10.1145/3468264.3468537} in the tail. \revision{We also include two recent mitigation techniques, MAAT~\cite{10.1145/3540250.3549093} and STEALTH~\cite{10109333} in our experiments to evaluate our approach against more advanced methods. The results are reported in Table~\ref{tab:EVT-Bias-Mitigation} and ~\ref{tab:stealth_maat}.} The column Accuracy Loss shows the accuracy difference between the original and mitigated models with positive values indicating improved accuracy in the mitigated model, columns AOD, EOD, \revision{SPD, and DI} report the absolute values of average-based fairness measures, and the column ECD shows the amount of discrimination in the tail.
\revision{Darker gray shades indicate lower rankings, while lighter shades represent higher rankings (no shading indicates the top-ranked method).}

\revision{We use the Scott-Knott ranking outcomes to compare the four mitigation methods where we consider a statistically significant improvement over the original baseline model (the vanilla model) as a win. While the tables include all metrics, we explain the results for one average-based metric and one tail-based metric. Consider the AOD metric, we find that EG~\cite{agarwal2018reductions} outperforms other methods where it wins in 19 cases (out of 40). STEALTH~\cite{10109333}, MAAT~\cite{10.1145/3540250.3549093}, and Fair-SMOTE~\cite{10.1145/3468264.3468537} win in 14, 6, and 4 cases in reducing AOD biases. In terms of average AOD over all benchmarks; EG, STEALTH, MAAT, and SMOTE achieve an average of 0.03, 0.07, 0.07, and 0.08, respectively. In terms of number of cases with an AOD bias below or equal to 0.05; we observe that EG, STEALTH, MAAT, and SMOTE have 32, 22, 22, and 9 cases (out of 40), respectively.} 

\revision{When considering ECD metric, STEALTH demonstrates superior performance among the average-based mitigation methods in reducing tail discrimination. Specifically, we find that STEALTH wins in 31 cases (out of 40) whereas EG, MAAT, and Fair-SMOTE win in 15, 12, and 9 cases, respectively. In terms of average ECD over all benchmarks; STEALTH, EG, MAAT, and Fair-SMOTE achieve an average of 0.04, 0.20, 0.08, and 0.12, respectively. In terms of number of cases with an ECD bias below or equal to 0.05; STEALTH, EG, MAAT, and Fair-SMOTE have 32, 11, 15, and 9 cases (out of 40), respectively.
}

\begin{tcolorbox}[boxrule=1pt,left=1pt,right=1pt,top=1pt,bottom=1pt]
\textbf{Answer RQ3:} \revision{While the average-based mitigation methods~\cite{agarwal2018reductions,10.1145/3468264.3468537,10.1145/3540250.3549093,10109333} preserved or improved fairness based on metrics like AOD in 63\%, they increase unfairness in tail based on ECD metric in 35\% of cases. STEALTH~\cite{10109333} outperformed other mitigation methods significantly based on the ECD metric, failing only in 10\% of cases, while preserving/reducing the AOD bias in 65\%.
} 
\end{tcolorbox}

\subsection{Tail-aware Mitigation Algorithms (RQ4)}
\label{sec:rq4}
\setlength{\tabcolsep}{14.0pt} % Default value: 6pt
\begin{table*}[tbp!]
    \caption{Tail-aware bias mitigation.
    Legend is similar to Table~\ref{tab:EVT-Bias-Mitigation}
    %   Legend:
    % \textbf{P}: Protected Attribute,
    % \textbf{Acc$'$}: Accuracy of mitigated model,
    % \textbf{EOD$'$}: EOD of mitigated model,
    % \textbf{AOD$'$}: AOD of mitigated model,
    % \textbf{ACD$'$}: ACD$^2$ of mitigated model,
    % \textbf{ECD$'$}: ECD$^2$ of mitigated model, 
    % \textbf{RLD}$'_{n}$: RLD for the next $n$ queries for mitigated model.           
    }
    \label{tab:CVaR-Bias-Mitigation}
    \resizebox{\textwidth}{!}{
     {\color{black} \begin{tabular}{ | c | l  l |  c c c c c  c  | c  c c c c  c  | }
      \hline
      \multicolumn{3}{|c|}{\textbf{\color{black}{Name}}}  &  \multicolumn{6}{c|}{\color{black}{\textbf{Minimax-Fairness~\cite{10.1145/3461702.3462523}}}} & \multicolumn{6}{c|}{\color{black}{\textbf{ECD-Fair}}}
      \\
    \multicolumn{1}{|c}{\color{black}{Algorithm}}  & \color{black}{Dataset}  & \color{black}{P} & \color{black}{$Acc$ $loss$}  & \color{black}{$AOD$} & $EOD$ & $SPD$  & $DI$& \color{black}{$ECD$} & \color{black}{$Acc$ $loss$}  & \color{black}{$AOD$} & $EOD$ & $SPD$  &$DI$  & \color{black}{$ECD$} 
    \\ \hline
    \multirow{9}{*}{\color{black}{Tail-based}} & \multirow{2}{*}{\color{black}{Census}} & \color{black}{race} &$-0.07$ &$0.02$ & $0.04$ & $0.02$ & $0.74$ & \cellcolor{gray!75}$0.14$& $-0.04$&\cellcolor{gray!50}$0.08$ & \cellcolor{gray!25}$0.09$ & \cellcolor{gray!75}$0.14$ & \cellcolor{gray!50}$1.65$ & $0.02$\\
     & & \color{black}{sex} & $-0.04$&\cellcolor{gray!50}$0.03$ & \cellcolor{gray!50}$0.05$ & $0.06$ & \cellcolor{gray!50}$0.61$ & $0.03$  & $0.02$&\cellcolor{gray!25}$0.02$ & \cellcolor{gray!25}$0.03$ & \cellcolor{gray!25}$0.06$ & \cellcolor{gray!75}$0.68$ & \cellcolor{gray!25}$0.04$\\ \cline{2-3}
    & \color{black}{Credit} & \color{black}{sex} & $-0.09$ &$0.02$ & $0.04$ & $0.02$ & \cellcolor{gray!25}$0.79$ & $0.02$  &$-0.06$&\cellcolor{gray!25}$0.04$ & \cellcolor{gray!25}$0.08$ & \cellcolor{gray!25}$0.05$ & $0.21$ & $0.02$\\ \cline{2-3}
    \color{black}{DNN} & \color{black}{Bank} & \color{black}{age} &$-0.08$ &\cellcolor{gray!50}$0.06$ & \cellcolor{gray!25}$0.05$ & \cellcolor{gray!75}$0.07$ & \cellcolor{gray!25}$0.36$ & \cellcolor{gray!25}$0.05$  &$0.02$&$0.02$ & $0.04$ & \cellcolor{gray!25}$0.01$ & $0.18$ & $0.02$\\ \cline{2-3}
    
     & \multirow{1}{*}{\color{black}{Compas}} & \color{black}{race} & $-0.02$&$0.02$ & \cellcolor{gray!75}$0.03$ & $0.04$ & $0.08$ & \cellcolor{gray!100}$0.17$ &$0.02$&$0.02$ & $0.0$ & \cellcolor{gray!25}$0.06$ & \cellcolor{gray!25}$0.11$ & $0.03$\\ \cline{2-3}

    % & \multirow{3}{*}{Compas} & sex & .0&.01&-.02& -.07&.0  & .02 & -.01 & \cellcolor{gray!45}.12 \\
    % &  &  race &  .0&.01&-.03& -.40&-.01 & .02 & .01 & \cellcolor{gray!45}.12 \\
    % & &  age & .0&.01&.04& \cellcolor{gray!45}.45&.0  & .02 & .0 & -.08 \\ \cline{2-3}
     & \color{black}{Default} & \color{black}{sex} &$0.02$ &$0.01$ & \cellcolor{gray!25}$0.02$ & $0.02$ & $0.17$ & \cellcolor{gray!50}$0.08$ &$-0.01$&\cellcolor{gray!50}$0.04$ & \cellcolor{gray!75}$0.05$ & \cellcolor{gray!50}$0.04$ & \cellcolor{gray!50}$0.32$ & \cellcolor{gray!25}$0.06$\\ \cline{2-3}
     & \color{black}{heart} & \color{black}{sex} &$0.06$ &\cellcolor{gray!25}$0.11$ & \cellcolor{gray!25}$0.22$ & $0.06$ & $0.29$ & $0.01$ &$0.04$&\cellcolor{gray!25}$0.1$ & \cellcolor{gray!25}$0.26$ & \cellcolor{gray!50}$0.16$ & \cellcolor{gray!25}$0.43$ & $0.01$\\ \cline{2-3}
     & \color{black}{Meps15} & \color{black}{sex} & $0.0$&$0.01$ & \cellcolor{gray!25}$0.02$ & $0.04$ & $0.36$ & \cellcolor{gray!25}$0.04$  &$0.03$& \cellcolor{gray!25}$0.03$ & \cellcolor{gray!25}$0.03$ & \cellcolor{gray!25}$0.06$ & \cellcolor{gray!25}$0.63$ & $0.04$\\ \cline{2-3}
     & \color{black}{Meps16} & \color{black}{sex} & $0.04$& \cellcolor{gray!25}$0.02$ & \cellcolor{gray!25}$0.03$ & $0.04$ & $0.69$ & \cellcolor{gray!25}$0.02$  &$0.0$&\cellcolor{gray!50}$0.04$ & \cellcolor{gray!50}$0.06$ & \cellcolor{gray!25}$0.05$ & \cellcolor{gray!25}$0.93$ & \cellcolor{gray!75}$0.08$ \\ \cline{2-3}
     & \color{black}{Students} & \color{black}{sex} &$0.02$ &\cellcolor{gray!25}$0.06$ & $0.02$ & $0.03$ & $0.04$ & \cellcolor{gray!75}$0.21$&$-0.03$&\cellcolor{gray!25}$0.06$ & $0.02$ & $0.04$ & $0.04$ & $0.03$\\ \cline{2-3}
    \hline
    \multirow{9}{*}{\color{black}{Tail-based}} & \multirow{2}{*}{\color{black}{Census}} & \color{black}{race} &$-0.06$ &\cellcolor{gray!50}$0.1$ & \cellcolor{gray!50}$0.13$ & \cellcolor{gray!50}$0.12$ & \cellcolor{gray!25}$3.01$ & \cellcolor{gray!25}$0.17$ &$-0.01$&$0.03$ & $0.05$ & $0.03$ & \cellcolor{gray!25}$3.38$ & $0.03$\\
     & & \color{black}{sex} & $0.04$&\cellcolor{gray!50}$0.1$ & \cellcolor{gray!50}$0.13$ & \cellcolor{gray!75}$0.15$ & \cellcolor{gray!75}$0.77$ & \cellcolor{gray!25}$0.07$ &$-0.03$&$0.02$ & \cellcolor{gray!25}$0.04$ & $0.03$ & \cellcolor{gray!50}$0.73$ & $0.05$\\ \cline{2-3}
    & \color{black}{Credit} & \color{black}{sex} &$-0.03$ &\cellcolor{gray!25}$0.1$ & \cellcolor{gray!25}$0.17$ & \cellcolor{gray!50}$0.11$ & $0.62$ & \cellcolor{gray!100}$0.13$  &$-0.05$&$0.05$ & $0.09$ & $0.04$ & $0.15$ & $0.0$ \\ \cline{2-3}
    \color{black}{LR} & \color{black}{Bank} & \color{black}{age} & $-0.05$&\cellcolor{gray!25}$0.05$ & \cellcolor{gray!50}$0.09$ & \cellcolor{gray!25}$0.01$ & $0.16$ & $0.02$  &$0.02$&\cellcolor{gray!25}$0.06$ & \cellcolor{gray!50}$0.11$ & \cellcolor{gray!50}$0.02$ & \cellcolor{gray!50}$0.85$ & \cellcolor{gray!25}$0.04$\\ \cline{2-3}
    
     & \multirow{1}{*}{\color{black}{Compas}} & \color{black}{race} & $-0.01$&\cellcolor{gray!25}$0.02$ & $0.01$ & \cellcolor{gray!25}$0.06$ & $0.11$ & $0.04$ &$0.04$&\cellcolor{gray!50}$0.04$ & \cellcolor{gray!25}$0.01$ & \cellcolor{gray!25}$0.08$ & $0.15$ & $0.04$\\ \cline{2-3}
    % & \multirow{3}{*}{Compas} & sex & .0&.02&-.03&-.08&.01 & .02 & -.03 & -.04 \\  
    % &  &  race &  -.02&.02&-.08& -.44&.01  & .01 & -.03 & -.05 \\
    % & &  age & .0&.01&-.02 &-.05&.0 & .01 & .02 & .01 \\ \cline{2-3} 
     & \color{black}{Default} & \color{black}{sex} &$-0.03$ &\cellcolor{gray!25}$0.03$ & \cellcolor{gray!25}$0.05$ & \cellcolor{gray!50}$0.03$ & \cellcolor{gray!50}$0.46$ & $0.03$  &$-0.01$&\cellcolor{gray!25}$0.03$ & \cellcolor{gray!25}$0.05$ & \cellcolor{gray!50}$0.03$ & \cellcolor{gray!50}$0.43$ & $0.03$ \\ \cline{2-3}
     & \color{black}{Heart} & \color{black}{sex} &$-0.07$ &$0.12$ & $0.2$ & \cellcolor{gray!50}$0.21$ & \cellcolor{gray!25}$0.63$ & \cellcolor{gray!75}$0.1$ &$-0.04$&$0.09$ & $0.27$ & \cellcolor{gray!25}$0.13$ & $0.42$ & $0.02$\\ \cline{2-3}
     & \color{black}{Meps15} & \color{black}{sex} &$0.0$ &\cellcolor{gray!50}$0.06$ & \cellcolor{gray!50}$0.09$ & \cellcolor{gray!100}$0.07$ & \cellcolor{gray!50}$1.07$ & \cellcolor{gray!25}$0.06$ &$-0.01$&$0.03$ & $0.03$ & \cellcolor{gray!25}$0.04$ & \cellcolor{gray!25}$0.77$ & $0.02$\\ \cline{2-3}
     & \color{black}{Meps16} & \color{black}{sex} & $-0.04$&\cellcolor{gray!50}$0.07$ & \cellcolor{gray!50}$0.1$ & \cellcolor{gray!25}$0.07$ & \cellcolor{gray!50}$1.16$ & \cellcolor{gray!25}$0.07$ & $-0.01$ &\cellcolor{gray!25}$0.03$ & \cellcolor{gray!25}$0.04$ & $0.04$ & $0.68$ & $0.03$\\ \cline{2-3}
     & \color{black}{Students} & \color{black}{sex} &$0.03$ &$0.04$ & $0.02$ & $0.04$ & $0.04$ & \cellcolor{gray!75}$0.08$   &$0.0$&$0.04$ & $0.02$ & $0.03$ & $0.04$ & $0.01$\\ \cline{2-3}
     \hline

    \end{tabular}}
 }
 \vspace{-1.0 em}
\end{table*}

We first evaluate the effectiveness of  MiniMax-Fairness~\cite{10.1145/3461702.3462523}, which serves as our baseline, alongside our proposed in-process mitigator \revision{(ECD-Fair)}. 
Results in Table~\ref{tab:CVaR-Bias-Mitigation} follow a similar format to Table~\ref{tab:EVT-Bias-Mitigation} where we only include the DNN and Logistic regression models since the MiniMax-Fairness only supports these models among our base models.
\revision{Considering ECD metric, our approach significantly outperforms MiniMax-Fairness. Specifically, ECD-Fair wins in 18 cases (out of 20), while MiniMax-Fairness wins in 10 cases (out of 20).
When considering EOD and AOD metrics, ECD-Fair outperforms MiniMax-Fairness with 10 and 9 win cases vs. 7 and 5 win cases (out of 20). In terms of absolute values over all benchmarks, ECD-Fair achieves a average AOD and ECD of 0.04 and 0.03, respectively. The number of cases with AOD and ECD below 0.05 are 15 and 18 (out of 20), respectively.}

\revision{We also compare ECD-Fair to STEALTH~\cite{10109333} method  over the DNN and LR benchmarks as STEALTH outperformed other baseline methods. Based on the EOD and AOD metrics, we find that ECD-Fair wins in 10 and 9 cases (out of 20) vs. STEALTH wins in 6 and 8 cases (out of 20), respectively. When considering the ECD metric, ECD-Fair and STEALTH win in 18 and 16 cases (out of 20), respectively. STEALTH degrades unfairness in tail for 2 benchmarks, while ECD-fair does not increase the unfairness in the tail for any benchmark. Overall, while STEALTH demonstrates a competitive result, ECD-Fair slightly outperforms it for both tail and average metrics.
}    

\begin{tcolorbox}[boxrule=1pt,left=1pt,right=1pt,top=1pt,bottom=1pt]
\textbf{Answer RQ4:}
\revision{ECD-Fair significantly outperformed MiniMax-Fairness~\cite{10.1145/3461702.3462523}, a state-of-the-art tail-aware mitigator. When compared to STEALTH~\cite{10109333}, a competitive baseline, we found that ECD-Fair and STEALTH improved fairness in the tail for 90\% and 80\% of cases, respectively. ECD-Fair and STEALTH reduced the AOD bias in 45\% and 40\% of cases, respectively. 
}
\end{tcolorbox}

\section{Discussions}
\label{sec:discuss}
\noindent \textbf{Limitations.} 
\revision{One limitation is the lack of ground truth regarding the tail of ML outcome distributions. We can use the maximum individual discrimination in the validation dataset as it gives a lower-bound on the ground truth.}
Our approach requires the presence of protected attributes during inference. Therefore, it cannot be used to study worst-case fairness for notions such as fairness through unawareness, which requires the removal of protected attributes~\cite{dwork2012fairness}.
Our approach also depends on the representative individuals sampled from the same training distribution, and may not be valid for out-of-distribution queries. Finally, our approach assumes that flipping the sensitive values leads to valid representations to measure the sensitivity of ML models to the protected attributes.

\vspace{0.25 em}
\noindent \textbf{Threat to Validity.}
To address the internal validity and ensure our finding does not lead to invalid conclusions, we follow established guidelines and report the statistical significance of measures with the \revision{exponential and Scott-Knott statistical testing}. To ensure that our results are generalizable, we perform our experiments on three well-established training algorithms from \texttt{scikit-learn} and \texttt{TensorFlow} libraries with a popular mitigation algorithm from the \texttt{Fairlearn} library over \revision{160} fairness-sensitive tasks that have been widely used in the fairness research. It is an open problem whether the algorithms, hyperparameters, and datasets are sufficiently representative to cover challenging fairness scenarios.

\section{Related Work}
\label{sec:related}

\vspace{0.25 em}
\noindent \textbf{Fairness Testing of Data-Driven Software.}
Individual discrimination is a major fairness debugging method~\cite{agarwal2018automated,udeshi2018automated,zhang2020white,9793943,10.1145/3468264.3468537,10.1145/3650212.3680380}.
\textsc{Themis}~\cite{10.1145/3106237.3106277} is the closest approach.
While \textsc{Themis}~\cite{10.1145/3106237.3106277} focuses on the 
average causal discrimination between two subgroups via \textit{counterfactual} queries
with prevalent statistical guarantees of normal distributions, we introduce
the notion of extreme causal discrimination between two subgroups with exponentially
statistical guarantees of extreme value distributions. Rather than randomly sampling
data from the domain of variables, we leveraged generative AI models
to produce realistic test cases from the tail.

\vspace{0.25 em}
\noindent \textbf{Fairness in the Tail.} Multiple works consider the worst-case group fairness~\cite{williamson2019fairness,10.1145/3461702.3462523,shekhar2021adaptive}. Williamson and Menon~\cite{williamson2019fairness} leveraged conditional value at risk (CVaR) to minimize the expected loss and the worst-case loss of any group in the upper quantile. We found that CVaR might miss discrimination in the tail and cannot reason about the shape of tail. Diana et al.~\cite{10.1145/3461702.3462523} proposed a constrained optimization objective where the goal is to minimize the expected overall loss for all data instances subject to the hard constraints wherein no group loss can be more than a threshold.  We propose an in-process bias mitigator that significantly outperforms this technique as shown in RQ4.

\vspace{0.25 em}
\noindent \revision{
\noindent \textbf{Intersectional Fairness.} The keyword ``worst-case fairness'' has been also used in the relevant fairness literature~\cite{ghosh2021characterizing,zhang2022adaptive,chen2024fairness}. However, their notion of fairness still relies on regular ``average" fairness metrics like the rate of favorable outcomes per each subgroup. In particular, intersectional fairness concerns about the summary of fairness statistics when there are fairness measures for n subgroups. For example, Ghosh et al.~\cite{ghosh2021characterizing} suggests a min-max ratio that takes the maximum for average favorable outcomes of all subgroups and divides it by the min for average favorable outcomes of all subgroups. On the other hand, our fairness measure looks at the tail of ML outcome distributions per each subgroup via EVT and compares the tail distributions between groups to quantify the amounts of discrimination. 
}

\vspace{0.25 em}
\noindent \textbf{Other Application of EVT for Fairness.}
In addition to its technical applications~\cite{10.1145/3644815.3644989,abella2017measurement}
Extreme value theory has been significantly used to study income and wealth inequalities around the world~\cite{piketty2003income,saez2004income,10.1093/cesifo/49.4.479}. Piketty and Saez~\cite{piketty2003income} used the Generalized Pareto Distribution to study the distribution of income in the US between 1913 and 1998. 
Wang~\cite{wang2022fairness} studied the concept of Degree of Matthew Effect in recommendation systems via extreme value theory
whereas we consider social bias (discrimination against protected groups) in decision-making systems (based on classifications).

\section{Conclusion and Future Work}
\label{sec:conclusion}
We studied fairness through the lens of extreme value theory. Our proposed approach 
fitted well to model the worst-cases counterfactual bias with statistical guarantees and revealed the limitations of a state-of-the-art bias reduction algorithm in the worst-case. There are multiple exciting future directions. One direction is to
leverage EVT to provide a notion of AI harms to understand if automated decision-support software systematically
harms a vulnerable community. 

\vspace{1.0 em}
\noindent \textbf{Acknowledgement.}
The authors thank the anonymous ICSE reviewers for their time and
invaluable feedback to improve this paper. Monjezi and Tizpaz-Niari were also affiliated with UT El Paso in the completion of this work. 
Tizpaz-Niari and Trivedi have been partially supported by NSF under grants
CCF-2317206 and CCF-2317207.

\bibliography{reference}

\begin{thebibliography}{10}
\providecommand{\url}[1]{#1}
\csname url@samestyle\endcsname
\providecommand{\newblock}{\relax}
\providecommand{\bibinfo}[2]{#2}
\providecommand{\BIBentrySTDinterwordspacing}{\spaceskip=0pt\relax}
\providecommand{\BIBentryALTinterwordstretchfactor}{4}
\providecommand{\BIBentryALTinterwordspacing}{\spaceskip=\fontdimen2\font plus
\BIBentryALTinterwordstretchfactor\fontdimen3\font minus
  \fontdimen4\font\relax}
\providecommand{\BIBforeignlanguage}[2]{{%
\expandafter\ifx\csname l@#1\endcsname\relax
\typeout{** WARNING: IEEEtran.bst: No hyphenation pattern has been}%
\typeout{** loaded for the language `#1'. Using the pattern for}%
\typeout{** the default language instead.}%
\else
\language=\csname l@#1\endcsname
\fi
#2}}
\providecommand{\BIBdecl}{\relax}
\BIBdecl

\bibitem{goodfellow2016deep}
I.~Goodfellow, Y.~Bengio, and A.~Courville, \emph{Deep learning}.\hskip 1em
  plus 0.5em minus 0.4em\relax MIT press, 2016.

\bibitem{sutton2018reinforcement}
R.~S. Sutton and A.~G. Barto, \emph{Reinforcement learning: An
  introduction}.\hskip 1em plus 0.5em minus 0.4em\relax MIT press, 2018.

\bibitem{devlin2018bert}
J.~Devlin, M.-W. Chang, K.~Lee, and K.~Toutanova, ``Bert: Pre-training of deep
  bidirectional transformers for language understanding,'' \emph{arXiv preprint
  arXiv:1810.04805}, 2018.

\bibitem{chatgpt}
O.~ChatGPT, ``Chatgpt: Optimizing language models for dialogue,''
  \url{https://openai.com/blog/chatgpt/}, 2022, online.

\bibitem{compas-article}
S.~M. Julia~Angwin, Jeff~Larson and L.~Kirchne, ``Machine bias,''
  \url{https://www.propublica.org/article/machine-bias-risk-assessments-in-criminal-sentencing},
  2021, online.

\bibitem{elyounes2020computer}
D.~A. Elyounes, ``" computer says no!": The impact of automation on the
  discretionary power of public officers,'' \emph{Vand. J. Ent. \& Tech. L.},
  vol.~23, p. 451, 2020.

\bibitem{Ranchords2021AutomatedGF}
S.~Ranchord{\'a}s and L.~Scarcella, ``Automated government for vulnerable
  citizens: Intermediating rights,'' \emph{SSRN Electronic Journal}, 2021.

\bibitem{dorothyIRS}
\BIBentryALTinterwordspacing
D.~A. Brown, ``The {IRS} is targeting the poorest americans,'' August 2021,
  [Online; posted 27-July-2021]. [Online]. Available:
  \url{https://www.theatlantic.com/ideas/archive/2021/07/how-race-plays-tax-policing/619570/}
\BIBentrySTDinterwordspacing

\bibitem{tizpaz2023metamorphic}
S.~Tizpaz-Niari, V.~Monjezi, M.~Wagner, S.~Darian, K.~Reed, and A.~Trivedi,
  ``Metamorphic testing and debugging of tax preparation software,'' in
  \emph{2023 IEEE/ACM 45th International Conference on Software Engineering:
  Software Engineering in Society (ICSE-SEIS)}.\hskip 1em plus 0.5em minus
  0.4em\relax IEEE, 2023, pp. 138--149.

\bibitem{petrongolo2019gender}
B.~Petrongolo, ``The gender gap in employment and wages,'' \emph{Nature Human
  Behaviour}, vol.~3, no.~4, pp. 316--318, 2019.

\bibitem{anderson1996racial}
D.~Anderson and D.~Shapiro, ``Racial differences in access to high-paying jobs
  and the wage gap between black and white women,'' \emph{ILR Review}, vol.~49,
  no.~2, pp. 273--286, 1996.

\bibitem{dwork2012fairness}
C.~Dwork, M.~Hardt, T.~Pitassi, O.~Reingold, and R.~Zemel, ``Fairness through
  awareness,'' in \emph{Proceedings of the 3rd innovations in theoretical
  computer science conference}, 2012, pp. 214--226.

\bibitem{coles2001introduction}
S.~Coles, J.~Bawa, L.~Trenner, and P.~Dorazio, \emph{An introduction to
  statistical modeling of extreme values}.\hskip 1em plus 0.5em minus
  0.4em\relax Springer, 2001, vol. 208.

\bibitem{10.1145/3106237.3106277}
\BIBentryALTinterwordspacing
S.~Galhotra, Y.~Brun, and A.~Meliou, ``Fairness testing: testing software for
  discrimination,'' in \emph{Proceedings of the 2017 11th Joint Meeting on
  Foundations of Software Engineering}, ser. ESEC/FSE 2017.\hskip 1em plus
  0.5em minus 0.4em\relax New York, NY, USA: Association for Computing
  Machinery, 2017, p. 498–510. [Online]. Available:
  \url{https://doi.org/10.1145/3106237.3106277}
\BIBentrySTDinterwordspacing

\bibitem{williamson2019fairness}
R.~Williamson and A.~Menon, ``Fairness risk measures,'' in \emph{International
  Conference on Machine Learning}.\hskip 1em plus 0.5em minus 0.4em\relax PMLR,
  2019, pp. 6786--6797.

\bibitem{diebolt2007goodness}
J.~Diebolt, M.~Garrido, and S.~Girard, ``A goodness-of-fit test for the
  distribution tail,'' 2007.

\bibitem{abella2017measurement}
J.~Abella, M.~Padilla, J.~D. Castillo, and F.~J. Cazorla, ``Measurement-based
  worst-case execution time estimation using the coefficient of variation,''
  \emph{ACM Transactions on Design Automation of Electronic Systems (TODAES)},
  vol.~22, no.~4, pp. 1--29, 2017.

\bibitem{CTGAN}
L.~Xu, M.~Skoularidou, A.~Cuesta-Infante, and K.~Veeramachaneni, \emph{Modeling
  tabular data using conditional GAN}.\hskip 1em plus 0.5em minus 0.4em\relax
  Red Hook, NY, USA: Curran Associates Inc., 2019.

\bibitem{8285168}
Z.~Wan, Y.~Zhang, and H.~He, ``Variational autoencoder based synthetic data
  generation for imbalanced learning,'' in \emph{2017 IEEE Symposium Series on
  Computational Intelligence (SSCI)}, 2017, pp. 1--7.

\bibitem{agarwal2018reductions}
A.~Agarwal, A.~Beygelzimer, M.~Dud{\'\i}k, J.~Langford, and H.~Wallach, ``A
  reductions approach to fair classification,'' in \emph{International
  Conference on Machine Learning}.\hskip 1em plus 0.5em minus 0.4em\relax PMLR,
  2018, pp. 60--69.

\bibitem{10.1145/3468264.3468537}
\BIBentryALTinterwordspacing
J.~Chakraborty, S.~Majumder, and T.~Menzies, ``Bias in machine learning
  software: Why? how? what to do?'' in \emph{Proceedings of the 29th ACM Joint
  Meeting on European Software Engineering Conference and Symposium on the
  Foundations of Software Engineering}, ser. ESEC/FSE 2021.\hskip 1em plus
  0.5em minus 0.4em\relax New York, NY, USA: Association for Computing
  Machinery, 2021, p. 429–440. [Online]. Available:
  \url{https://doi.org/10.1145/3468264.3468537}
\BIBentrySTDinterwordspacing

\bibitem{10.1145/3540250.3549093}
\BIBentryALTinterwordspacing
Z.~Chen, J.~M. Zhang, F.~Sarro, and M.~Harman, ``Maat: a novel ensemble
  approach to addressing fairness and performance bugs for machine learning
  software,'' in \emph{Proceedings of the 30th ACM Joint European Software
  Engineering Conference and Symposium on the Foundations of Software
  Engineering}, ser. ESEC/FSE 2022.\hskip 1em plus 0.5em minus 0.4em\relax New
  York, NY, USA: Association for Computing Machinery, 2022, p. 1122–1134.
  [Online]. Available: \url{https://doi.org/10.1145/3540250.3549093}
\BIBentrySTDinterwordspacing

\bibitem{10109333}
L.~Alvarez and T.~Menzies, ``Don’t lie to me: Avoiding malicious explanations
  with stealth,'' \emph{IEEE Software}, vol.~40, no.~3, pp. 43--53, 2023.

\bibitem{10.1145/3461702.3462523}
\BIBentryALTinterwordspacing
E.~Diana, W.~Gill, M.~Kearns, K.~Kenthapadi, and A.~Roth, ``Minimax group
  fairness: Algorithms and experiments,'' in \emph{Proceedings of the 2021
  AAAI/ACM Conference on AI, Ethics, and Society}, ser. AIES '21.\hskip 1em
  plus 0.5em minus 0.4em\relax New York, NY, USA: Association for Computing
  Machinery, 2021, p. 66–76. [Online]. Available:
  \url{https://doi.org/10.1145/3461702.3462523}
\BIBentrySTDinterwordspacing

\bibitem{leadbetter2012extremes}
M.~R. Leadbetter, G.~Lindgren, and H.~Rootz{\'e}n, \emph{Extremes and related
  properties of random sequences and processes}.\hskip 1em plus 0.5em minus
  0.4em\relax Springer Science \& Business Media, 2012.

\bibitem{Dua:2019-census}
\BIBentryALTinterwordspacing
D.~Dua and C.~Graff, ``{UCI} machine learning repository,'' 2017. [Online].
  Available: \url{https://archive.ics.uci.edu/ml/datasets/census+income}
\BIBentrySTDinterwordspacing

\bibitem{zhang2020white}
P.~Zhang, J.~Wang, J.~Sun, G.~Dong, X.~Wang, X.~Wang, J.~S. Dong, and T.~Dai,
  ``White-box fairness testing through adversarial sampling,'' in
  \emph{Proceedings of the ACM/IEEE 42nd International Conference on Software
  Engineering}, 2020, pp. 949--960.

\bibitem{9793943}
H.~Zheng, Z.~Chen, T.~Du, X.~Zhang, Y.~Cheng, S.~Ti, J.~Wang, Y.~Yu, and
  J.~Chen, ``Neuronfair: Interpretable white-box fairness testing through
  biased neuron identification,'' in \emph{2022 IEEE/ACM 44th International
  Conference on Software Engineering (ICSE)}, 2022, pp. 1519--1531.

\bibitem{10.1145/3460319.3464820}
\BIBentryALTinterwordspacing
L.~Zhang, Y.~Zhang, and M.~Zhang, ``Efficient white-box fairness testing
  through gradient search,'' in \emph{Proceedings of the 30th ACM SIGSOFT
  International Symposium on Software Testing and Analysis}, ser. ISSTA 2021,
  2021, p. 103–114. [Online]. Available:
  \url{https://doi.org/10.1145/3460319.3464820}
\BIBentrySTDinterwordspacing

\bibitem{10.1145/3510003.3510137}
\BIBentryALTinterwordspacing
M.~Fan, W.~Wei, W.~Jin, Z.~Yang, and T.~Liu, ``Explanation-guided fairness
  testing through genetic algorithm,'' in \emph{Proceedings of the 44th
  International Conference on Software Engineering}, ser. ICSE '22.\hskip 1em
  plus 0.5em minus 0.4em\relax New York, NY, USA: Association for Computing
  Machinery, 2022, p. 871–882. [Online]. Available:
  \url{https://doi.org/10.1145/3510003.3510137}
\BIBentrySTDinterwordspacing

\bibitem{Heart-disease}
\BIBentryALTinterwordspacing
``{UCI}:heart disease data set,'' 2001. [Online]. Available:
  \url{https://archive.ics.uci.edu/ml/datasets/Heart+Disease}
\BIBentrySTDinterwordspacing

\bibitem{10.1145/3238147.3238165}
\BIBentryALTinterwordspacing
S.~Udeshi, P.~Arora, and S.~Chattopadhyay, ``Automated directed fairness
  testing,'' in \emph{Proceedings of the 33rd ACM/IEEE International Conference
  on Automated Software Engineering}, ser. ASE '18.\hskip 1em plus 0.5em minus
  0.4em\relax New York, NY, USA: Association for Computing Machinery, 2018, p.
  98–108. [Online]. Available: \url{https://doi.org/10.1145/3238147.3238165}
\BIBentrySTDinterwordspacing

\bibitem{10.1145/3510003.3510123}
\BIBentryALTinterwordspacing
H.~Zheng, Z.~Chen, T.~Du, X.~Zhang, Y.~Cheng, S.~Ji, J.~Wang, Y.~Yu, and
  J.~Chen, ``Neuronfair: interpretable white-box fairness testing through
  biased neuron identification,'' in \emph{Proceedings of the 44th
  International Conference on Software Engineering}, ser. ICSE '22.\hskip 1em
  plus 0.5em minus 0.4em\relax New York, NY, USA: Association for Computing
  Machinery, 2022, p. 1519–1531. [Online]. Available:
  \url{https://doi.org/10.1145/3510003.3510123}
\BIBentrySTDinterwordspacing

\bibitem{pmlr-v157-zhao21a}
\BIBentryALTinterwordspacing
Z.~Zhao, A.~Kunar, R.~Birke, and L.~Y. Chen, ``Ctab-gan: Effective table data
  synthesizing,'' in \emph{Proceedings of The 13th Asian Conference on Machine
  Learning}, ser. Proceedings of Machine Learning Research, V.~N.
  Balasubramanian and I.~Tsang, Eds., vol. 157.\hskip 1em plus 0.5em minus
  0.4em\relax PMLR, 17--19 Nov 2021, pp. 97--112. [Online]. Available:
  \url{https://proceedings.mlr.press/v157/zhao21a.html}
\BIBentrySTDinterwordspacing

\bibitem{math11040977}
\BIBentryALTinterwordspacing
E.~Nazari, P.~Branco, and G.-V. Jourdan, ``Autogan: An automated
  human-out-of-the-loop approach for training generative adversarial
  networks,'' \emph{Mathematics}, vol.~11, no.~4, 2023. [Online]. Available:
  \url{https://www.mdpi.com/2227-7390/11/4/977}
\BIBentrySTDinterwordspacing

\bibitem{10.1007/978-3-031-35891-3_26}
\BIBentryALTinterwordspacing
A.~Rajabi and O.~O. Garibay, ``Distance correlation gan: Fair tabular data
  generation with generative adversarial networks,'' in \emph{Artificial
  Intelligence in HCI: 4th International Conference, AI-HCI 2023, Held as Part
  of the 25th HCI International Conference, HCII 2023, Copenhagen, Denmark,
  July 23–28, 2023, Proceedings, Part I}.\hskip 1em plus 0.5em minus
  0.4em\relax Berlin, Heidelberg: Springer-Verlag, 2023, p. 431–445.
  [Online]. Available: \url{https://doi.org/10.1007/978-3-031-35891-3_26}
\BIBentrySTDinterwordspacing

\bibitem{Zhang2015LearningCF}
\BIBentryALTinterwordspacing
X.~Zhang, Y.~Fu, A.~Zang, L.~Sigal, and G.~Agam, ``Learning classifiers from
  synthetic data using a multichannel autoencoder,'' \emph{ArXiv}, vol.
  abs/1503.03163, 2015. [Online]. Available:
  \url{https://api.semanticscholar.org/CorpusID:8164829}
\BIBentrySTDinterwordspacing

\bibitem{ISLAM2021105950}
\BIBentryALTinterwordspacing
Z.~Islam, M.~Abdel-Aty, Q.~Cai, and J.~Yuan, ``Crash data augmentation using
  variational autoencoder,'' \emph{Accident Analysis and Prevention}, vol. 151,
  p. 105950, 2021. [Online]. Available:
  \url{https://www.sciencedirect.com/science/article/pii/S000145752031770X}
\BIBentrySTDinterwordspacing

\bibitem{xiao2023latent}
Y.~Xiao, A.~Liu, T.~Li, and X.~Liu, ``Latent imitator: Generating natural
  individual discriminatory instances for black-box fairness testing,'' in
  \emph{Proceedings of the 32nd ACM SIGSOFT international symposium on software
  testing and analysis}, 2023, pp. 829--841.

\bibitem{castillo2014methods}
J.~D. Castillo, J.~Daoudi, and R.~Lockhart, ``Methods to distinguish between
  polynomial and exponential tails,'' \emph{Scandinavian Journal of
  Statistics}, vol.~41, no.~2, pp. 382--393, 2014.

\bibitem{sokal1995biometry}
R.~Sokal and F.~Rohlf, ``Biometry: The principles and practice of statistics in
  biological research 3rd edition wh freeman and co,'' \emph{New York}, 1995.

\bibitem{tizpaz2022fairness}
S.~Tizpaz-Niari, A.~Kumar, G.~Tan, and A.~Trivedi, ``Fairness-aware
  configuration of machine learning libraries,'' in \emph{Proceedings of the
  44th International Conference on Software Engineering}, 2022, pp. 909--920.

\bibitem{bird2020fairlearn}
\BIBentryALTinterwordspacing
S.~Bird, M.~Dud{\'i}k, R.~Edgar, B.~Horn, R.~Lutz, V.~Milan, M.~Sameki,
  H.~Wallach, and K.~Walker, ``Fairlearn: A toolkit for assessing and improving
  fairness in {AI},'' Microsoft, Tech. Rep. MSR-TR-2020-32, May 2020. [Online].
  Available:
  \url{https://www.microsoft.com/en-us/research/publication/fairlearn-a-toolkit-for-assessing-and-improving-fairness-in-ai/}
\BIBentrySTDinterwordspacing

\bibitem{Dua:2019-credit}
\BIBentryALTinterwordspacing
``{UCI} machine learning repository (german credit),'' 2017. [Online].
  Available:
  \url{https://archive.ics.uci.edu/ml/datasets/statlog+(german+credit+data)}
\BIBentrySTDinterwordspacing

\bibitem{Dua:2019-bank}
\BIBentryALTinterwordspacing
``{UCI} machine learning repository (bank marketing),'' 2017. [Online].
  Available: \url{https://archive.ics.uci.edu/ml/datasets/bank+marketing}
\BIBentrySTDinterwordspacing

\bibitem{compas-dataset}
ProPublica, ``Compas software ananlysis,''
  \url{https://github.com/propublica/compas-analysis}, 2021, online.

\bibitem{Default-Credit}
\BIBentryALTinterwordspacing
``{UCI}:default of credit card clients data set,'' 2009. [Online]. Available:
  \url{https://archive.ics.uci.edu/ml/datasets/default+of+credit+card+clients}
\BIBentrySTDinterwordspacing

\bibitem{MEP}
\BIBentryALTinterwordspacing
``Medical expenditure panel survey,'' 2014. [Online]. Available:
  \url{https://meps.ahrq.gov/mepsweb/}
\BIBentrySTDinterwordspacing

\bibitem{Student-performance}
\BIBentryALTinterwordspacing
``Student performance data set,'' 2014. [Online]. Available:
  \url{https://archive.ics.uci.edu/ml/datasets/Student+Performance}
\BIBentrySTDinterwordspacing

\bibitem{tensorflow2015-whitepaper}
\BIBentryALTinterwordspacing
M.~Abadi, A.~Agarwal, P.~Barham, E.~Brevdo, Z.~Chen, C.~Citro, G.~S. Corrado,
  A.~Davis, J.~Dean, M.~Devin, S.~Ghemawat, I.~Goodfellow, A.~Harp, G.~Irving,
  M.~Isard, Y.~Jia, R.~Jozefowicz, L.~Kaiser, M.~Kudlur, J.~Levenberg,
  D.~Man\'{e}, R.~Monga, S.~Moore, D.~Murray, C.~Olah, M.~Schuster, J.~Shlens,
  B.~Steiner, I.~Sutskever, K.~Talwar, P.~Tucker, V.~Vanhoucke, V.~Vasudevan,
  F.~Vi\'{e}gas, O.~Vinyals, P.~Warden, M.~Wattenberg, M.~Wicke, Y.~Yu, and
  X.~Zheng, ``{TensorFlow}: Large-scale machine learning on heterogeneous
  systems,'' 2015, software available from tensorflow.org. [Online]. Available:
  \url{https://www.tensorflow.org/}
\BIBentrySTDinterwordspacing

\bibitem{scikit-learn}
F.~Pedregosa, G.~Varoquaux, A.~Gramfort, V.~Michel, B.~Thirion, O.~Grisel,
  M.~Blondel, P.~Prettenhofer, R.~Weiss, V.~Dubourg, J.~Vanderplas, A.~Passos,
  D.~Cournapeau, M.~Brucher, M.~Perrot, and E.~Duchesnay, ``Scikit-learn:
  Machine learning in {P}ython,'' \emph{Journal of Machine Learning Research},
  vol.~12, pp. 2825--2830, 2011.

\bibitem{udeshi2018automated}
S.~Udeshi, P.~Arora, and S.~Chattopadhyay, ``Automated directed fairness
  testing,'' in \emph{Proceedings of the 33rd ACM/IEEE International Conference
  on Automated Software Engineering}, 2018, pp. 98--108.

\bibitem{chakraborty2020fairway}
J.~Chakraborty, S.~Majumder, Z.~Yu, and T.~Menzies, ``Fairway: a way to build
  fair ml software,'' in \emph{Proceedings of the 28th ACM Joint Meeting on
  European Software Engineering Conference and Symposium on the Foundations of
  Software Engineering}, 2020, pp. 654--665.

\bibitem{bellamy2019ai}
R.~K. Bellamy, K.~Dey, M.~Hind, S.~C. Hoffman, S.~Houde, K.~Kannan, P.~Lohia,
  J.~Martino, S.~Mehta, A.~Mojsilovi{\'c} \emph{et~al.}, ``Ai fairness 360: An
  extensible toolkit for detecting and mitigating algorithmic bias,'' \emph{IBM
  Journal of Research and Development}, vol.~63, no. 4/5, pp. 4--1, 2019.

\bibitem{gilleland2013software}
E.~Gilleland, M.~Ribatet, and A.~G. Stephenson, ``A software review for extreme
  value analysis,'' \emph{Extremes}, vol.~16, no.~1, pp. 103--119, 2013.

\bibitem{Hess_cliff}
M.~Hess and J.~Kromrey, ``Robust confidence intervals for effect sizes: A
  comparative study of cohen's d and cliff's delta under non-normality and
  heterogeneous variances,'' \emph{Paper Presented at the Annual Meeting of the
  American Educational Research Association}, 01 2004.

\bibitem{6235961}
N.~Mittas and L.~Angelis, ``Ranking and clustering software cost estimation
  models through a multiple comparisons algorithm,'' \emph{IEEE Transactions on
  Software Engineering}, vol.~39, no.~4, pp. 537--551, 2013.

\bibitem{8983215}
F.~Yang, Z.~Yu, Y.~Liang, X.~Gan, K.~Lin, Q.~Zou, and Y.~Zeng, ``Grouped
  correlational generative adversarial networks for discrete electronic health
  records,'' in \emph{2019 IEEE International Conference on Bioinformatics and
  Biomedicine (BIBM)}, 2019, pp. 906--913.

\bibitem{Theis2015ANO}
\BIBentryALTinterwordspacing
L.~Theis, A.~van~den Oord, and M.~Bethge, ``A note on the evaluation of
  generative models,'' \emph{CoRR}, vol. abs/1511.01844, 2015. [Online].
  Available: \url{https://api.semanticscholar.org/CorpusID:2187805}
\BIBentrySTDinterwordspacing

\bibitem{chundawat2022tabsyndex}
V.~S. Chundawat, A.~K. Tarun, M.~Mandal, M.~Lahoti, and P.~Narang, ``Tabsyndex:
  a universal metric for robust evaluation of synthetic tabular data,''
  \emph{arXiv preprint arXiv:2207.05295}, 2022.

\bibitem{FID}
M.~Heusel, H.~Ramsauer, T.~Unterthiner, B.~Nessler, and S.~Hochreiter, ``Gans
  trained by a two time-scale update rule converge to a local nash
  equilibrium,'' in \emph{Proceedings of the 31st International Conference on
  Neural Information Processing Systems}, ser. NIPS'17.\hskip 1em plus 0.5em
  minus 0.4em\relax Red Hook, NY, USA: Curran Associates Inc., 2017, p.
  6629–6640.

\bibitem{SDV}
N.~Patki, R.~Wedge, and K.~Veeramachaneni, ``The synthetic data vault,'' in
  \emph{IEEE International Conference on Data Science and Advanced Analytics
  (DSAA)}, Oct 2016, pp. 399--410.

\bibitem{borg2005modern}
I.~Borg and P.~J. Groenen, \emph{Modern multidimensional scaling: Theory and
  applications}.\hskip 1em plus 0.5em minus 0.4em\relax Springer Science \&
  Business Media, 2005.

\bibitem{agarwal2018automated}
A.~Agarwal, P.~Lohia, S.~Nagar, K.~Dey, and D.~Saha, ``Automated test
  generation to detect individual discrimination in ai models,'' \emph{arXiv
  preprint arXiv:1809.03260}, 2018.

\bibitem{10.1145/3650212.3680380}
\BIBentryALTinterwordspacing
V.~A. Dasu, A.~Kumar, S.~Tizpaz-Niari, and G.~Tan, ``Neufair: Neural network
  fairness repair with dropout,'' ser. ISSTA 2024.\hskip 1em plus 0.5em minus
  0.4em\relax New York, NY, USA: Association for Computing Machinery, 2024, p.
  1541–1553. [Online]. Available:
  \url{https://doi.org/10.1145/3650212.3680380}
\BIBentrySTDinterwordspacing

\bibitem{shekhar2021adaptive}
S.~Shekhar, G.~Fields, M.~Ghavamzadeh, and T.~Javidi, ``Adaptive sampling for
  minimax fair classification,'' \emph{Advances in Neural Information
  Processing Systems}, vol.~34, pp. 24\,535--24\,544, 2021.

\bibitem{ghosh2021characterizing}
A.~Ghosh, L.~Genuit, and M.~Reagan, ``Characterizing intersectional group
  fairness with worst-case comparisons,'' in \emph{Artificial Intelligence
  Diversity, Belonging, Equity, and Inclusion}.\hskip 1em plus 0.5em minus
  0.4em\relax PMLR, 2021, pp. 22--34.

\bibitem{zhang2022adaptive}
M.~Zhang and J.~Sun, ``Adaptive fairness improvement based on causality
  analysis,'' in \emph{Proceedings of the 30th ACM Joint European Software
  Engineering Conference and Symposium on the Foundations of Software
  Engineering}, 2022, pp. 6--17.

\bibitem{chen2024fairness}
Z.~Chen, J.~M. Zhang, F.~Sarro, and M.~Harman, ``Fairness improvement with
  multiple protected attributes: How far are we?'' in \emph{Proceedings of the
  IEEE/ACM 46th International Conference on Software Engineering}, 2024, pp.
  1--13.

\bibitem{10.1145/3644815.3644989}
\BIBentryALTinterwordspacing
S.~Tizpaz-Niari and S.~Sankaranarayanan, ``Worst-case convergence time of ml
  algorithms via extreme value theory,'' in \emph{Proceedings of the IEEE/ACM
  3rd International Conference on AI Engineering - Software Engineering for
  AI}, ser. CAIN '24.\hskip 1em plus 0.5em minus 0.4em\relax New York, NY, USA:
  Association for Computing Machinery, 2024, p. 211–221. [Online]. Available:
  \url{https://doi.org/10.1145/3644815.3644989}
\BIBentrySTDinterwordspacing

\bibitem{piketty2003income}
T.~Piketty and E.~Saez, ``Income inequality in the united states, 1913--1998,''
  \emph{The Quarterly journal of economics}, vol. 118, no.~1, pp. 1--41, 2003.

\bibitem{saez2004income}
E.~Saez, ``Income and wealth concentration in a historical and international
  perspective, uc berkeley and nber, forthcoming in john quigley,'' in
  \emph{Poverty, the Distribution of Income, and Public Policy, A conference in
  honor of Eugene Smolensky}, 2004.

\bibitem{10.1093/cesifo/49.4.479}
\BIBentryALTinterwordspacing
A.~B. Atkinson, ``{Income Inequality in OECD Countries: Data and
  Explanations},'' \emph{CESifo Economic Studies}, vol.~49, no.~4, pp.
  479--513, 12 2003. [Online]. Available:
  \url{https://doi.org/10.1093/cesifo/49.4.479}
\BIBentrySTDinterwordspacing

\bibitem{wang2022fairness}
H.~Wang, ``Fairness metrics for recommender systems,'' in \emph{2022 9th
  international Conference on Wireless Communication and Sensor Networks
  (ICWCSN)}, 2022, pp. 89--92.

\end{thebibliography}
\bibliographystyle{IEEEtran}

\end{document}